\documentclass[pra,twocolumn,amsmath,amssymb,amsthm,superscriptaddress]{revtex4-1}
\usepackage{graphicx,amsmath,relsize,epstopdf,upgreek,color,mathtools,bm,mathptmx,enumitem}
\usepackage{thmtools}
\usepackage{thm-restate}

\usepackage{multirow}
\usepackage[table,xcdraw]{xcolor}
\usepackage{soul}

\newtheorem{theorem}{Theorem}[section]

\newtheorem{lemma}[theorem]{Lemma}

\newcommand{\Haar}[1]{\left<#1\right>_\mathrm{Haar}}
\newcommand{\poly}{\mathrm{poly}}
\newcommand{\polylog}{\mathrm{polylog}}
\newcommand{\ket}[1]{|{#1}\rangle}
\newcommand{\bra}[1]{\langle{#1}|}

\newcommand{\opinner}[3]{\langle #1|#2|#3\rangle}

\newcommand{\rvec}[1]{\pmb{#1}}
\newcommand{\dyadic}[1]{{\bf#1}}
\newcommand{\tr}[1]{\mathrm{tr}\!\left\{#1\right\}}
\newcommand{\Tr}[1]{\mathrm{Tr}\{#1\}}
\newcommand{\D}{\mathrm{d}}
\newcommand{\DF}{d_\mathrm{F}}
\newcommand{\I}{\mathrm{i}}

\newcommand{\E}[1]{\mathrm{e}^{\mbox{\footnotesize$#1$}}}

\newcommand{\VAR}[2]{\mathrm{Var}_{#1}\!\left[#2\right]}
\newcommand{\MEAN}[2]{\langle{#1}\rangle_{#2}}

\newcommand{\Uenc}[2]{U^{(#1)}_{\mathrm{enc}}(#2)}

\newcommand{\fCQ}{f_\mathrm{C/Q}}
\newcommand{\fC}{f_\mathrm{C}}
\newcommand{\fQ}{f_\mathrm{Q}}
\newcommand{\cC}{\rvec{c}_\mathrm{C}}
\newcommand{\cQ}{\rvec{c}_\mathrm{Q}}
\newcommand{\SC}{S_\mathrm{C}}
\newcommand{\SQ}{S_\mathrm{Q}}

\newcommand{\Ngt}{N_\mathrm{gt}}
\newcommand{\Ntp}{N_\mathrm{tp}}
\newcommand{\resrc}{\mathrm{resrc}}
\newcommand{\resrcC}{\mathrm{resrc}_\mathrm{C}}
\newcommand{\resrcQ}{\mathrm{resrc}_\mathrm{Q}}
\newcommand{\cmplx}{\mathrm{resrc}}
\newcommand{\cmplxC}{\mathrm{resrc}_\mathrm{C}}
\newcommand{\cmplxQ}{\mathrm{resrc}_\mathrm{Q}}

\begin{document}

\title{Exponential data encoding for quantum supervised learning}

\author{S.~Shin}
\affiliation{Department of Physics and Astronomy, 
	Seoul National University, 08826 Seoul, South Korea}

\author{Y.~S.~Teo}
\email{yong.siah.teo@gmail.com}
\affiliation{Department of Physics and Astronomy, 
	Seoul National University, 08826 Seoul, South Korea}

\author{H.~Jeong}
\email{h.jeong37@gmail.com}
\affiliation{Department of Physics and Astronomy,
	Seoul National University, 08826 Seoul, South Korea}

\begin{abstract}
  Reliable quantum supervised learning of a multivariate function mapping depends on the expressivity of the corresponding quantum circuit and measurement resources. We introduce exponential-data-encoding strategies that are hardware-efficient and optimal amongst all non-entangling Pauli-encoded schemes, which is sufficient for a quantum circuit to express general functions having very broad Fourier frequency spectra using only exponentially few encoding gates. We show that such an encoding strategy not only reduces the quantum resources, but also exhibits practical resource advantage during training in contrast with known efficient classical strategies when polynomial-depth training circuits are also employed. When computation resources are constrained, we numerically demonstrate that even exponential-data-encoding circuits with single-layer training modules can generally express functions that lie outside the classically-expressible region, thereby supporting the practical benefits of such a resource advantage. Finally, we illustrate the performance of exponential encoding in learning the potential-energy surface of the ethanol molecule and California's housing prices
\end{abstract}

\maketitle

\section {Introduction} In the current noisy intermediate-scale quantum~(NISQ) era~\cite{Preskill2018quantumcomputingin}, a variety of NISQ algorithms have already been proposed to exploit the computing potentials of noisy quantum devices that are currently at our disposal~\cite{Bromley:2020applications,Bharti:2022noisy,Finnila:1994quantum,Kadowaki:1998quantum,Aaronson:2011computational,Aaronson:2011linear-optical,Hamilton:2017gaussian,Trabesinger:2012quantum,Georgescu:2014quantum}. Among them, quantum machine learning (QML)~\cite{Schuld:2015introduction,Schuld:2019quantum,Carleo:2019machine,date2020quantum,Perez-Salinas:2020aa,dutta2021singlequbit,Goto:2021universal} garnered a huge attention in conjunction with the long-standing reputation of classical machine learning~(CML). A natural question arises: ``Under which circumstances is QML advantageous over CML?'' Owing to the apparent stability and widely-regarded success of CML, the quest to search for avenues where QML outperforms CML becomes challenging. Practicality issues aside, the richness of quantum-computing techniques in QML makes them an interesting subject in its own right that deserves deeper exploration~\cite{Schuld.2203.01340}. Many techniques in CML have since been converted to the QML versions~\cite{RebentrostQSVM.2014, cong2019quantum, lloyd2018qgan, Liu2018anomaly, dunjko2016qrl}. Several performance aspects of QML have also been analyzed, such as expressive power \cite{Du:2020aa,Du:2022.expressivity}, generalization properties \cite{Banchi:2021generalization,Caro2021encodingdependent, Huang2021general}, and sample complexity \cite{Arunachalam:2017}.

The goal of ML is to learn a function mapping $f(\rvec{x})$ by training a particular computational model from a dataset $\{\rvec{x}_j\}$. The training accuracy relies on the function space the model generates~(\emph{expressivity}). Classical neural networks, for instance, exhibits a ``universal approximation property'' that ensures the universality for even single-hidden-layer models~\cite{hornik1989multilayer}. By the same token, universality of quantum models has also been investigated~\cite{Goto:2021universal,Perez-Salinas:2020aa}. We shall focus on the important paradigm of \emph{variational quantum machine learning}, where a variational quantum circuit is optimized to approximate unknown $f(\rvec{x})$ using a training dataset. In~\cite{Schuld:2021aa}, QML models are expressed as finite Fourier series of a frequency spectrum determined by the generator eigenvalues of the encoding gates. The authors showed that a sufficiently large circuit and arbitrary observable measurements can approximate any $f(\rvec{x})$ well, as finite Fourier series can approximate $L^2$ functions given sufficiently large frequency and coefficient spans~\cite{Weisz:2012aa}. A QML model may hence be treated as a Fourier-featured linear model~(FFLM)~\cite{Rahimi:2007random,Tancik:2020Fourier}. 

In this article, based on the above \emph{FFLM computed by a variational NISQ circuit}~(QFFLM)~\cite{Biamonte:2021universal,Cerezo:2021variational,Cao:2019quantum,Endo:2021hybrid,McArdle:2020quantum}, we propose a much more efficient quantum (training-)data-encoding strategy that generates an exponentially large frequency spectrum in the number of encoding gates employed. For a fixed Fourier degree $\DF$ (the maximum frequency over all $M$ variables), we show that such an \emph{exponential encoding} requires the least number of encoding gates \mbox{$N=\log_3(2\DF+1)$} \emph{per encoded variable} amongst all non-entangling Pauli-encoded circuits, allowing exponential reduction of quantum resources compared to naive data-reuploading. Using FFLMs for ML, QFFLMs can possess a resource advantage in training compared to their classical counterparts~(CFFLMs). For supervised learning, we derive the criterion $N_\mathrm{gt} < O\!\left(\epsilon K^{M/2}\right)$ for the resource advantage, involving the number of single-qubit and CNOT gates $\Ngt$, FFLM dimension~$K^M$, and quantum-circuit sampling precision~$\epsilon$. When exponential encoding and a hardware-efficient {\it ansatz} are used, we show that this criterion is satisfied with $\Ngt=O(\poly(MN))$. Numerical demonstrations concerning superior expressivity and learning advantage with exponentially-encoded QFFLMs are also discussed.

\section{Optimal Pauli-data-encoded QML model with exponential encoding} 
\begin{figure}[t]
	\centering
	\includegraphics[width = 1\columnwidth]{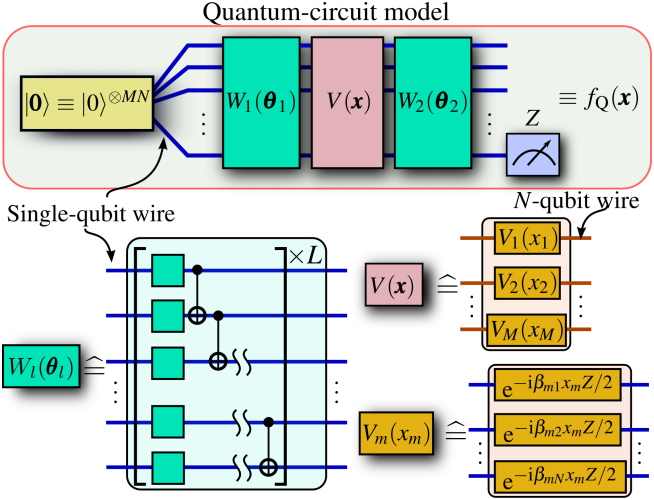}
	\caption{\label{fig:QSL_model}Schematic QML model used to express an $M$-variate bounded $\fQ(\rvec{x})$ [$|\fQ(\rvec{x})|\leq1$]. Each training unitary $W_l(\rvec{\theta}_l)$ comprises $L$ layers of single-qubit~(each carrying three independent training parameters) and the full nearest-neighbor CNOT array. The unitary $V(\rvec{x})$ encodes each element $x_m$ of a training datum $\rvec{x}$ onto an $N$-qubit Pauli-$Z$ gate with prechosen weights $\beta_{m1},\ldots,\beta_{mN}$.}
\end{figure}

The unitary $U_{\rvec{x};\rvec{\theta}}=W_2(\rvec{\theta}_2)V(\rvec{x})W_1(\rvec{\theta}_1)$ acting on an initialized pure state ket $\ket{\rvec{0}}$, followed by a single-qubit Pauli-$Z$ measurement, describes a rather general bounded QML model for large circuit depths~$L$ (elaborated in Fig.~\ref{fig:QSL_model}), defined by training parameters $\rvec{\theta}=(\rvec{\theta}_1^\top\,\,\rvec{\theta}_2^\top)^\top$ and training datum~$\rvec{x}$. By assigning $V(\rvec{x}) = \bigotimes_{m=1}^{M}\bigotimes_{n=1}^{N}\E{-\I\,\beta_{mn}x_m\,Z/2} = \bigotimes_{m=1}^{M}\sum_{\rvec{k}_m\in\{0,1\}^N} \ket{\rvec{k}_m}\E{-\I\,\lambda_{\rvec{k}_m}x_m}\bra{\rvec{k}_m}$ as a diagonal encoding unitary in the $N$-qubit standard basis~$\{\ket{\rvec{k}_m}\equiv\ket{\rvec{k}}\}$, where $\lambda_{\rvec{k}_m}$ are eigenvalues of $\sum_{n=1}^N \beta_{mn}Z_n/2$, the function $\fQ(\rvec{x})$ expressible by the QML model is a finite $M$-variate Fourier series: 
\begin{equation}
	\fQ(\rvec{x})= \bra{\rvec{0}}U_{\rvec{x};\rvec{\theta}}^{\dag}\,Z_N\,U_{\rvec{x};\rvec{\theta}}\ket{\rvec{0}}=\!\!\!\!\!\!\!\!\!\!\sum_{n_1\in \Omega_1,n_2 \in \Omega_2,\ldots, n_M \in \Omega_M} \!\!\!\!\!\!\!\!\!\!\!\!\!\!\!\widetilde{c}_{n_1,n_2,\ldots,n_M}\,\E{-\I\,\rvec{n}\bm{\cdot}\rvec{x}}\,,
	\label{eq:quantum_model}
\end{equation}
where $\widetilde{c}_{n_1,n_2,\ldots,n_M}$ are linear combinations involving $Z_N$, $W_2$ and $D_1=\sum_{\rvec{k}'}\ket{\rvec{k}'}\opinner{\rvec{k}'}{W_1}{\rvec{0}}\bra{\rvec{k}'}$~(see Appendix~\ref{app:deriv_1}). For sufficiently deep $W_l$s and extensive \emph{frequency spectra} $\Omega_m$s, $\fQ(\rvec{x})$ is a finite Fourier series~[$|\fQ(\rvec{x})|\leq1$] that approximates any target function $f(\rvec{x})$~\cite{Schuld:2021aa,Weisz:2012aa} well up to amplitude and period rescalings. Throughout the article, we focus on QML models of the \emph{parallel} kind shown in Fig.~\ref{fig:QSL_model}. As $\Omega_m$s are only determined by the type of encoding generators regardless of their positions in the circuit~\cite{Schuld:2021aa, Caro2021encodingdependent}, all analyses about the $\Omega_m$s extend to different \emph{ansatz} topologies such as serial circuits with alternating trainable and encoding modules.

To analyze how the choices of $\beta_{mn}>0$ affect the extensiveness of $\Omega_m$, we note that  $\lambda_{\rvec{k}_m}=\sum_{n=1}^{N}a^{(\rvec{k}_m)}_{n}$, with $a^{(\rvec{k}_m)}_{n}=\pm\beta_{mn}/2$, are potentially $2^N$ distinct eigenvalues in the absence of degeneracy. Thus, the spectrum $\Omega_m=\{\lambda_{\rvec{k}_m}-\lambda_{\rvec{k}'_m}|\rvec{k}_m,\rvec{k}'_m\in\{0,1\}^N\}$ must contain \emph{at most} $4^N-2^N+1$ distinct frequencies~(including~zero). If we \emph{naively encode} all \mbox{$\beta_{mn}=1$}, then there will only be $2N+1$ distinct values: $\Omega_m = \{-N,-N+1,\ldots,-1,0,1,\ldots,N-1,N\}$; that is, we need $N=\DF$ qubits per encoded variable to express a degree-$\DF$ Fourier series. 

With $N$ qubits, we can, instead, maximize the coverage of $\Omega_m$ by noting that each frequency element is a sum of $N$ numbers picked from the set $\{\beta_{mn},-\beta_{mn},0\}$, giving \emph{at most} $3^N$ distinct frequencies. We show in Appendix~\ref{app:exp_enc} that the
\begin{equation}
\text{\bf exponential encoding scheme}:\quad\beta_{mn} = 3^{n-1}
\label{eq:main1} 
\end{equation}
supply us a dense and exponentially-extensive spectrum $\Omega_m\equiv\Omega_\mathrm{exp}=\{-(3^{N}-1)/2, -(3^{N}-3)/2, \ldots, -1, 0 ,1 , \ldots, (3^{N}-3)/2, (3^{N}-1)/2\}$ for all $1\leq m\leq M$, with $\DF=(3^N-1)/2$. Our first result~\eqref{eq:main1} implies that \mbox{$N=\log_3(2\DF+1)$} qubits is sufficient to realize a degree-$\DF$ Fourier series with this optimal encoding. The limitation to $3^N$ dense and distinct frequencies stems from the fact that Pauli rotations have eigenvalues that differ only in the sign. One may improve the coverage to $O(4^N)$ using nonlocal Hermitian-generator encodings, although hardware feasibility may be called into question~\cite{Caro2021encodingdependent}. 

\section{QFFLMs and resource advantage} As~\eqref{eq:quantum_model} gives rise to a finite Fourier series, we write $\fQ(\rvec{x})=\cQ(\rvec{\theta})\bm{\cdot}\rvec{\phi}(\rvec{x})$, with $\rvec{\phi}(\rvec{x})=\bigotimes^M_{m=1}\rvec{\phi}(x_m)$ and $\rvec{\phi}(x_m)=\sqrt{2}(2^{-1/2},\cos x_m,\sin x_m,\ldots,\cos(n_{(K-1)/2}x_m),\sin(n_{(K-1)/2}x_m))^\top$
is a $K$-dimensional feature column containing all Fourier-basis functions covering a frequency spectrum $\Omega_m$ per variable as training features $(K \equiv 2\DF+1)$. Thus, we immediately recognize that $\fQ(\rvec{x})$ is a QFFLM~\cite{Schuld:2019quantum,Schuld:2021supervised}.

If an $M$-variate function $f(\rvec{x})\cong\rvec{c}\bm{\cdot}\rvec{\phi}(\rvec{x})$ is well-approximated by a Fourier series defined by $\rvec{c}$, we can train a $K^M$-dimensional FFLM either classically~(CFFLM) or with a quantum circuit~(QFFLM) to learn it. With a QFFLM, we may train $U_{\rvec{x};\rvec{\theta}}$ using training datasets $\{\rvec{x}_j,y_j= f(x_j)\}$. This involves minimizing the loss function, say the mean squared-error~(MSE) $\mathcal{L}_{\rvec{\theta}}\propto\sum_j[\fQ(\rvec{x}_j)-y_j]^2$. To clearly analyze the \emph{computational resources} (denoted by $\resrc$) needed in training QFFLMs, we shall consider general gradient-based optimization methods for minimizing $\mathcal{L}_{\rvec{\theta}}$, which entail the computation $\partial_{\theta_k}\mathcal{L}_{\rvec{\theta}}\propto\sum_j[\fQ(\rvec{x}_j) - y_j]\partial_{\theta_k}\fQ(\rvec{x}_j)$ for every $\theta_k$. On NISQ devices, we may carry out the parameter-shift rule~\cite{Mitarai:2018quantum,Schuld:2019evaluating} (see also ) $\partial_{\theta_k}\fQ(\rvec{x}_j)=[\fQ(\rvec{x}_j;\theta_k+\pi/2)-\fQ(\rvec{x}_j;\theta_k-\pi/2)]/2$, so that all $\fQ$s constituting $\partial_{\theta_k}\mathcal{L}_{\rvec{\theta}}$ (evaluated at \emph{different} circuit parameters) are independently sampled from the circuit. 

Suppose that every model function $\fQ(\rvec{x}_j)$ is sampled with some assigned precision $\epsilon$ from $\Ngt$ single-qubit and CNOT gates, then \emph{each $\fQ$ sampling} incurs $O(\Ngt/\epsilon^2)$ gate operations. Tracking all function samplings leads to the overall $\nabla_{\rvec{\theta}}\mathcal{L}_{\rvec{\theta}}$ sampling resources  $\resrcQ=O(\Ngt^2/\epsilon^2)$. We next compare $\resrcQ$ with the resources~$\resrcC$ for computing loss-function gradients with a CFFLM defined as $\fC(\rvec{x})=\cC\bm{\cdot}\rvec{\phi}(\rvec{x})$, which amounts to evaluating dot products. In terms of classical scalar addition and multiplication operations, we show that $\resrcC=\Omega(K^M)$ for known efficient classical strategies~\cite{gudenberg:inria-00074262,Johnson:1982extensions,Griewank:2008,margossianADreview,wu2019efficient,NIPS2014_310ce61c,guo2016quantization,Murphy:2012machine,Pearson:1901lines,Hotelling:1936relations}, which encompass both exact and approximate function computations. Note that while $\resrcC$ depends linearly on $K^M$, $\resrcQ$ relates to quantum-circuit properties and can depend logarithmically in $K^M$. A resource advantage requires $\resrcQ<\resrcC$, or
\begin{equation}
	N_\mathrm{gt} < O\!\left(\epsilon K^{M/2}\right)\,.
	\label{eq:qadvantage}
\end{equation}
That basic operations for both CFFLMs and QFFLMs are equivalent is assumed. Although in practice, basic quantum operations lag classical ones in time, the resulting complexity prefactor is a constant regardless of the model size, and is therefore negligible in \emph{resource-scaling} comparisons. Importantly, we observe that computation bottlenecks during training originate from evaluating $\fQ$ and $\fC$. Hence, \eqref{eq:qadvantage} holds for any $\mathcal{L}_{\rvec{\theta}}$ which takes $f_{Q/C}$ as input, and even for gradient-free optimization methods. All relevant technical details are found in Appendix~\ref{app:resrc_qfflm}.

\section{Resource advantage with exponential encoding}

Care has to be taken in assigning $\epsilon$ for QFFLMs. In computing $\mathcal{L}_{\rvec{\theta}}$, fixing the circuit-sampling repetitions to $O(1/\epsilon^2)$ yields the \emph{same} precision $\epsilon$ for \emph{both} the estimators $\widehat{\fQ(\rvec{x}_j)}$ and $\widehat{\partial_{\theta_k}f_\mathrm{Q}(\rvec{x}_j)}$. On the other hand, their respective \emph{desired} precisions $\epsilon_{f}$ and $\epsilon_{\partial f}$ should \emph{at most} scale with typical orders of $|\fQ(\rvec{x}_j)|$ and $|\partial_{\theta_k}f_\mathrm{Q}(\rvec{x}_j)|$ so that estimations are not mere random guesses~\cite{McClean:2018barren}. This suggests the conservative choice   $\epsilon=\min\{\epsilon_f,\epsilon_{\partial f}\} =\min\left\{\sqrt{\MEAN{|f_\mathrm{Q}(\rvec{x}_j)|^2}{}},\sqrt{\MEAN{|\partial_{\theta_k}f_\mathrm{Q}(\rvec{x}_j)|^2}{}}\right\}$.

When the training parameters $\rvec{\theta}_1$ and $\rvec{\theta}_2$ are randomly initialized to start the minimization of some loss function $\mathcal{L}_{\rvec{\theta}}$, the \emph{barren-plateau phenomenon} refers to the statements $\MEAN{\partial_{\theta_k} \mathcal{L}_{\rvec{\theta}}}{}=0$ (averaged over randomly-chosen $\rvec{\theta}$) and that $\VAR{}{\partial_{\theta_k} \mathcal{L}_{\rvec{\theta}}}$ vanishes exponentially with the size of circuit. Intuitively, the gradient landscape of $\mathcal{L}_{\rvec{\theta}}$ quickly becomes extremely flat with increasing qubit number~\cite{McClean:2018barren,Arrasmith:2021effect,Cerezo:2021cost,Holmes:2022connecting}. When barren plateaus exist, $\min\left\{\sqrt{\MEAN{|f_\mathrm{Q}(\rvec{x}_j)|^2}{}},\sqrt{\MEAN{|\partial_{\theta_k}f_\mathrm{Q}(\rvec{x}_j)|^2}{}}\right\}=O(\alpha^{-MN/2})$ for some \mbox{$\alpha>1$}~\cite{Thanasilp:2021subtleties}, so that $\epsilon=O(\alpha^{-MN/2})$. In our context, since the measurement observable is a Pauli operator, and the {\it ansatz} for $W_l$ in~Fig.~\ref{fig:QSL_model} tends to a two-design with $L=O(\poly (MN))$~\cite{harrow_random_2009,Puchala_Z._Symbolic_2017}, we derive in Appendix~\ref{app:BPP} that $\MEAN{f_\mathrm{Q}^2}{}=O(2^{-{MN}})$ and $\MEAN{(\partial_{\theta_k} f_\mathrm{Q})^2}{}\leq O(2^{-{MN}})$, where $\left<\,\,\bm{\cdot}\,\,\right>$ is an average over $W_l$s. This implies that $\alpha=2$.

When exponential encoding is used in the presence of barren plateaus, $K=3^N$ and criterion \eqref{eq:qadvantage} becomes $N_\mathrm{gt} < O\!\left((3/\alpha)^{MN/2}\right)$, telling us that a feasible resource advantage requires \mbox{$\alpha<3$}. From the previous section, as $\alpha=2$ for QFFLMs of polynomial-depth $W_l$s, an exponentially-encoded circuit of $\Ngt=O(\poly(MN))$ indeed permits a resource advantage as $O(\poly(MN))\ll O\!\left((3/2)^{MN/2}\right)$ for large~$MN$. The existence of such an advantage, even under the influence of barren plateaus, is \emph{only} possible with encodings that generate exponentially-large models.

All $M$-variate FFLM functions $f_\mathrm{model}(\rvec{x})$ are characterized by the convex subspace $C_{K^M}=\{\rvec{c}\,\,|\,\,|\rvec{c}\bm{\cdot}\rvec{\phi}(\rvec{x})|\leq1\text{ for all }\rvec{x}\in[0,2\pi)^M\}$ of $K^M$-dimensional $\rvec{c}$. In the simplest case where $K=3$ and $M=1$, we show in Appendix~\ref{sec:CKM_struct} that the geometry of $C_{3}$ is that of a bicone. Generating \emph{any} $\cQ(\rvec{\theta}) \in C_{K^M}$ with a QFFLM needs at least $K^{M}$ free (training-circuit) parameters~($\rvec{\theta}$). However, when the number of free parameters $\Ntp = O(\Ngt)\geq K^{M}$, QFFLMs holds no resource advantage as criterion~{eq:qadvantage} can never be satisfied. This means that resource-advantageous QFFLMs possess coefficients  $\rvec{c}_{Q}(\rvec{\theta})$ that necessarily cover a smaller subspace $\SQ\subset C_{K^M}$. In this situation, where $\Ntp < K^M$, we say that the model is \emph{underparametrized}, and this is the only situation where QFFLMs can be resource-advantageous. Underparametrized models are thus crucial for practical implementations, especially in realistic scenarios where $\Ngt$ is limited.
 \begin{figure}[t]
	\centering
	\includegraphics[width = 1\columnwidth]{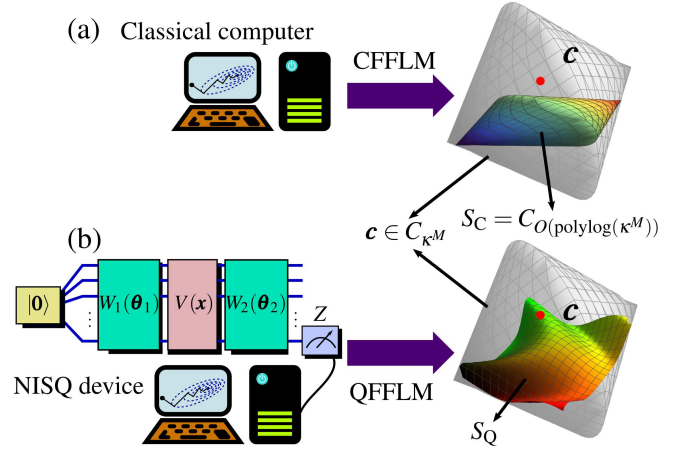}
	\caption{\label{fig:model_geom} Visualization of the full convex space $C_{\kappa^M}$ of bounded $\kappa^M$-dimensional functions. The subspace $S_{C/Q} \subseteq C_{\kappa^M}$ and has a dimension of at most $\Ntp$. (a)~For a fixed $\resrc=O(\polylog(\kappa^M))$, an optimal CFFLM generates $\SC=C_{O(\polylog(\kappa^M))}$. (b)~With the \emph{same} resrc constraint and dense exponential encoding, the QFFLM gives a subspace $\SQ$ that can exceed the boundaries of $C_{O(\polylog(\kappa^M))}$. Therefore, while a classically-expressible $\rvec{c}$ strictly lies in $C_{O(\polylog(\kappa^M))}$, QFFLMs can express $\rvec{c}\notin C_{O(\polylog(\kappa^M))}$.}	
\end{figure}

 \begin{figure}[t]
	\centering
	\includegraphics[width = 1\columnwidth]{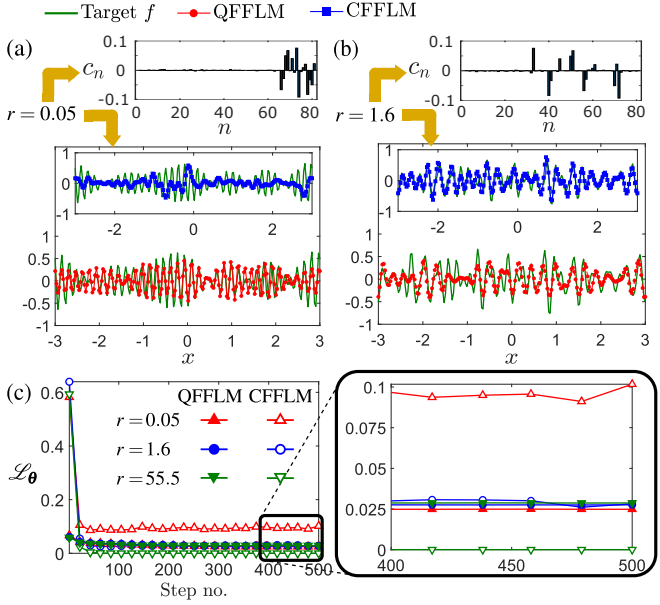}
	\caption{\label{fig:sim}QFFLM~($L=1$) and CFFLM performances [of similar order in computational-resources~($\resrc$)] on learning two univariate functions $-1\leq f(x)\leq1$ of period $[-\pi,\pi)$ different characteristic $r$. (a)~It is clear that the trained QFFLM expressivity for $r=0.05$ is higher than that for the CFFLM~(inset). (b)~For $r=1.6$, where the Fourier coefficients $c_n$ within and outside the classically-expressible region are comparable, so are the expressivities of both models. The $\mathcal{L}_{\rvec{\theta}}$ graphs (averaged over five runs) in~(c) reflect training accuracies for various $r$, where that for the CFFLM at $r=0.05$, in particular, saturates at a much larger value of around 0.08 than its QFFLM counterpart. The CFFLM expressivity is naturally high for $r=55.5$, where $f$ is classically expressible. QFFLM performs similarly for all regimes of $r$, with improved expressivities for larger $L$. Appendix~\ref{sec:last_sec} further states the technical details behind these simulations.}	
\end{figure} 

\section{Practical benefits of underparametrized QFFLMs}
\subsection{Different expressible function spaces for two models} \label{sec:VA} 
The subspace of functions expressible by a Q(C)FFLM is given by $\SQ$~($\SC$) of a dimension \emph{no greater than} $\Ntp^\mathrm{Q(C)FFLM}$, which is the number of free (trainable) parameters, so that underparametrized models always generate subspaces of smaller dimensions than $C_{K^M}$. For example, if subspace $S \subseteq C_3$ is characterized by $\Ntp=1$ free parameter such that  $\rvec{c}(\theta)=2^{-1}(\sin\theta, \cos\theta, \sin\theta\,\cos\theta)^\top$, then $S$ itself is a one-dimensional curve lying in $C_3$. For a QFFLM, the interplay between $\Ntp^\mathrm{QFFLM}$, {\it circuit-ansatz} choice [which fixes the form of $\rvec{c}(\rvec{\theta})$, and consequently $\SQ$], and model dimension~$K^M$ determines if a target $f(\rvec{x})$ can be accurately expressed. 

Suppose that $f(\rvec{x})$ is characterized by a $\rvec{c}\in C_{\kappa^M}$ of dimension $\kappa^M$, which is generally different from $K^M$, the model dimension, and assume that $\kappa^M\gg1$ so that $\resrc$ is restricted to $O(\polylog(\kappa^M))$. Without \emph{a priori} information about $\rvec{c}$, there are two options to do machine learning on $f(\rvec{x})$ with an FFLM. Going by the classical route~[Fig.~\ref{fig:model_geom}(a)], as $\resrcC$ depends on the model dimension, we may construct a CFFLM of dimension $O(\polylog(\kappa^M))$ that is maximally permitted by the resrc constraint. Since any underparametrized CFFLM offers a worse expressivity than the fully-parametrized one, we choose the fully-parametrized CFFLM which utilizes {the same number of free parameters as the dimension---$\Ntp^{\mathrm{CFFLM}} = O(\polylog(\kappa^M))$. Such a fully-parametrized model can express the entire function space $\SC=C_{O(\polylog(\kappa^M))}$ defined by its dimension. Going by the quantum route~[Fig.~\ref{fig:model_geom}(b)], under the \emph{same} $\resrc$ constraint, an $O(\kappa^M)$-dimensional QFFLM can be constructed with $O(\log(\kappa^M))$ encoding gates using the dense exponential encoding strategy in \eqref{eq:main1}. Polylogarithmic resrc constrains $\Ntp^{\mathrm{QFFLM}} = O(\polylog(\kappa^M)\epsilon)$, where $\epsilon$ depends on the circuit \emph{ansatz}. With $L=O(\polylog(\kappa^M))$, $\epsilon \sim \alpha^{-\log_3(\kappa^M)}$~($\alpha > 1$) in view of barren plateaus. 

In the absence of barren plateaus, $\epsilon$ decays as a power law with the model size. Then, the dimensions of $\SQ$ and $\SC$ are both $O(\polylog(\kappa^M))$, and thus comparable in order. However, the QFFLM dimension $O(\kappa^M)$ results in an $O(\Ntp^{\mathrm{QFFLM}})$-dimensional $\SQ$ that can exceed the boundaries of $\SC = C_{O(\polylog(\kappa^M))}$, so that $\rvec{c}\notin C_{O(\polylog(\kappa^M))}$ may still be expressible by the QFFLM~(see Fig.~\ref{fig:model_geom}). 

\begin{figure}[t]
	\centering
	\includegraphics[width = 1\columnwidth]{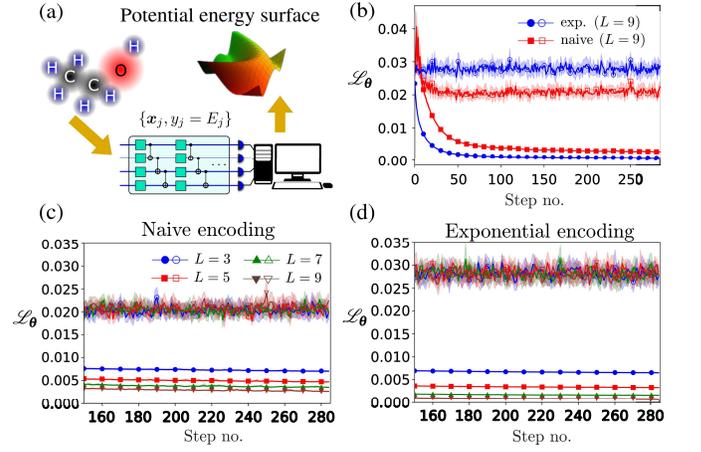}
	\caption{\label{fig:ethanol}(a)~Learning the potential-energy surface with the naively (serial) reuploading~\cite{Perez-Salinas:2020aa} and exponentially-encoded QFFLMs with otherwise identical training settings. Filled and unfilled markers refer to training and test losses. Training and testing shown respectively for (b)~$\Ntp^{\mathrm{QFFLM}}=1134$~($L=9$), and $\Ntp^{\mathrm{QFFLM}}=126L$ using (c)~naive and (d)~exponential models. All graphs are averaged over 10 runs and all corresponding 95\% confidence regions are shown.}
\end{figure}

\subsection{Numerical simulations }
Figure~\ref{fig:sim} supports the arguments of Section \ref{sec:VA} with numerical simulations on learning univariate functions $f(x)$ of various $\rvec{c}$ distributions~($M=1,\kappa=81$). We chose a CFFLM corresponding to a $64$-dimensional model using $\Ntp^{\mathrm{CFFLM}}=64$ parameters. The dimension of this CFFLM is less than $\kappa^M$, and $\SC = C_{64} \subset C_{81}$. We compare this with a shallow $[L=O(\log(MN))]$ 4-qubit exponentially-encoded QFFLM that is barren-plateau-free~\cite{Cerezo:2021cost} and offers a similar $\resrc$ order: this has $\Ntp^{\mathrm{QFFLM}}=16$ parameters and the shallowest possible $W_1$ and $W_2$ of $L=1$, where each single-qubit rotation in $W_l$ is encoded with two training parameters corresponding to $Y$ and $Z$ gates. This QFFLM is of dimension~$81$, identical to that of $f(x)$, but possesses a \emph{16-dimensional} $\SQ \subset C_{81}$ that can exceed $\SC = C_{64}$ in certain boundaries. We define $r=\sqrt{\sum_{n=0}^{63} \vert c_n\vert^2 }/\sqrt{ \sum_{n=64}^{80} \vert c_n\vert^2}$ that measures $f(x)$'s classical-expressibility, where $c_n$ is the target Fourier coefficient labeled with odd indices for cosines and even ones for sines. Figure~\ref{fig:sim} shows that when $r\ll1$, QFFLM with exponential encoding expresses $f$ much more accurately, revealing the limitations of CFFLM whenever $\Ntp^{\mathrm{CFFLM}}<\kappa^M$. Based on this general observation, QFFLMs with exponential encoding are attractive alternatives for learning general $f(\rvec{x})$ with high-frequency Fourier components.

Figure~\ref{fig:ethanol} presents a machine-learning example in physics where high-dimensional QFFLMs with exponential data encoding are applied to express the unknown target function. The \texttt{revised-MD17 dataset}~\cite{Christensen2020,MD17,Rupp:2012fast} is used to train QFFLMs to learn the potential-energy surface of the ethanol molecule based on its atomic positions and nuclear charges, preprocessed into a ($M=36$)-variate function learning problem. To cope with such a large $M$, serial data-reuploading topologies are employed. The naive and exponential underparametrized QFFLMs respectively correspond to dimensions $7^{36}$ and $27^{36}$. Figure~\ref{fig:cali} compares the naive- and exponential-encoding schemes for learning California's housing prices~($M=8$). The quantum circuits employed for this problem take on a different \emph{ansatz}, namely the strongly-entangled-layers \emph{ansatz}~\cite{pennylane}. The Reader may consult Appendix~\ref{app:four_examples} for more details regarding the simulation of these problems.

We see that exponential encoding shows increasingly better training-loss performances as $L$ increases. On the other hand, one observes the contrasted test-loss performances in the two different machine-learning problems. For the potential-energy-surface learning problem, averaging over numerical experiments of various randomized seed values show a consistently higher test losses using exponential encoding. We speculate that this could arise from a much larger model-frequency spectrum relative to that of the actual target function, resulting in aliasing and overfitting~\cite{Peters:2022generalization}. In marked contrast, the test losses for exponential encoding in the housing-price learning problem coincide faithfully with the corresponding training losses, which are much lower than those of naive encoding for $L=3$. This could suggest the existence of other target-function learning problems that do require the assistance of exponential encoding for resource efficiency. Further studies on exponential-model generalizability is an interesting follow-up research that is beyond the scope of this article.

\begin{figure}[t]
	\centering
	\includegraphics[width = \columnwidth]{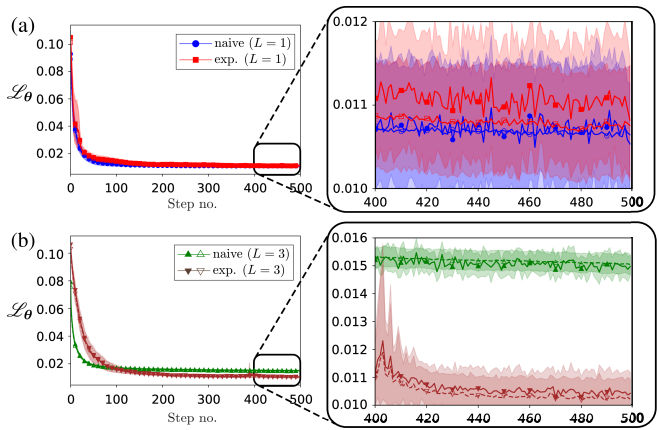}
	\caption{\label{fig:cali}Numerical results for the California Housing Prices dataset. Here, we set $\Ntp^{\mathrm{QFFLM}}=96L$ for (a)~$L = 1$ and (b)~$L = 3$. With increasing $L$, the gap between training and test losses of both types of encoding grows larger. All corresponding 95\% confidence regions over 10 runs are shown.}
\end{figure}

\section{Discussions} We proposed the exponential data encoding scheme for variational quantum machine learning that generates a Fourier-featured linear model of exponentially-large frequency spectra given a small number of encoding gates or qubits. This not only reduces computational resources compared to existing data reuploading models, but also offers a training-resource advantage over their classical counterparts using only polynomial-depth circuits. Exponential encoding is an important element for a quantum resource advantage as frequency spectra that are polynomially large with respect to the qubit or encoding-gate number can be efficiently constructed with classical models resource-wise, as in~Ref.~\cite{classicalsurrogate}. For the same resource order, quantum and classical models exhibit very different function expressivities, such that the former can express functions outside the classically-expressible region.

In Refs.~\cite{Mitarai2018QCL,Caro:2021encoding-dependent}, exponential number of basis functions had been discussed, and possibility for a resource advantage was also hinted. Here, we provided an explicit hardware-efficient methodology to control the presence of Fourier basis functions and show the existence of a resource advantage under physically-realistic constraints. Basis transformations between Fourier and polynomial types are briefly discussed in Appendix~\ref{app:bases}.
	
While exponential encoding allows any quantum supervised-learning model to flexibly acquire an arbitrary Fourier-frequency spectrum just by adjusting the training-data encoding weights, the overall function expressivity still depends on the coverage of the Fourier coefficients, which is limited to the training-gate number and circuit-{\it ansatz} universality. In the NISQ era with limited quantum resources, one is restricted to variational-quantum models with either highly extensive spectra and under-parameterized {\it ans{\"a}tze}, or non-extensive spectra and fully-parametrized {\it ans{\"a}tze}. As potentially advantageous quantum supervised-learning models are underparametrized ones, we believe that deeper studies of such models in terms of their generalizability and expressivity are pertinent to NISQ applications. Furthermore, our work suggests that the exploration of classical Fourier-featured models will help bridge concepts between classical and quantum learning methods, thereby unraveling the latter's hidden potentials.\\

\begin{acknowledgments}
	The authors are grateful for the insightful and beneficial discussions with C.~Oh. This work is supported by HMC, the National Research Foundation of Korea (NRF) grants funded by the Korea government~(Grant Nos.~NRF-2020R1A2C1008609, NRF-2020K2A9A1A06102946, NRF-2019R1A6A1A10073437 and NRF-2022M3E4A1076099) \emph{via} the Institute of Applied Physics at Seoul National University, and the Institute of Information \& Communications Technology Planning \& Evaluation (IITP) grant funded by the Korea government (MSIT) (IITP-2021-0-01059 and IITP-2022-2020-0-01606).
\end{acknowledgments}

\appendix

\begin{figure*}[t]
	\centering
	\includegraphics[width = 1.7\columnwidth]{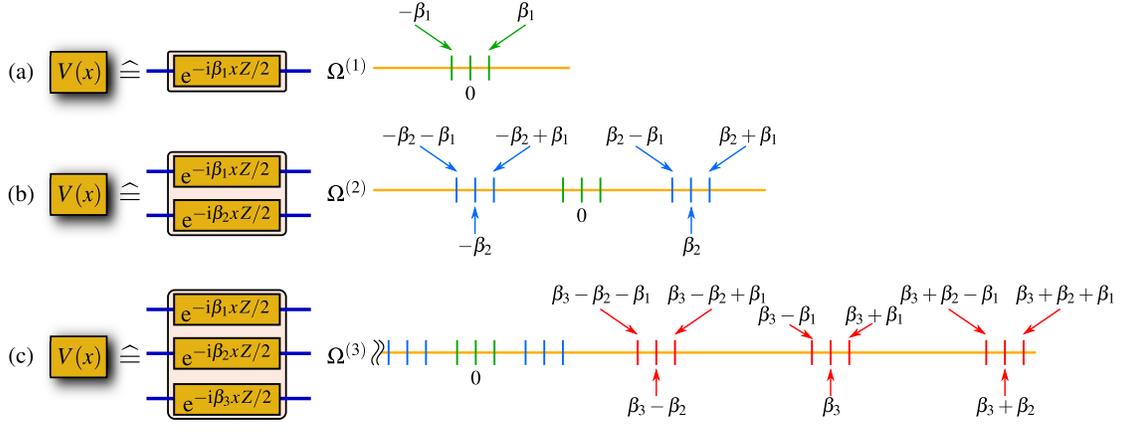}
	\caption{Visualization of the frequency spectra. Elements of the spectrum $\Omega$ are represented on the real line. Adding a $k$th $Z$-encoding gate with weight $\beta_k$ to $V(x)$ corresponding to the spectrum $\Omega^{(k-1)}$ introduces new frequency elements, which are $\{\alpha\pm\beta_k \,\vert\, \alpha \in \Omega^{(k-1)} \}$}
	\label{fig:freq_spect}
\end{figure*}

\section{Exponential encoding schemes}

\subsection{Derivation of Eq.~\eqref{eq:quantum_model}}
\label{app:deriv_1}

Beginning with the $M$-variate encoding unitary operator for $N$-qubit systems,
\begin{equation}
	V(\rvec{x}) = \bigotimes_{m=1}^{M}\bigotimes_{n=1}^{N}\E{-\I\,\beta_{mn}x_m\,Z/2} = \bigotimes_{m=1}^{M}\sum_{\rvec{k}_m\in\{0,1\}^N} \ket{\rvec{k}_m}\E{-\I\,\lambda_{\rvec{k}_m}x_m}\bra{\rvec{k}_m}\,,
\end{equation}
which comes from the product encoding in the single-qubit $Z$-Pauli rotation, the multivariate circuit-unitary action on the initial state ket $\ket{\rvec{0}}$ simplifies to
\begin{widetext}
\begin{align}
	U_{\rvec{x};\rvec{\theta}}\ket{\rvec{0}}=&\,W_2(\rvec{\theta}_2)V(\rvec{x})W_1(\rvec{\theta}_1)\ket{\rvec{0}}\nonumber\\
	=&\,W_2(\rvec{\theta}_2)\sum_{\rvec{k}'_1\in\{0,1\}^N}\ldots\sum_{\rvec{k}'_M\in\{0,1\}^N}\ket{\rvec{k}'_1,\rvec{k}'_2,\ldots,\rvec{k}'_M}\E{-\I\left(\lambda_{\rvec{k}_1'}x_1+\ldots+\lambda_{\rvec{k}_M'}x_M\right)}\bra{\rvec{k}'_1,\rvec{k}'_2,\ldots,\rvec{k}'_M}W_1(\rvec{\theta}_1)\ket{\rvec{0}}\nonumber\\
	=&\,\sum_{\rvec{k}'_1\in\{0,1\}^N}\ldots\sum_{\rvec{k}'_M\in\{0,1\}^N}W_2D_1\ket{\rvec{k}'_1,\rvec{k}'_2,\ldots,\rvec{k}'_M}\E{-\I\sum^M_{m=1}\lambda_{\rvec{k}_m'}x_m}\,,
\end{align}
\end{widetext}
where
\begin{align}
	D_1=&\,\sum_{\rvec{k}'_1\in\{0,1\}^N}\ldots\sum_{\rvec{k}'_M\in\{0,1\}^N}\ket{\rvec{k}'_1,\rvec{k}'_2,\ldots,\rvec{k}'_M}\nonumber\\
	&\,\qquad\qquad\times\bra{\rvec{k}'_1,\rvec{k}'_2,\ldots\rvec{k}'_M}W_1\ket{\rvec{0}}\bra{\rvec{k}'_1,\rvec{k}'_2,\ldots,\rvec{k}'_M}
\end{align}
is diagonal in the multivariate computational basis $\{\ket{\rvec{k}'_1,\rvec{k}'_2,\ldots\rvec{k}'_M}\}$, and all arguments are dropped for the sake of notational simplicity. Therefore
\begin{align}
	\fQ(\rvec{x})=&\,\sum_{\rvec{k}_1,\rvec{k}'_1\in\{0,1\}^N}\ldots\sum_{\rvec{k}_M,\rvec{k}'_M\in\{0,1\}^N}\E{-\I\sum^M_{m=1}\left(\lambda_{\rvec{k}_m'}-\lambda_{\rvec{k}_m}\right)x_m}\nonumber\\
	&\,\qquad\times\bra{\rvec{k}_1,\rvec{k}_2,\ldots,\rvec{k}_M}W_2^\dag D_1^\dag Z_NW_2D_1\ket{\rvec{k}'_1,\rvec{k}'_2,\ldots,\rvec{k}'_M}\,.
\end{align}

The integral differences $\lambda_{\rvec{k}_m'}-\lambda_{\rvec{k}_m}$ constitute the entire frequency spectrum $\Omega_m$ for the $m$th variable $x_m$. The explicit set $\Omega_m$ would then depend on the weights $\beta_{mn}$ attributed to the encoding of each qubit. More generally, we may write
\begin{equation}
	\fQ(\rvec{x})=\sum_{n_1 \in \Omega_1}\sum_{n_2 \in \Omega_2}\ldots\sum_{n_M \in \Omega_M} \widetilde{c}_{n_1,n_2,\ldots,n_M}\,\E{-\I\,\rvec{n}\bm{\cdot}\rvec{x}}
\end{equation}
as a partial Fourier series in terms of the spectra $\Omega_1$, $\Omega_2$, \ldots, $\Omega_M$, where the coefficients $\widetilde{c}_{n_1,n_2,\ldots,n_M}$ are indeed linear combinations of the amplitudes $\bra{\rvec{k}_1,\rvec{k}_2,\ldots,\rvec{k}_M}W_2^\dag D_1^\dag Z_NW_2D_1\ket{\rvec{k}'_1,\rvec{k}'_2,\ldots,\rvec{k}'_M}$ with constituents $\rvec{k}_m,\rvec{k}'_m$ corresponding specifically to the correct integer $n_m$ for all $1\leq m\leq M$ simultaneously.

\begin{figure*}[t]
	\centering
	\includegraphics[width = 2\columnwidth]{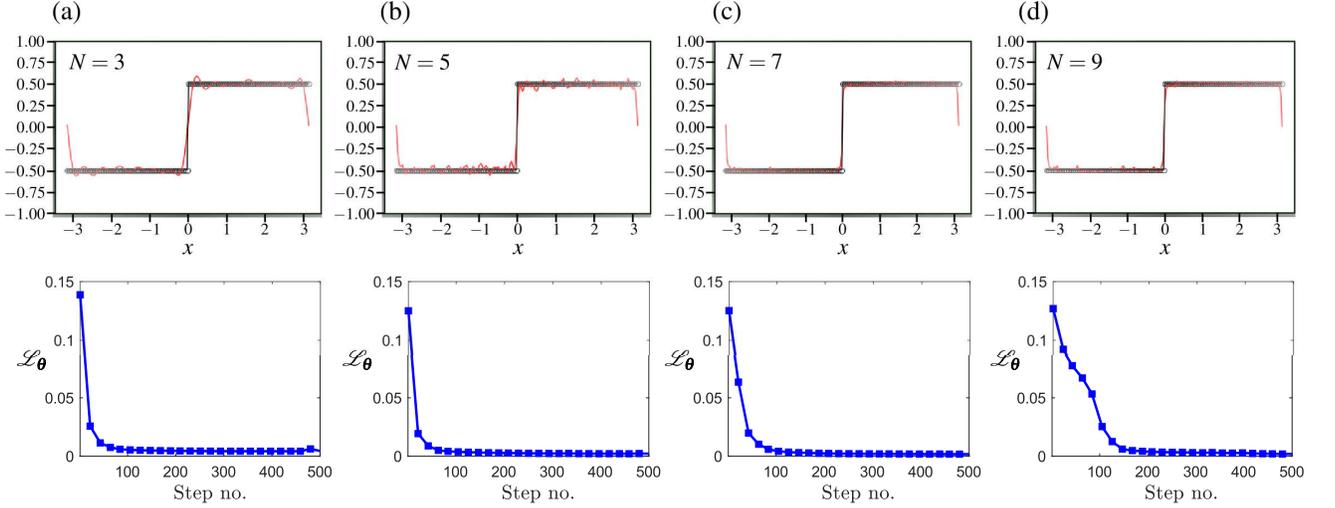}
	\caption{\label{fig:step}Expressivity for $f_{\mathrm{step}}(x)$, where QFFLM predictions~(red curve) are compared with the target function values~(circular markers). In all panels (a)--(d), plots corresponding to the minimized mean-squared-error loss function are shown. We set $L=N+1$ for each trainable unitary $W_l$. QFFLM training is carried out with the Adam gradient optimizer of learning rate~0.05, for 500 iterative steps.}
\end{figure*}

\subsection{Recurrence relation in Eq.~(2)}
\label{app:exp_enc}

We shall investigate the possible frequencies of finite Fourier series that is induced by the QFFLM when choosing different $\beta_{mn}$s. Here, we look for integer-valued $\beta_{mn}$s since we only consider, without loss of generality, the input domain $[-\pi,\pi)^M$. Only the $M=1$ case is necessary to understand the situation, as the \emph{frequency spectra} of $M$-variate Fourier series are just Cartesian products of univariate spectra. We start with only a single Pauli encoding ($N=1$). With $V(x) = \E{-\I\beta_1 x\,Z/2}$ and upon setting $\beta_1 = 1$, we have $\Omega^{(1)}=\{-1, 0 ,1\}$ as the frequency spectrum~[see Fig.~\ref{fig:freq_spect}(a)].

If we, next, append one more encoding gate $\E{-\I \beta_2 x\,Z/2}$ to $V(x)$, we have $\Omega^{(2)}=\{-\beta_2 - 1, -\beta_2 , -\beta_2 + 1, - 1, 0 , 1,\beta_2 - 1, \beta_2, \beta_2 + 1\}$ as illustrated in Fig.~\ref{fig:freq_spect}(b). More generally, let $\Omega^{(k)}$ as the $k$th frequency spectrum as a result of using $k$ encoding gates. Then $\Omega^{(k)}$ has all elements from $\Omega^{(k-1)}$, along with the new ones generated by adding $\pm\beta_k$ to all elements of $\Omega^{(k-1)}$:
\begin{equation}
	\Omega^{(k)} = \{ \Omega^{(k-1)} - \beta_k , \Omega^{(k-1)} , \Omega^{(k-1)} + \beta_k\},
	\label{eq:Omega}
\end{equation}
where $\Omega^{(k-1)} + \beta_k$ denotes the set containing all elements of $\Omega^{(k-1)}$ added to $\beta_k$. As $\Omega^{(k)}$ always contains elements that are symmetrically distributed about zero, to generate a maximally non-degenerate integer-valued frequency spectrum, the inequality
\begin{align}
	\max\left\{\alpha \in \Omega^{(k-1)}\right\}  &< \beta_k - \max\left\{\alpha \in \Omega^{(k-1)}\right\}\nonumber\\
	\text{or}\quad2\,\max\left\{\alpha \in \Omega^{(k-1)}\right\} &< \beta_k
	\label{eq:nondegencondition}
\end{align}
is to be satisfied, where we can easily deduce that $\max\left\{\alpha \in \Omega^{(k-1)}\right\} = \sum_{j=1}^{k-1}\beta_j$.

If we want to construct a dense integer-valued spectrum that starts from 0, the following recursive equation 
\begin{equation}
	2\sum_{j=1}^{k-1}\beta_j + 1 = \beta_k
\end{equation} 
must be satisfied, where setting $\beta_1 = 1$ gives  $\beta_k=3^{k-1}$, giving the dense exponential encoding scheme in the main text. We can also construct maximally non-degenerate frequency spectra by choosing largely-spaced $\beta_k$s satisfying the inequality~\eqref{eq:nondegencondition}. For example, instead of using $\beta_k = 3^{k-1}$, the definition $\beta_k = l^{k-1}$ with $l>3$ would generate a frequency spectrum of cardinality $3^N$ as well. Such a choice, however, would result in a sparse frequency distribution and possibly a larger maximum frequency. 

We can rather flexibly control the elements in $\Omega$ using Eq.~\eqref{eq:Omega}. For example, if we want to set the maximum frequency to some value $\beta^*$, then we may simply choose $N$ $\beta_j$s that satisfy $\sum_{j=1}^{N} \beta_j = \beta^*$. If $\beta^*$ is smaller than $3^N$ and $\beta_j$s are all integers, then the cardinality of the frequency spectrum will be smaller than $3^N$. This leads to a degenerate frequency spectrum that contains repeated integral frequency values, which is equivalent to assigning larger Fourier-coefficient weights to these values. Therefore, if we have some prior knowledge about the target function, such as some intuition about its Fourier frequencies, we can flexibly tweak the weights $\beta_j$ in the QFFLM so that it possesses the desired frequencies, and intentionally introduce degeneracy in its frequency spectrum to enhance the expressive power.

\begin{figure*}[t]
	\centering
	\includegraphics[width = 2\columnwidth]{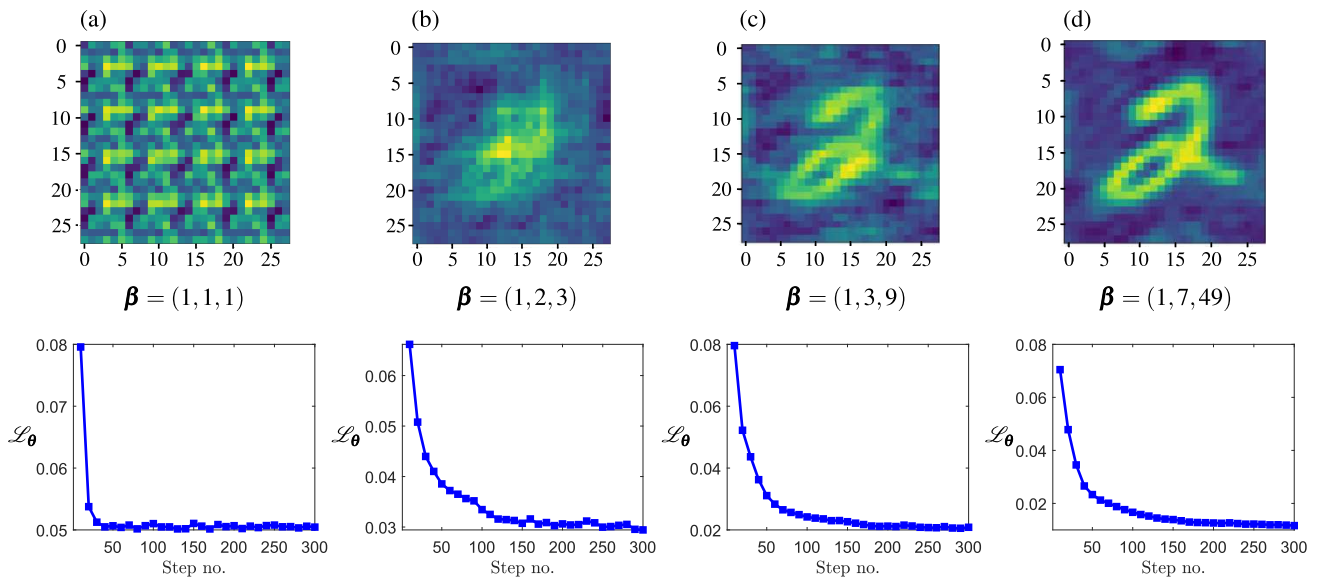}
	\caption{\label{fig:mnist}Quality of 2D image expression of the handwritten digit ``2'', conveyed through the training accuracy in image regression. We consider $L=10$ layers for each trainable unitary $W_1$ and $W_2$, and only vary the training-data encoding scheme. QFFLM training is carried out with the Adam gradient optimizer of learning rate~0.03, for 300 iterative steps. To compare the expressivity performances of QFFLMs with different encoded schemes, we choose the three-tuple of encoded weights $\rvec{\beta}=(\beta_{11},\beta_{12},\beta_{13})=(\beta_{21},\beta_{22},\beta_{23})$ as (a)~$(1,1,1)$, (b)~$(1,2,3)$, (c)~$(1,3,9)$ and (d)~$(1,7,49)$. Reconstructed images~(c) and (d) are predictions of degree $\DF=3^3 = 27$ bivariate Fourier series that are respectively dense and sparse in frequency spectra.}
\end{figure*}

\subsection{Additional examples with (non)dense exponential encoding}
\label{app:four_examples}

Two additional important demonstrations of function expressivity of QFFLMs shall be presented in this subsection. All simulations for QFFLMs are performed with the \texttt{Pennylane} Python library~\footnote{Visit \url{https://pennylane.ai/}.}. The loss function $\mathcal{L}_{\rvec{\theta}}$ chosen to quantify the model training accuracy is the mean squared-error $\mathcal{L}_{\rvec{\theta}}\propto\sum_{j=1}[f_{\mathrm{Q}}(\rvec{x}_j)-y_j]^2$ between the QFFLM predictions $f_{\mathrm{Q}}(\rvec{x}_j)$ and the target outputs $y_j$ for the dataset $\{\rvec{x}_j\}$. Since the training dataset is large and covers the complete function period uniformly, the training accuracy of the QFFLM defined by $\mathcal{L}_{\rvec{\theta}}$ directly equivalates to model expressivity. Hence, lowly-expressive models naturally result in large nonzero $\mathcal{L}_{\rvec{\theta}}$ bias. For the moment, when target functions are unknown, we propose training datasets using a bottom up approach by gradually increasing the number of layers $L$ until both the saturated training and test loss values reach the respective minimum values, whilst keeping all aspects of training constant.

\subsubsection{Step function reconstruction}

We employ the parallel QFFLM possessing a ``hardware-efficient'' \emph{ansatz} for $W_l$, as illustrated in Fig.~\ref{fig:QSL_model}, to investigate other examples of function expression. As the first example, we look at model expressivity for the univariate step function defined as 
\begin{equation}
	f_{\mathrm{step}}(x) = \begin{cases}
		\,\dfrac{1}{2}\, &\text{if } \, 0 \leq x \leq \pi\,, \\
		-\dfrac{1}{2}\, &\text{if } \, -\pi < x < 0\,.
	\end{cases}
\end{equation}
Since expressing plateaued functions requires Fourier-series models of very extensive frequency spectra $\Omega$, exponentially-encoded QFFLMs are ideal for this purpose. In Fig.~\ref{fig:step}, we see that increasing the qubit number quickly improves expressivity.

\subsubsection{Two-dimensional image regression}

We shall now discuss the second example of bivariate function expression. For this, we consider a 2D image of the handwritten digit~``2'' extracted from the MNIST dataset~\footnote{Visit the official MNIST website at \url{http://yann.lecun.com/exdb/mnist/}.} as the target image for demonstrating the expressive power of exponentially-encoded QFFLMs. The target image array has a resolution of $28 \times 28$, where each of the 764 array values has been normalized to have a magnitude bounded by~1. This array, therefore, corresponds to the set of outputs of a bivariate target function $|f(\rvec{x})|\leq1$ that is to be learnt with a QFFLM. We define this bivariate QFFLM using a ``hardware-efficient'' \emph{ansatz} with six qubits (three qubits for each feature variable $x_1$ and $x_2$).

Figure~\ref{fig:mnist} shows the expressivity of various QFFLMs of different $\beta_{mn}$ encoding schemes according to the resulting predicted images upon model training. Both the naive~[Fig.~\ref{fig:mnist}(a)] and linear-step~[Fig.~\ref{fig:mnist}(b)] encoding schemes show poor expressivities. The exponential-encoding schemes corresponding to Figs.~\ref{fig:mnist}(c) and~(d) result in $K^M=(3^3)^2=729$-dimensional QFFLMs. Owing to the predominantly uniform target image-array values representing the image background, which is expressible only by a Fourier series of a large frequency spectrum, it turns out that a nondense exponential-encoding scheme ($\beta_{mn}=7^{n-1}$) gives a much more efficient QFFLM for expressing such handwritten images than the dense one ($\beta_{mn}=3^{n-1}$). Thus, if one has some prior information about the target function, one can control the weights $\beta_{mn}$ to achieve better expressivity.

\subsubsection{Molecular-dynamics dataset of ethanol}

\begin{figure*}[t]
	\centering
	\includegraphics[width = 2\columnwidth]{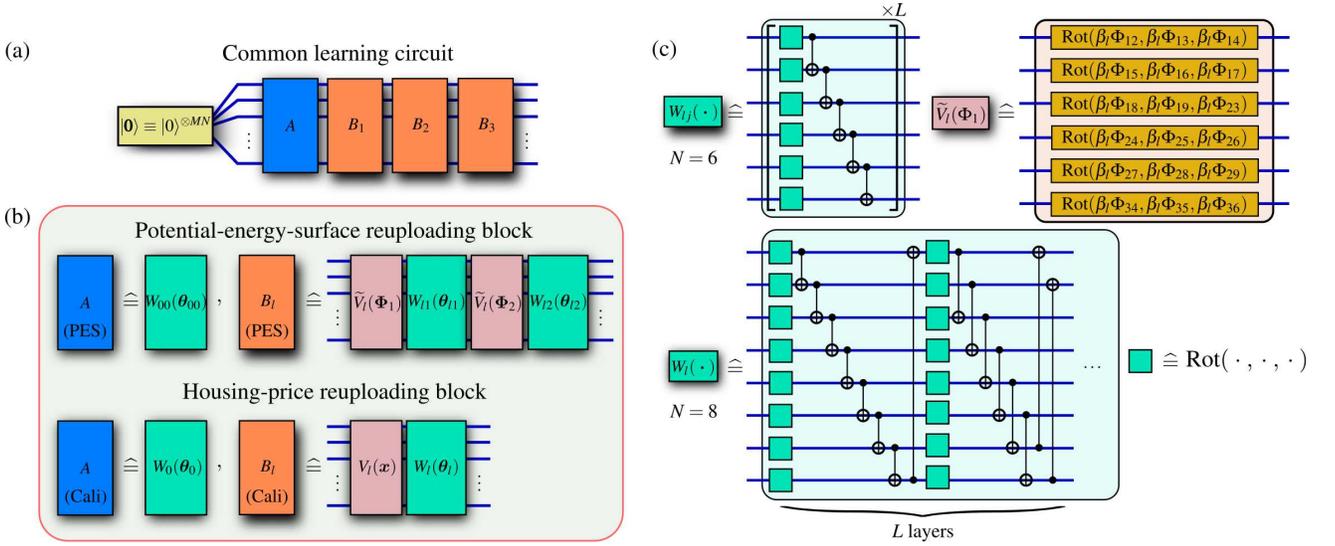}
	\caption{\label{fig:SMcircuits}Circuit diagrams for training with the revised MD17 and California housing-price datasets. (a)~Both circuits share a very similar structure that involves an initial trainable block $A$ and three classical-data encoding (or reuploading) blocks $B_1$, $B_2$ and $B_3$. (b)~The specific details of $A$ and $B_l$ that depend on the task at hand differ by the number of encoding unitary operators ($\widetilde{V}_l$ or $V_l$ depending on which of the two tasks is executed) and trainable modules $W$, (c)~where the latter respectively involve $N=6$ and $N=8$ qubits. The structure of $V_l$ follows that given in Fig.~\ref{fig:QSL_model}. In~(c), the structure of $\widetilde{V}_l(\Phi_1)$ is explicitly shown as an example.}
\end{figure*}

The complete revised-MD17~(rMD17) data bank contains molecular-dynamics training datasets of ten molecules~ \cite{Christensen2020}. The dataset of each molecule consists of 4 types of data, namely the atomic numbers, atomic spatial coordinates, atomic force vectors~(not considered in our context), and the energy scalars. Each dataset allows us to learn the potential-energy surface of the corresponding molecule. As an example, we select the dataset of the ethanol molecule that contains nine atoms, where each input training datum comprises the set of nine atomic numbers $\{Z_j\}^9_{j=1}$ and the set of nine spatial-coordinate columns $\{\rvec{r}_j\}^9_{j=1}$ (each three-dimensional). This is paired with the corresponding energy scalar as the output training datum.

The raw input data are further preprocessed into two-body Coulomb-potential functions. More specifically, given the datum $\{\rvec{r}_j,Z_j\}^9_{j=1}$, we define $ \Phi_{j_1\,\,j_2>j_1} = Z_{j_1}Z_{j_2}/ |\rvec{r}_{j_1} - \rvec{r}_{j_2}|$, where $Z_j$ is the atomic number of the $j$th atom, so that one obtains $M=9\cdot8/2 = 36$ new features per datum. As, we are encoding these features into single-qubit gates, we further normalize each of them such that they are within the interval [$0,2\pi$]. Each input datum is therefore now a 36-dimensional column $\rvec{\Phi}$. This preprocessing procedure suits this learning task since the potential-energy surface originates from Coulomb interactions among the atomic charges~\cite{MD17,Rupp:2012fast}. The output data were also standardized and normalized to be in range $[-1,1]$. We have chosen $2000$ data from the rMD17 ethanol dataset and split them into $1000$ training data and $1000$ test data. Training is done with the full batch using the Adam optimizer and the test loss is computed from randomly-sampled $200$ output test data out of the $1000$. 

For this problem, because of the large number $M=36$ of features, we employed a six-qubit circuit and reuploaded the data three times~[see Fig.~\ref{fig:SMcircuits}]. Single-qubit rotation gates $\mathrm{Rot}(\theta_1,\theta_2,\theta_3) = R_{Z}(\theta_1)R_Y(\theta_2)R_Z(\theta_3) = R_{Z}(\theta_1)F^{\dag}R_Z(\theta_2)F R_Z(\theta_3) $ to encode three features, where $Y = F^{\dag}ZF$ and $R_G$ is the Pauli-$G=X,Y,Z$ rotation gate. Each reuploading is done with a reuploading block $B_{l=1,2,3}$. One such block is composed of two encoding layers (both equal to $V_l$) and two trainable layers ($W_{l1},W_{l2}$) which are interspersed. A trainable layer $W_{lj}$ is in turn composed of $L$ layers that are hardware-efficient. These encoding layers are used to encode a \emph{single} training input datum comprising all $M=36$ features packed into the column $\rvec{\Phi}$ using a total of 12 $\mathrm{Rot}$ gates. If we split this column into two---$\rvec{\Phi}=(\rvec{\Phi}_1^\top\,\,\,\rvec{\Phi}_2^\top)^\top$---each with 18 elements, then one $V_l$ uses six gates encode the first $18$ elements of $\rvec{\Phi}$ ($\rvec{\Phi}_1$), and another $V_l$ uses the other six gates to encode the remaining 18 elements ($\rvec{\Phi}_2$). Such a split encoding is done three times, where each time a different weight $\beta_l$ is used. Remember that each such triple-encoding procedure is carried out on a \emph{single} datum each time, which is to be repeated for all data. 

Therefore, in this serial (triple-reuploading) configuration, we effectively have three encoding gates supplying three different encoding weights $\beta_1$, $\beta_2$ and $\beta_3$ for every feature. The total number of free trainable circuit parameters is $L\times6\times3+3 \times L \times 6 \times 2\times3 = 126 L$.

We compared the performance between naive-encoding and our exponential-encoding models through the aforementioned data-reuploading scheme. Both models differ only by the encoding type, and all other settings are identical. They are trained with the same input/output data and evaluated with exactly the same test batch data. We also tested the performance by increasing $L$ and witnessed a significant enhancement in learning with exponential encoding. We note that naively- and exponential-encoded QFFLMs respectively generate $(2\cdot3+1)^{36}=7^{36}$ and $(3^3)^{36}=27^{36}$ Fourier basis functions, which are very architecturally challenging for CFFLMs to handle.

\subsubsection{California housing-price dataset}

This is one of the basic machine-learning benchmarking dataset given by the \texttt{scikit-learn} Python package. It includes 20640 data, with each input datum being an eight-dimensional vector that stores information about each house and the corresponding output datum recording the house price. The main task is to predict the housing price using an eight-featured model. We train both the naively- and exponentially-encoded QFFLMs with 16512 training data, and evaluate them with the rest as test data. Training is done with randomized batches of 8000 data using the Adam optimizer. Model testing is performed on the full set of test data. All input features are standardized and normalized to $[-\pi,\pi]$, and output data are also normalized to $[0.03,1.0]$. 

Here, we used exactly same \emph{ansatz} as in Ref.~\cite{classicalsurrogate}, where the trainable unitary operators $W_l$ is defined using the \texttt{StronglyEntanglingLayers} Python subpackage from \texttt{Pennylane}~\cite{pennylane}. The Reader may refer to Fig.~\ref{fig:SMcircuits}(c) for a visual representation of the \emph{ansatz}. A total of four trainable unitary modules and three data-encoding unitary modules, which supply a total of $L\times8\times3+3\times L\times8\times3=96L$ free trainable-circuit parameters. As always, both QFFLMs are trained with exactly the same data: only the encoding strategy is different. Here, we again observe a better learning performance with exponential encoding~(see Fig.~\ref{fig:cali}).

\subsection{From trigonometric basis to another}
\label{app:bases}

The trigonometric bases that are inherent to FFLMs are canonical to Fourier-series representations. However, depending on the way training data are encoded, one is free to employ a different basis to express functions. For certain function classes that may be more naturally represented by polynomial functions, there is a reason to invoke a polynomial-type basis instead. If we rescale the (finite) domain of a univariate function $f(x)$ to $-1\leq x\leq 1$, we may make the connection between trigonometric and polynomial functions with the Chebyshev polynomials:
\begin{align}
	\mathrm{T}_n(\cos x) =&\, \cos(nx)\quad\quad\quad\,\,(\text{First kind})\nonumber\,,\\
	\mathrm{U}_n(\cos x)\sin x =&\, \sin((n+1)x)\quad(\text{Second kind})\,.
\end{align}
Then, a straightforward encoding $x\mapsto\cos^{-1}x$ of training data $x$ allows us to map the trigonometric basis $\{1, \cos{x}, \sin{x}, \ldots, \cos{(nx)}, \sin{(nx)}\}$ into the new set of overcomplete basis $\{1, x, \sqrt{1-x^2}, 2x^2-1, x\sqrt{1-x^2}, \ldots, \mathrm{T}_n(x), \mathrm{U}_{n-1}(x)\sqrt{1-x^2}\}$, where the Chebyshev polynomials of each kind are themselves a complete basis. Clearly, if we adopt the exponential-encoding scheme $\beta_n=3^{n-1}$, we can construct an exponentially large polynomial basis for a given number of encoding gates. Similarly, we can also construct an exponentially large nonlinear function basis by using such exponential weights and the encoding $\beta_n \cos^{-1}(g(x))$ of some nonlinear function $g(x)$. Similar kinds of nonlinear data encodings have been used to prove certain universality properties of QML models in Refs.~\cite{Perez-Salinas:2020aa, Goto:2021universal}.

\section{Barren-plateau phenomenon}
\label{app:BPP}

We explicitly derive the corresponding data-encoding-independent barren-plateau statements for the mean squared-error loss function
\begin{equation}
	\mathcal{L}_{\rvec{\theta}}=\int(\D\rvec{x})[\fQ(\rvec{x})-f(\rvec{x})]^2\,,
	\label{eq:MSEcontinuous}
\end{equation}
where for simplicity we shall assume that the training dataset is sufficiently large, so that the discrete average over this dataset conveniently becomes an integral average with respect to the $M$-variate normalized measure $(\D\rvec{x})$ over the complete periods. The lengthy, yet straightforward, calculations may be broken down into several stages.

\subsection{Useful identities}

We generalize the analysis of the barren-plateau phenomenon to any arbitrary $N$-qubit serial circuits containing ${L_{\mathrm{train}}}$ trainable unitary modules $\{W_l\}^{L_{\mathrm{train}}}_{l=1}$ that each have poly($N,2$) circuit depth, so that a randomized $W_l$ for many classes of circuit \emph{ans{\"a}tze} (including those consisting of regularly repeated arrangements of randomized single-qubit and CNOT gates) may be approximated as a two-design~\cite{harrow_random_2009}. Such a generalization covers arbitrary data reuploading cases, where classical-data encoding occurs at multiple instances throughout the quantum circuit. Without loss of generality, numerical examples presented in the main text refer to ${L_{\mathrm{train}}}=2$.

In view of this, the following integration result
\begin{align}
	&\,\int(\D U)_\mathrm{Haar}\, U^*_{j'_1k'_1}U^*_{j'_2k'_2}U_{j_1k_1}U_{j_2k_2}\nonumber\\	=&\,\dfrac{\delta_{j_1,j'_1}\delta_{j_2,j'_2}\delta_{k_1,k'_1}\delta_{k_2,k'_2}+\delta_{j_1,j'_2}\delta_{j_2,j'_1}\delta_{k_1,k'_2}\delta_{k_2,k'_1}}{d^2-1}\nonumber\\
	&\,-\dfrac{\delta_{j_1,j'_1}\delta_{j_2,j'_2}\delta_{k_1,k'_2}\delta_{k_2,k'_1}+\delta_{j_1,j'_2}\delta_{j_2,j'_1}\delta_{k_1,k'_1}\delta_{k_2,k'_2}}{d(d^2-1)}
	\label{eq:Weingarten2}
\end{align}
in terms of the computational matrix elements $U_{jk}=\opinner{j}{U}{k}$ for a $d=2^{N}$-dimensional random unitary operator $U$ distributed according to the Haar measure $(\D U)_\mathrm{Haar}$, and the basic identity
\begin{equation}
	\Haar{U\,O\,U^\dag}=\int(\D U)_\mathrm{Haar}\,U\,O\,U^\dag=\frac{1}{d}\tr{O}\,,
\end{equation}
are crucial~\cite{Puchala_Z._Symbolic_2017} and all we need to derive all statistical statements. By tracking all indices, it is possible to derive another useful integral identity
\begin{align}
	&\,\Haar{U^{\otimes2}\,O\, U^{\dag\,\otimes2}}=\int(\D U)_\mathrm{Haar}\,U^{\otimes2}\,O\, U^{\dag\,\otimes2}\nonumber\\
	=&\,\left[\dfrac{\tr{O}}{d^2-1}-\dfrac{\tr{O\tau}}{d(d^2-1)}\right]1+\left[\dfrac{\tr{O\tau}}{d^2-1}-\dfrac{\tr{O}}{d(d^2-1)}\right]\tau\,,
	\label{eq:HaarV2}
\end{align}
where $\tau$ is the swap operator with the trace property $\tr{O_1\otimes O_2\,\tau}=\tr{O_1O_2}=\tr{U'^{\otimes2}\,O_1\otimes O_2\,U'^{\dag\otimes2}\,\tau}$ for any two observables $O_1$ and $O_2$, and unitary operator $U'$. When $O$ is a Pauli observable, we have
\begin{equation}
	\Haar{U^{\otimes2}\,O^{\otimes2}\, U^{\dag\,\otimes2}}=\dfrac{d\,\tau-1}{d^2-1}\,.
	\label{eq:pauliV2}
\end{equation} 

As we are discussing gradients, the forms of $W_l^{(1)}$ and $W_l^{(2)}$ in the derivative $\partial_{\mu l} W_l=-\frac{\I}{2}\,W^{(1)}_l\sigma_{\mu l} W^{(2)}_l$with respect to the $\mu$th parameter $\theta_{\mu l}$ in the $l$th training module need not fulfill the architectural constraint of a two-design. This is the case when $\theta_{\mu l}$ to which the gradient is taken lies near the edges of the trainable module $W_l$, so that either $W_l^{(1)}$ or $W_l^{(2)}$ are thin. Furthermore, for the majority of parameters in the bulk trainable modules $(1<l<{L_{\mathrm{train}}})$, both $W_l^{(1)}$ and $W_l^{(2)}$ need not possess Haar~first and second moments in order to obtain analytical statements. The subsequent calculations are also made \emph{independent} of the encoding unitary operators.

In subsequent discussions, we shall categorize the barren-plateau statements into three separate cases, namely
\begin{align*}
	\textbf{Case I:} & \,\,\,1<l<{L_{\mathrm{train}}}\,,\\
	\textbf{Case II:} & \,\,\,l=1\,,\\
	\textbf{Case III:} & \,\,\,l={L_{\mathrm{train}}}\,.
\end{align*}

\subsection{A simple manifestation of two-design averages on $\fQ(\rvec{x})$}

By nature of the ``hardware-efficient'' {\it ansatz}, random circuit initialization implies that
\begin{widetext}
\begin{align}
	\Haar{\fQ(\rvec{x})}=&\,\Haar{\opinner{\rvec{0}}{W_1^{\dag}V_1(\rvec{x})^{\dag}\ldots W_{{L_{\mathrm{train}}}-1}^{\dag}V_{{L_{\mathrm{train}}}-1}(\rvec{x})^{\dag} W_{L_{\mathrm{train}}}^{\dag}\,O\, W_{L_{\mathrm{train}}}V_{{L_{\mathrm{train}}}-1}(\rvec{x})W_{{L_{\mathrm{train}}}-1}\ldots V_1(\rvec{x})W_{1}}{\rvec{0}}}=0\,,\nonumber\\
	\Haar{\fQ^2(\rvec{x})}=&\,\Haar{\opinner{\rvec{0}}{W_1^{\dag\otimes2}V_1(\rvec{x})^{\dag\otimes2}\ldots W_{{L_{\mathrm{train}}}-1}^{\dag\otimes2}V_{{L_{\mathrm{train}}}-1}(\rvec{x})^{\dag\otimes2} W_{L_{\mathrm{train}}}^{\dag\otimes2}\,O^{\otimes2}\, W_{L_{\mathrm{train}}}^{\otimes2}V_{{L_{\mathrm{train}}}-1}(\rvec{x})^{\otimes2}\ldots V_1(\rvec{x})^{\otimes2}W_1^{\otimes2}}{\rvec{0}}}=\dfrac{1}{d+1}\,,
\end{align}
\end{widetext}
upon recalling Eq.~\eqref{eq:pauliV2}. We now see a first manifestation of random circuit initialization on two-design approximable circuits, namely $\Haar{\fQ(\rvec{x})}$ and $\Haar{\fQ^2(\rvec{x})}=\VAR{}{\fQ}\sim O(1/d)$. This is a manifestation of the barren-plateau phenomenon. We see that the analogous barren-plateau phenomenon gives rise to similar statements for the loss function $\mathcal{L}_{\rvec{\theta}}$.

\subsection{$\MEAN{\partial_{\mu l} \mathcal{L}_{\rvec{\theta}}}{}=0$ for all cases}

The average of the derivative of $ \mathcal{L}_{\rvec{\theta}}$ with respect to the parameter $\theta_{\mu l}$ corresponding to the $\mu$th qubit for the $l$th trainable module,
\begin{equation}
	\MEAN{\partial_{\mu l} \mathcal{L}_{\rvec{\theta}}}{}=2\int(\D\rvec{x})\,\MEAN{\left[\fQ(\rvec{x})-f(\rvec{x})\right]\partial_{\mu l}\fQ(\rvec{x})}{}\,,
\end{equation}
comprises the average of two terms, namely
\begin{widetext}
	\begin{align}
		\partial_{\mu l}f_\mathrm{Q}(\rvec{x})=&\,\dfrac{\I}{2}\opinner{\rvec{0}}{B(\rvec{x})^\dag W^{(2)\dag}_l\sigma_\mu W^{(1)\dag}_l\Uenc{l}{\rvec{x}}^\dag A(\rvec{x})^\dag O A(\rvec{x})\Uenc{l}{\rvec{x}}W_lB(\rvec{x})}{\rvec{0}}+\mathrm{c.c.}\,,\nonumber\\
		f_\mathrm{Q}(\rvec{x})\,\partial_{\mu l}f_\mathrm{Q}(\rvec{x})=&\,\dfrac{\I}{2}\opinner{\rvec{0}}{B(\rvec{x})^{\dag\otimes2}W^{(2)\dag\otimes2}_l1\otimes\sigma_\mu W^{(1)\dag\otimes2}_l\Uenc{l}{\rvec{x}}^{\dag\otimes2} A(\rvec{x})^{\dag\otimes2} O^{\otimes2} A(\rvec{x})^{\otimes2}\Uenc{l}{\rvec{x}}^{\otimes2}W^{\otimes2}_lB(\rvec{x})^{\otimes2}}{\rvec{0}}+\mathrm{c.c.}\,,
	\end{align}
\end{widetext}
where $A(\rvec{x})=\prod^{l+1}_{l'={L_{\mathrm{train}}}}[\Uenc{l'}{\rvec{x}}W_{l'}]$ and $B(\rvec{x})=\prod^{1}_{l'=l-1}[\Uenc{l'}{\rvec{x}}W_{l'}]$ such that the unitary operator $U_{\rvec{x};\rvec{\theta}}=A(\rvec{x})\Uenc{l}{\rvec{x}}W_lB(\rvec{x})$ represents the complete ${L_{\mathrm{train}}}$-layered quantum circuit. As usual, the arguments $\rvec{\theta}_l$ are suppressed. The average of $\partial_{\mu l}f_\mathrm{Q}(\rvec{x})$ is easiest to treat.

\begin{flushleft}
	\boxed{\MEAN{\partial_{\mu l}f_\mathrm{Q}(\rvec{x})}{}\text{for {\bf Case I}}\text{, and {\bf Case II}}}
\end{flushleft}

As $\Haar{W_{L_{\mathrm{train}}}^{\dag}\Uenc{{L_{\mathrm{train}}}}{\rvec{x}}^{\dag}\, O\,\Uenc{{L_{\mathrm{train}}}}{\rvec{x}}W_{L_{\mathrm{train}}}}=0$ when $O$ is a Pauli operator,
\begin{align}
	\MEAN{\partial_{\mu l}f_\mathrm{Q}(\rvec{x})}{}=0\quad\text{for any $W^{(1)}_l$ and $W^{(2)}_l$}\,.
\end{align}

\begin{flushleft}
	\boxed{\MEAN{\partial_{\mu l}f_\mathrm{Q}(\rvec{x})}{}\text{for {\bf Case III}}}
\end{flushleft}

For parameters in the edge training module $W_{L_{\mathrm{train}}}$ next to $O$,
we inspect the operator 
\begin{equation}
	Q_1=W^{(2)\dag}_{L_{\mathrm{train}}}\sigma_\mu W^{(1)\dag}_{L_{\mathrm{train}}}\Uenc{{L_{\mathrm{train}}}}{\rvec{x}}^{\dag} O\, \Uenc{{L_{\mathrm{train}}}}{\rvec{x}}W^{(1)}_{L_{\mathrm{train}}}W^{(2)}_{L_{\mathrm{train}}}\,.
\end{equation}
Now, note that $\tr{\Uenc{{L_{\mathrm{train}}}-1}{\rvec{x}}^{\dag} Q_1 \Uenc{{L_{\mathrm{train}}}-1}{\rvec{x}}}=\tr{Q_1}=\tr{\sigma_\mu \sigma'(\rvec{x})}$, where 
\begin{equation}
	\sigma'(\rvec{x})= W^{(1)\dag}_{L_{\mathrm{train}}}\Uenc{{L_{\mathrm{train}}}}{\rvec{x}}^{\dag} O\, \Uenc{{L_{\mathrm{train}}}}{\rvec{x}}W^{(1)}_{L_{\mathrm{train}}}
	\label{eq:sigprime}
\end{equation}
is yet another (rotated) Pauli operator parametrized by $\rvec{x}$, so that $\sigma'^2=1$ and $\tr{\sigma'}=0$. As $\tr{\sigma_\mu\sigma'}$ is real,
\begin{align}
	\MEAN{\partial_{\mu l}f_\mathrm{Q}(\rvec{x})}{}=&\,\dfrac{\I}{2}\left<\opinner{\rvec{0}}{B(\rvec{x})^\dag Q_1B(\rvec{x})}{\rvec{0}}\right>	+\,\mathrm{c.c.}\nonumber\\
	=&\,\dfrac{\I}{2d}\left<\tr{\sigma_\mu\sigma'}\right>-\dfrac{\I}{2d}\left<\tr{\sigma_\mu\sigma'}\right>=0\,.
\end{align}

\begin{flushleft}
	\boxed{\MEAN{f_\mathrm{Q}(\rvec{x})\,\partial_{\mu l}f_\mathrm{Q}(\rvec{x})}{}\text{for {\bf Case I}}\text{, and {\bf Case II}}}
\end{flushleft}

Upon taking the average over $W_{{L_{\mathrm{train}}}}$ using Eq.~\eqref{eq:pauliV2}, we have 
\begin{align}
	&\,\MEAN{f_\mathrm{Q}(\rvec{x})\,\partial_{\mu l}f_\mathrm{Q}(\rvec{x})}{}\nonumber\\
	=&\,\dfrac{\I}{2(d+1)}\opinner{\rvec{0}}{B(\rvec{x})^{\dag\otimes2} W^{(2)\dag\otimes2}_l1\otimes\sigma_\mu W^{(2)\otimes2}_lB(\rvec{x})^{\otimes2}}{\rvec{0}}+\mathrm{c.c.}
\end{align}
Since $W^{(2)\dag\otimes2}_l1\otimes\sigma_\mu W^{(2)\otimes2}_l$ is yet another Pauli operator, these resulting expectation values are real regardless of whether $l=1$ or not (that is, whether $B(\rvec{x})=1$ correspondingly or not), so that $\MEAN{f_\mathrm{Q}(\rvec{x})\,\partial_{\mu l}f_\mathrm{Q}(\rvec{x})}{}=0$ for both cases.

\begin{flushleft}
	\boxed{\MEAN{f_\mathrm{Q}(\rvec{x})\,\partial_{\mu l}f_\mathrm{Q}(\rvec{x})}{}\text{for {\bf Case III}}}
\end{flushleft}

For this case, we look at the operator
\begin{align}
	Q_2=&\,W^{(2)\dag\otimes2}_{L_{\mathrm{train}}}\sigma'(\rvec{x})\otimes\sigma_\mu\sigma'(\rvec{x})\,W^{(2)\otimes2}_{L_{\mathrm{train}}}\,,
\end{align}
From the realization that $\sigma'(\rvec{x})$ in Eq.~\eqref{eq:sigprime} is a Pauli operator, the following two trace properties
\begin{align}
	&\,\tr{\Uenc{{L_{\mathrm{train}}}-1}{\rvec{x}}^{\dag\otimes2} Q_2\,\Uenc{{L_{\mathrm{train}}}-1}{\rvec{x}}^{\otimes2}}\nonumber\\
	=&\,\tr{\sigma'(\rvec{x})\otimes\sigma_\mu\sigma'(\rvec{x})}=0\,,\nonumber\\
	&\,\tr{\Uenc{{L_{\mathrm{train}}}-1}{\rvec{x}}^{\dag\otimes2} Q_2\,\Uenc{{L_{\mathrm{train}}}-1}{\rvec{x}}^{\otimes2}\tau}\nonumber\\
	=&\,\tr{\sigma'(\rvec{x})\sigma_\mu\sigma'(\rvec{x})}=\tr{\sigma_\mu}=0
\end{align} 
become apparent. The proofs for $\MEAN{\partial_{\mu l} \mathcal{L}_{\rvec{\theta}}}{}=0$ are therefore complete.

\subsection{$\VAR{}{\partial_{\mu l} \mathcal{L}_{\rvec{\theta}}}\leq O(1/d)$ for all cases}

We hereby show that $\VAR{}{\partial_{\mu l} \mathcal{L}_{\rvec{\theta}}}\lesssim O(1/d)$ for two-design training circuit modules. Using the shorthand $\overline{g(\rvec{x})}\equiv\int\,(\D\rvec{x})\,g(\rvec{x})$ to denote the training-data average, we first make use of the Cauchy--Schwarz inequality to obtain
\begin{align}
	|\partial_{\mu l} \mathcal{L}_{\rvec{\theta}}|^2=&\,4\left|\overline{\left[\fQ(\rvec{x})-f(\rvec{x})\right]\partial_{\mu l}\fQ(\rvec{x})}\right|^2\nonumber\\
	\leq&\, 4\,\mathcal{L}_{\rvec{\theta}}\,\overline{\left|\partial_{\mu l}\fQ(\rvec{x})\right|^2}\leq16\,\overline{\left|\partial_{\mu l}\fQ(\rvec{x})\right|^2}\,,
\end{align}
or
\begin{equation}
	\VAR{}{\partial_{\mu l} \mathcal{L}_{\rvec{\theta}}}\leq 16\,\left<\overline{\left|\partial_{\mu l}\fQ(\rvec{x})\right|^2}\right>\,,
\end{equation}
since $\MEAN{\partial_{\mu l} \mathcal{L}_{\rvec{\theta}}}{}=0$ as demonstrated previously. The crucial quantity is now the average 
\begin{widetext}
	\begin{align}		
		\MEAN{|\partial_{\mu l}f_\mathrm{Q}(\rvec{x})|^2}{}=&\,-\dfrac{1}{4}\opinner{\rvec{0}}{B(\rvec{x})^{\dag\otimes2}W^{(2)\dag\otimes2}_l\Uenc{l}{\rvec{x}}^{\dag\otimes2}\sigma_\mu^{\otimes2} W^{(1)\dag\otimes2}_l A(\rvec{x})^{\dag\otimes2} O^{\otimes2} A(\rvec{x})^{\otimes2}\Uenc{l}{\rvec{x}}^{\otimes2}W^{\otimes2}_lB(\rvec{x})^{\otimes2}}{\rvec{0}}\nonumber\\
		&\,+\dfrac{1}{4}\opinner{\rvec{0}}{B(\rvec{x})^{\dag\otimes2}W^{(2)\dag\otimes2}_l\Uenc{l}{\rvec{x}}^{\dag\otimes2}1\otimes\sigma_\mu W^{(1)\dag\otimes2}_l A(\rvec{x})^{\dag\otimes2} O^{\otimes2}\nonumber\\
			&\,\qquad\qquad\qquad\qquad\qquad\qquad\times A(\rvec{x})^{\otimes2}\Uenc{l}{\rvec{x}}^{\otimes2}W^{(1)\otimes2}_l\sigma_\mu\otimes1W^{(2)\otimes2}_lB(\rvec{x})^{\otimes2}}{\rvec{0}}+\mathrm{c.c.}\,.
	\end{align}
\end{widetext}

\begin{flushleft}
	\boxed{\MEAN{|\partial_{\mu l}f_\mathrm{Q}(\rvec{x})|^2}{}\text{for {\bf Case I}}}
\end{flushleft}

For general bulk modules, averaging over $W_{L_{\mathrm{train}}}$ yields 
\begin{align}		
	&\,\MEAN{|\partial_{\mu l}f_\mathrm{Q}(\rvec{x})|^2}{}\nonumber\\
	=&\,-\dfrac{d}{4(d^2-1)}\left<\opinner{\rvec{0}}{B(\rvec{x})^{\dag\otimes2} W^{(2)\dag\otimes2}_l\sigma_\mu^{\otimes2}W^{(2)\otimes2}_l B(\rvec{x})^{\otimes2}}{\rvec{0}}\right>\nonumber\\
	&\,+\dfrac{d}{4(d^2-1)}+\mathrm{c.c.}\,.
\end{align}
When $l>1$,
\begin{equation}		
	\MEAN{|\partial_{\mu l}f_\mathrm{Q}(\rvec{x})|^2}{\mathrm{I}}=\dfrac{d^2}{2(d+1)(d^2-1)}\,.
\end{equation}

\begin{flushleft}
	\boxed{\MEAN{|\partial_{\mu l}f_\mathrm{Q}(\rvec{x})|^2}{}\text{for {\bf Case II}}}
\end{flushleft}

If $l=1$, we note that 
\begin{align}
	\gamma_\mathrm{II}=&\,\left<\opinner{\rvec{0}}{W^{(2)\dag\otimes2}_1\sigma_\mu^{\otimes2}W^{(2)\otimes2}_1}{\rvec{0}}\right>\nonumber\\
	=&\,\left<\opinner{\rvec{0}}{W^{(2)\dag}_1\sigma_\mu W^{(2)}_1}{\rvec{0}}^2\right>\leq1\,,
	\label{eq:gammaII}
\end{align}
such that
\begin{equation}		
	\MEAN{|\partial_{\mu,1}f_\mathrm{Q}(\rvec{x})|^2}{\mathrm{II}}=\dfrac{d(1-\gamma_\mathrm{II})}{2(d^2-1)}\leq\dfrac{d}{2(d^2-1)}\,.
\end{equation}

\begin{flushleft}
	\boxed{\MEAN{|\partial_{\mu l}f_\mathrm{Q}(\rvec{x})|^2}{}\text{for {\bf Case III}}}
\end{flushleft}

For this case, properties of the operators

\begin{align}
	Q_{3\mathrm{a}}=&\,W^{(2)\dag\otimes2}_{L_{\mathrm{train}}}\sigma_\mu^{\otimes2} \sigma'(\rvec{x})^{\otimes2}\,W^{(2)\otimes2}_{L_{\mathrm{train}}}\,,\nonumber\\
	Q_{3\mathrm{b}}=&\,W^{(2)\dag\otimes2}_{L_{\mathrm{train}}}1\otimes\sigma_\mu \sigma'(\rvec{x})^{\otimes2}\sigma_\mu\otimes1W^{(2)\otimes2}_{L_{\mathrm{train}}}
\end{align}
are necessary, where $\sigma'(\rvec{x})$ is defined in Eq.~\eqref{eq:sigprime}. To start off,
\begin{align}
	\tr{Q_{3\mathrm{a}}}=&\,\tr{\sigma_\mu\sigma'(\rvec{x})}^2=\tr{Q_{3\mathrm{b}}}\,.
\end{align}
For the trace properties with the swap operator, they are
\begin{align}
	&\,\tr{\Uenc{{L_{\mathrm{train}}}-1}{\rvec{x}}^{\dag\otimes2}Q_{3\mathrm{a}}\Uenc{{L_{\mathrm{train}}}-1}{\rvec{x}}^{\otimes2}\tau}\nonumber\\
	=&\,\tr{\sigma_\mu\sigma'(\rvec{x})\sigma_\mu\sigma'(\rvec{x})}\equiv\gamma^{(2)}_\mathrm{III}(\rvec{x})\,,\nonumber\\
	&\,\tr{\Uenc{{L_{\mathrm{train}}}-1}{\rvec{x}}^{\dag\otimes2}Q_{3\mathrm{b}}\Uenc{{L_{\mathrm{train}}}-1}{\rvec{x}}^{\otimes2}\tau}\nonumber\\
	=&\,\tr{\sigma'(\rvec{x})\sigma_\mu \sigma_\mu\sigma'(\rvec{x})}=d\,.
\end{align}
These are critical in evaluating the average over $W_{{L_{\mathrm{train}}}-1}$ by invoking Eq.~\eqref{eq:pauliV2}:
\begin{align}
	\gamma^{(1)}_\mathrm{III}(\rvec{x})\equiv&\,\left<\tr{\sigma_\mu\sigma'(\rvec{x})}^2\right>\,,\nonumber\\	
	\gamma^{(2)}_\mathrm{III}(\rvec{x})\equiv&\,\left<\tr{[\sigma_\mu\sigma'(\rvec{x})]^2}\right>\,,\nonumber\\
	\MEAN{|\partial_{\mu l}f_\mathrm{Q}(\rvec{x})|^2}{\mathrm{III}}
	=&\,-\dfrac{\gamma^{(1)}_\mathrm{III}(\rvec{x})+\gamma^{(2)}_\mathrm{III}(\rvec{x})}{2d(d+1)}+\dfrac{\gamma^{(1)}_\mathrm{III}(\rvec{x})+d}{2d(d+1)}\nonumber\\
	=&\,\dfrac{d-\gamma^{(2)}_\mathrm{III}(\rvec{x})}{2d(d+1)}\leq\dfrac{1}{d+1}\,.
	\label{eq:gammaIII}
\end{align}
The final inequality is obtained from the fact that
\begin{align}
	\tr{[\sigma_\mu\sigma'(\rvec{x})]^2}^2\leq\tr{\sigma_\mu^2}\tr{[\sigma'(\rvec{x})\sigma_\mu\sigma'(\rvec{x})]^2}=d^2\,,
\end{align}
or $-d\leq\gamma^{(2)}_\mathrm{III}(\rvec{x})\leq d$, where we remind the Reader that $\left<\,\,\bm{\cdot}\,\,\right>$ is the average over random $W_{L_{\mathrm{train}}}^{(1)}$s.

Collecting all results, we have
\begin{align}	
	\VAR{}{\partial_{\mu l} \mathcal{L}_{\rvec{\theta}}}\leq&\,
	\begin{cases}
		\dfrac{8d^2}{(d+1)(d^2-1)}	&\mathrm{for}~\textbf{Case I}\,,\\[2ex]
		\dfrac{8d(1-\gamma_\mathrm{II})}{d^2-1}\leq\dfrac{8d}{d^2-1}	&\mathrm{for}~\textbf{Case II}\,,\\[2ex]
		\dfrac{8[d-\overline{\gamma_\mathrm{III}(\rvec{x})}]}{d(d+1)}\leq\dfrac{16}{d+1}	&\mathrm{for}~\textbf{Case III}\,,
	\end{cases}	
\end{align}
$\gamma_\mathrm{II}=\left<\opinner{\rvec{0}}{W^{(2)\dag}_1\sigma_\mu W^{(2)}_1}{\rvec{0}}^2\right>$ and $\gamma_\mathrm{III}(\rvec{x})=\left<\tr{[\sigma_\mu\sigma'(\rvec{x})]^2}\right>$. As special cases, one arrives at $\gamma_\mathrm{II}=1/(d+1)$ if $W_1^{(2)}$ is a two-design, and $\overline{\gamma_\mathrm{III}(\rvec{x})}=-d/(d^2-1)$ if $W_{L_{\mathrm{train}}}^{(1)}$ is a two-design, respectively, where the latter is obtained from the identity
\begin{align}
	&\,\Haar{VAV^\dag BVCV^\dag}=\dfrac{1}{d^2-1}\left(\tr{A}\tr{C}B+\tr{B}\tr{AC}\!\!1\right)\nonumber\\
	&\qquad\qquad\qquad-\dfrac{1}{d(d^2-1)}\left(\tr{A}\tr{B}\tr{C}1+\tr{AC}B\right)
\end{align}
that can be consequently derived from Eq.~\eqref{eq:Weingarten2}. These all give $\VAR{}{\partial_{\mu l} \mathcal{L}_{\rvec{\theta}}}\leq8d^2/[(d+1)(d^2-1)]$ for any case and arbitrary ${L_{\mathrm{train}}}$.

Finally, to obtain the results in the main text for ${L_{\mathrm{train}}}=2$, we simply substitute $N=\DF$ for the naive encoding strategy and $N=\log_3(2\DF+1)$ for the dense exponential encoding strategy.

\section{A resource advantage for the QFFLM}
\label{app:resrc_qfflm}

Our task at hand is to train a given $M$-variate Fourier-featured linear model~(FFLM) using a gradient-based optimization method that minimizes the MSE loss function $\mathcal{L}_{\rvec{\theta}} \propto  \sum_{j}(f_\mathrm{model}(\rvec{x}_j; \rvec{\theta}) - y_j)^2$. Explicitly, an FFLM takes the form
\begin{equation}
	f_\mathrm{model}(\rvec{x}_j ; \rvec{\theta}) = \rvec{c}_\mathrm{model}(\rvec{\theta})\cdot\rvec{\phi}(\rvec{x}_j)\,,
\end{equation}
where the $\Ntp$ trainable parameters are consolidated in $\rvec{\theta} \in \mathbb{R}^{\Ntp}$, and $\rvec{\phi}(\rvec{x})$ is the Fourier feature column. We may choose to train this computational model using either a classical computer or a variational NISQ device shown in Fig.~\ref{eq:quantum_model}. We denote the subscript ``model'' as  ``C'' for the former~(CFFLM), and ``Q'' for the latter~(QFFLM). 

For a CFFLM, its Fourier coefficient column $\rvec{c}_{\mathrm{model}}=\cC$ may generally depend on $\rvec{\theta}$ in a nonlinear fashion. If one parametrizes a $K^M$-dimensional $c_C$ such that it spans the entire convex space $C_{K^M}$ (see Sec.~\ref{sec:CKM_struct} in this SM), we obtain a universal $K^M$-dimensional CFFLM. If one considers a QFFLM, then $\rvec{\theta} \in [-\pi,\pi)^{\Ntp}$ is the parameter column that configures $O(\Ntp)$ trainable quantum gates. The corresponding $\rvec{c}_{\mathrm{model}}=\cQ$ contains elements that are nonlinear functions of $\rvec{\theta}$. Analytical forms of $\cQ$'s elements vary with different \emph{ansatz} used. For example, if the ``hardware-efficient'' \emph{ansatz} shown in Fig.~\ref{fig:QSL_model} of the main text is employed, then by considering $M=1$ for simplicity, we have $c_{\mathrm{Q},l}(\rvec{\theta}) =  \sum_{(\rvec{m},\rvec{n})\in I_l} \chi^l_{(\rvec{m},\rvec{n})} T^{\Ntp}_{(\rvec{m},\rvec{n})}(\rvec{\theta})$ for $1\leq l\leq K$, where $\chi^k_{(\rvec{m},\rvec{n})}$ is the weight of $T^{\Ntp}_{(\rvec{m},\rvec{n})}(\rvec{\theta}) = (\cos{\theta_1})^{m_1}(\sin{\theta_1})^{n_1}\ldots (\cos{\theta_{\Ntp}})^{m_{\Ntp}}(\sin{\theta_{\Ntp}})^{n_{\Ntp}}$, and $I_l$ is the set of nonnegative $(\rvec{m},\rvec{n})$s for the $l$th element $c_{\mathrm{Q},l}$. In the canonical case where each training parameter $\theta_j$ appears once in the quantum circuit, $m_l + n_l = 2$.

Model training involves computation of the MSE loss-function gradient with respect to $\rvec{\theta}$ given a training dataset $\{\rvec{x}_j,y_j\}$:
\begin{equation}
	\dfrac{\partial \mathcal{L}_{\rvec{\theta}}}{\partial \theta_{k}} \propto \sum_j\dfrac{\partial\mathcal{L}_j(\rvec{\theta})}{\partial\theta_k} = \sum_j[f_{\mathrm{model}}(\rvec{x}_j; \rvec{\theta}) - y_j]\dfrac{\partial f_{\mathrm{model}}(\rvec{x}_j;\rvec{\theta})}{\partial \theta_{k}}
	\label{eq:lossgradient}
\end{equation}
We analyze and compare the gradient computational resources~$\cmplx$ by counting the required number of basic computation elements (computational resources) to calculate Eq.~\eqref{eq:lossgradient} using both the CFFLM and QFFLM. The basic computation elements when using a CFFLM are scalar \textit{multiplication}, \textit{addition} and \textit{nonlinear operations}, which we shall consider to all be equivalent resource-wise; that is, computing $x\pm y$, $xy$, $\sqrt{y}$, $\cos(nx)$ or $\sin(nx)$ for scalars $x$ and $y$ amounts to the same resource usage~\cite{Griewank:2008}. We shall also disregard resource count originating from memory allocation, storage and reading, as they can vary significantly with different techniques and memory architectures. For the QFFLM, we take the elementary quantum gates, which are the single-qubit rotation and CNOT gates as basic computation elements.

From hereon, we shall only consider univariate models $f_{\mathrm{model}}(x_j;\rvec{\theta})$ without loss of generality, where the model ($\rvec{c}_{\mathrm{model}}$) dimension is set to $K$. For general $M$-variate models, all subsequent discussions still hold with $K \rightarrow K^M$. 

\subsection{Exact $\mathcal{L}_{\rvec{\theta}}$-gradient computation for a CFFLM}
\label{subsec:cfflm}

Fourier feature mapping [$x_j\mapsto\rvec{\phi}(x_j)$] is first performed \emph{once} on every training datum $x_j$. This preprocessing step requires $K = 2\DF + 1$ (nonlinear) operations to calculate $\{1, \cos x_j , \sin x_j, \ldots, \cos(n_{\DF}x_j), \sin(n_{\DF}x_j) \}$. This results can be stored and reused during the entire training duration. After that, exact computation of $\partial \mathcal{L}_{\rvec{\theta}}/\partial \theta_{k}$ in~\eqref{eq:lossgradient} for a CFFLM using a classical computer may be carried out according to the following steps:\\[1ex]

\noindent
For each $x_j$ and $\theta_k$,
\begin{enumerate}[label=\textbf{\arabic*.}]	
	\item Calculate $\fC(x_j; \rvec{\theta})$ given $\rvec{\theta}$.
	\item Subtract the training output $y_j$ from the calculated $\fC(x_j; \rvec{\theta})$.
	\item Calculate the partial derivative $\partial \fC(x_j; \rvec{\theta})/\partial\theta_k$.
	\item Multiply answers from the second and third steps together.
\end{enumerate}

{\bf Step~1} requires $K$ multiplications and $K$ additions to calculate the inner product $\fC(x_j;\rvec{\theta})=\cC(\rvec{\theta})\bm{\cdot} \rvec{\phi}(x_j)$, amounting to a total of $2K$ operations. If one uses a computationally nontrivial parametrization for $\rvec{c}_\mathrm{C}(\rvec{\theta})$, then additional computational resources $R_\mathrm{I}$ would be needed. {\bf Step~2} only requires one subtraction operation. In {\bf Step~3}, we denote by $R_\mathrm{II}$ the amount of computational resources for the partial derivative,
\begin{equation}
	\dfrac{\partial \fC(x_j;\rvec{\theta})}{\partial \theta_{k}} = \sum^K_{l=1}\phi_l(x_j)\dfrac{\partial c_{\mathrm{C},l}(\rvec{\theta})}{\partial \theta_{k}}\,.
	\label{eq:deriv_cfflm}
\end{equation}
The actual value of $R_\mathrm{II}$ depends on the parametrization of $\cC$. {\bf Step~4} involves just one multiplication operation.

For every single datum $x_j$, {\bf Steps~3} and~{\bf4} are repeated for each component of $\rvec{\theta}\in\mathbb{R}^{\Ntp}$. Therefore, the total number of basic operations $\cmplxC$ for calculating $\partial\mathcal{L}(\rvec{\theta})/\partial\theta_k$ is 
\begin{equation}
	\cmplxC\propto2K + R_{\mathrm{I}} + 1 + \Ntp (R_{\mathrm{II}} + 1)\,,
\end{equation}
with additional $K$ operations from Fourier-feature preprocessing mentioned in the beginning of this subsection. Suppose that, now, $\Ntp=K$ and we simply parametrize $\cC(\rvec{\theta}) = \rvec{\theta} \in \mathbb{R}^K$. Then, the CFFLM may be parametrized into a $K$-dimensional universal model such that \emph{both} $R_\mathrm{I}=0=R_\mathrm{II}$, since $\cC(\bm{\cdot})$ \emph{in this case} is the identity function and $\partial c_{\mathrm{C},l}/\partial\theta_k=\delta_{l,k}$. Hence, for such a CFFLM, $\cmplxC\propto3K+1$, or $\cmplxC=\Omega(K)$.

If $\Ntp < K$, we may reduce the number of repetitions of {\bf Steps~3} and~{\bf4}. In this \textit{under-parametrized} case, but $R_\mathrm{I}\geq0$ since $\cC$ is now a $K$-dimensional (nonlinear) vectorial function of the $\Ntp$ training parameters. The resources $R_{\mathrm{II}}$ for computing $\partial\fC(x_j;\rvec{\theta})/\partial\theta_k$, equivalently its components $\partial c_{\mathrm{C},l}(\rvec{\theta})/\partial\theta_k$, is also generally nonzero. The class of techniques known as \emph{automatic differentiation}~(AD)~\cite{margossianADreview} may be used to compute such a partial derivative. An advantage over conventional finite-difference methods, for instance, is the need for computing only a single function $c_{\mathrm{C},l}(\rvec{\theta})$ per $\theta_k$ instead of a pair of them displaced differently in $\rvec{\theta}$, since AD stores and reuses the computed $c_{\mathrm{C},l}(\rvec{\theta})$s. As a consequence, the number of basic operations for computing every $\partial c_{\mathrm{C},l}(\rvec{\theta})/\partial\theta_k$ is $O(R_{\mathrm{I}})$~\cite{margossianADreview}. Finally, the $2K$ basic operations arising from dot-product computation in {\bf Step~1} still exists. It is now clear that a positive $R_{\mathrm{I}}$ and $R_{\mathrm{II}}$ only increases $\cmplxC$, and the dot-product computation is the bottleneck even for under-parametrized cases. 

In other words, regardless of whether a given CFFLM is under-parametrized or not,
\begin{equation}
	\cmplxC=\Omega(K)\,.
	\label{eq:cmplxC}
\end{equation}

\subsection{Variational $\mathcal{L}_{\rvec{\theta}}$-gradient computation for a QFFLM}
\label{subsec:qfflm}

Analogous to Fourier-feature preprocessing for the CFFLM, preprocessing for QFFLMs involves the multiplication of weights $\beta_n$ to each training data input $x_j$, the output of which is then encoded onto a single-qubit gate. The number of multiplication is the same as that of the encoding gates, which is $\log_3(K)$ using exponential encoding, and never exceeds the total number of basic gates ($\Ngt$). This sets up the ``quantum equivalent'' of Fourier-feature mapping. As with CFFLMs, this data preprocessing is performed only once and the resulting data-encoded gate $V$ is repeatedly used in the QFFLM training. Neglecting the data preprocessing step, the procedure for computing $\partial \mathcal{L}_{\rvec{\theta}}/\partial \theta_{k}$ in \eqref{eq:lossgradient} of a QFFLM is as follows:\\[1ex]

\noindent
For each $x_j$ and $\theta_k$,
\begin{enumerate}[label=\textbf{\arabic*.}]	
	\item Sample $\fQ(x_j; \rvec{\theta})$ from the variational quantum circuit given $\rvec{\theta}$.
	\item Subtract the training output $y_j$ from the sampled $\fQ(x_j; \rvec{\theta})$.
	\item Sample the partial derivative $\partial \fQ(x_j; \rvec{\theta})/\partial\theta_k$ from the variational quantum circuit.
	\item Multiply answers from the second and third steps together.
\end{enumerate}

While {\bf Steps~2} and~{\bf4} are the same as in the case of CFFLMs, the key differences lie in {\bf Steps~1} and~{\bf3}. In {\bf Step~1}, $\fQ(x_j;\rvec{\theta})$ is sampled from the variational NISQ circuit that contains $\Ngt$ basic gates, where $\Ngt$ is completely dependent on our choice of training-circuit \emph{ansatz}, and does not need to scale with the model dimension $K$. In order to sample $\fQ(x_j;\rvec{\theta})$ up to a desired precision $\epsilon_f$, a total of $O(\Ngt/\epsilon_f^2)$ gate operations are needed. In {\bf Step~3}, the partial derivative
\begin{equation}
	\dfrac{\partial \fQ(x_j; \rvec{\theta})}{\partial \theta_k} = \dfrac{1}{2}\left[\fQ\!\left(x_j; \rvec{\theta} + \dfrac{\pi}{2} \rvec{e}_k\right) - \fQ\!\left(x_j; \rvec{\theta} - \dfrac{\pi}{2} \rvec{e}_k\right)\right] 
	\label{eq:paramshift}
\end{equation}
may be defined as a difference of two QFFLM functions of different circuit parameters using the parameter-shift~(PS) rule, where $\rvec{e}_k$ is represented by the unit vector that has $1$ in its $k$th component and $0$s otherwise. So, if $\partial \fQ(x_j; \rvec{\theta})/\partial\theta_k$ is to be sampled up to some desired precision $\epsilon_{\partial f}$, then $2\,O(\Ngt/\epsilon_{\partial f}^2)$ gate operations are needed to sample both $\fQ\!\left(x_j; \rvec{\theta} + \dfrac{\pi}{2} \rvec{e}_k\right)$ and $\fQ\!\left(x_j; \rvec{\theta} - \dfrac{\pi}{2} \rvec{e}_k\right)$ that make up $\partial \fQ(x_j; \rvec{\theta})/\partial\theta_k$ because the desired precision $\epsilon_{\partial f}$ for $\partial \fQ(x_j; \rvec{\theta})/\partial\theta_k$ is the same as the sampling precision of its component QFFLM functions from Eq.~\eqref{eq:paramshift}. Including the additional scalar subtraction and multiplication operations gives a total of $2\,O(\Ngt/\epsilon_{\partial f}^2) + 2 $ basic operations as necessary computational resources. Since for every datum $x_j$, we again need to repeat {\bf Steps~3} and~{\bf4} $\Ntp$ times, the overall amount of $\mathcal{L}_{\rvec{\theta}}$-gradient calculation resources employed reads 
\begin{equation}
	\cmplxQ\propto O\!\left(\dfrac{\Ngt}{\epsilon_f^2}\right) + 1 + \Ntp\!\left[2O\!\left(\dfrac{\Ngt}{\epsilon_{\partial f}^2}\right) + 3\right]\,,
\end{equation}
apart from the additional $\log_3(K)$ data-preprocessing operations. As $\Ntp = O(\Ngt)$, we have $\cmplxQ=O(\Ngt/\epsilon_f^2) + O(\Ngt^2/\epsilon_{\partial f}^2)$. 

We note that $\epsilon_f$ and $\epsilon_{\partial f}$ should not be arbitrarily chosen. In order to sample both $\fQ(x_j;\rvec{\theta})$ and $\partial\fQ(x_j;\rvec{\theta})/\partial\theta_k$ accurately, the desired errors $\epsilon_f$ and $\epsilon_{\partial f}$ should also scale \emph{at most} with these respective magnitudes. It turns out that these requirements would also result in an additive error of $\mathcal{L}_{\rvec{\theta}}$-gradient sampling that also scales at most with the magnitude of the $\mathcal{L}_{\rvec{\theta}}$-gradient. This is obvious from {\bf Step~4}, in which both sampled estimators $\widehat{\fQ(x_j;\rvec{\theta})}\sim\fQ(x_j;\rvec{\theta})+\epsilon_f$ and $\widehat{\partial\fQ(x_j;\rvec{\theta})/\partial\theta_k}\sim\partial\fQ(x_j;\rvec{\theta})/\partial\theta_k+\epsilon_{\partial f}$ have the respective additive errors $\epsilon_f$ and $\epsilon_{\partial f}$ are multiplied together:
\begin{align}
	\widehat{\dfrac{\partial \mathcal{L}_j}{\partial \theta_k}}&=[\widehat{\fQ(x_j;\rvec{\theta})}-y_j]\widehat{\dfrac{\partial \fQ(x_j ; \rvec{\theta})}{\partial \theta_k}} \nonumber\\
	&\sim[\fQ(x_j;\rvec{\theta})+\epsilon_f-y_j]\left[\dfrac{\partial \fQ(x_j ; \rvec{\theta})}{\partial \theta_k}+\epsilon_{\partial f}\right] \nonumber\\
	&\sim \dfrac{\partial \mathcal{L}_j}{\partial \theta_k} + \epsilon_f \dfrac{\partial \fQ(x_j ; \rvec{\theta})}{\partial \theta_k} + \epsilon_{\partial f}\fQ(x_j;\rvec{\theta})\,.
\end{align}
It is evident that if $\epsilon_f = O(\left\vert f(x_j;\rvec{\theta})\right\vert)$ and $\epsilon_{\partial f} = O(\left\vert \partial f(x_j ;\rvec{\theta})/\partial \theta_k \right\vert)$, then $\widehat{\partial \mathcal{L}_j/\partial\theta_k}\sim\partial \mathcal{L}_j/\partial\theta_k+O(\left\vert f(x_j;\rvec{\theta}) \partial f(x_j ;\rvec{\theta})/\partial \theta_k \right\vert)$. 

Since $\epsilon_f$ and $\epsilon_{\partial f}$ are respectively the model-function sampling precisions for estimating $\fQ(x_j;\rvec{\theta})$ and $\partial\fQ(x_j;\rvec{\theta})/\partial\theta_k$, we may take the more conservative route and replace $\epsilon_f,\epsilon_{\partial f}\rightarrow\epsilon=\min\{\epsilon_f,\epsilon_{\partial f}\}$. Finally, the overall gradient-calculation resources for $\mathcal{L}_{\rvec{\theta}}$ is therefore
\begin{equation}
	\cmplxQ=O\!\left(\dfrac{\Ngt^2}{\epsilon^2}\right)\,.
	\label{eq:cmplxQ}
\end{equation}
Comparing Eqs.~\eqref{eq:cmplxC} and \eqref{eq:cmplxQ} here gives Eq.~\eqref{eq:qadvantage}.

\subsection{Approximate $\mathcal{L}_{\rvec{\theta}}$-gradient computation for a CFFLM}

Given the approximate nature of QFFLM computation that is intrinsic to circuit sampling, we further investigate whether $\cmplxC$ can be improved if known approximation techniques are employed in $\mathcal{L}_{\rvec{\theta}}$-gradient computation. For a CFFLM, one may first consider approximating \emph{the form} of $\partial\fC(x_j;\rvec{\theta})/\partial\theta_k$ with the finite-difference method. However, doing so gives no reduction in computational resource scaling than AD~\cite{margossianADreview} that is utilized in the previous subsection. 

Therefore, we focus on approximating the calculation of vectorial inner products. In the literature, there do exist published works that discuss classical algorithms concerning high-dimensional inner-product search problems~\cite{wu2019efficient,NIPS2014_310ce61c,guo2016quantization}. These studies often rely on numerical routines related to the maximum inner product search~(MIPS) problem of maximizing the inner product between a query vector and a set of search data vectors. However, extensive studies were performed to enhance the computation rate of such search problems, and do not directly address the resources utilized for computing inner products themselves. 

Thus far, it appears that the only viable solution to reducing the $\cmplxC=\Omega(K)$ scaling is to perform inner products between $K$-dimensional $\rvec{\phi}(x_j)$s and $\cC$s using a projection method that projects these vectors onto an effective vector space of a smaller dimension. As usual, the $\mathcal{L}_{\rvec{\theta}}$-gradient computational resources using such a projection method shall neglect the data preprocessing step that computes $\rvec{\phi}(x_j)$ for all $x_j$ in the training set $X$ as described in Sec.~\ref{subsec:cfflm}. We highlight two popular projection methods that are each based on different numerical objectives.

\subsubsection{Random projection method}

We first discuss the \emph{random projection method}, where a randomized linear mapping $\mathcal{M}[\rvec{\phi}(x_j)]$ projects all $\rvec{\phi}(x_j)$s onto a $(\widetilde{d}<K)$-dimensional column vectors. A straightforward way to generate such a map is to define $\mathcal{M}[\rvec{\phi}(x_j)]=\widetilde{d}^{-1/2}\dyadic{A}\rvec{\phi}(x_j)\equiv\widetilde{\rvec{\phi}}(x_j)$, where the $\widetilde{d}\times K$ random matrix $\dyadic{A}$ has elements independently and identically distributed according to the standard Gaussian distribution. It is well-known that such random projections preserve the mutual distances between feature vectors $\rvec{\phi}(x_j)$ according to the following lemma~\cite{Johnson:1982extensions,JLnotes} that is satisfied by such a random linear map:

\begin{lemma}
	\label{lem:JL}
	\textit{[Johnson--Lindenstrauss~(JL)]} Given $0 < \widetilde{\epsilon} < 1$, the set $X$ of $\vert X \vert$ vectors in $\mathbb{R}^K$ and the effective dimension $\widetilde{d} \geq O(\log(\vert X \vert)/\widetilde{\epsilon}^2)$, there exists a linear map $\mathcal{M}: \mathbb{R}^K \rightarrow \mathbb{R}^{\widetilde{d}}$ such that 
	$$(1-\widetilde{\epsilon})\Vert \rvec{y}_1 - \rvec{y}_2 \Vert^2 \leq \Vert \mathcal{M}[\rvec{y}_1] - \mathcal{M}[\rvec{y}_2] \Vert^2 \leq (1+\widetilde{\epsilon})\Vert \rvec{y}_1 - \rvec{y}_2\Vert^2 $$
	for all $\rvec{y}_1,\rvec{y}_2\in X$.
\end{lemma}
\noindent
Additional mandatory \emph{precomputation steps} are needed to set up the random projection method before actual CFFLM training commences. These include constructing the $ \widetilde{d}\times K$ random matrix $\dyadic{A}$, which takes $O(K\widetilde{d})$ resources if we treat the generation of \emph{one} Gaussian random variable as \emph{one} basic computational operation, and multiplying $\dyadic{A}$ to all $\rvec{\phi}(x_j)$s, which demands $O(\vert X\vert K\widetilde{d})$ resources. Furthermore, since the Nyquist--Shannon theorem implies that a target function well-approximated by a $K$-dimensional FFLM of largest Fourier frequency $\DF=2K-1$ requires at least $O(K)$ equidistant training data points $\{x_j\}$ to avoid aliasing problems in expressing the function, we require $|X|=O(K)$. Therefore $\widetilde{d}=\Omega((\log K)/\widetilde{\epsilon}^2)$ is achieved with an inner-product complexity of $\Omega((\log K)/\widetilde{\epsilon}^2)$. 

Thus, a randomly-projected CFFLM is still a $K$-dimensional Fourier-featured model, 
\begin{equation}
	\fC^{\text{(rand-proj)}}(x_j) = \widetilde{\cC}^\top \widetilde{\rvec{\phi}}(x_j) = \widetilde{d}^{-1/2}\,\widetilde{\cC}^\top \dyadic{A}\rvec{\phi}(x_j)=(\widetilde{d}^{-1/2}\dyadic{A}^\top\widetilde{\cC})^\top\rvec{\phi}(x_j)\,,
\end{equation}
which is characterized by a $\widetilde{\cC}$ that is $[\widetilde{d}=\Omega((\log K)/\widetilde{\epsilon}^2)]$-dimensional. In other words, the dot-product computation resources may be reduced from $O(K)$ to $\Omega((\log K)/\widetilde{\epsilon}^2)$, \emph{provided} that precomputation steps of complexity $O(K^2(\log K)/\widetilde{\epsilon}^2)$ are carried out prior to CFFLM training. 

\subsubsection{Principal-component-analysis projection}

One may choose to perform a different kind of projection that is popular in machine learning. As there exist a few objectives that eventually lead to the same projection algorithm, we shall quote one exemplifying objective that is commonly considered. Suppose $\widetilde{\rvec{\phi}}(x_j)$ are the $(\widetilde{d}<K)$-dimensional projected feature column vectors of $\rvec{\phi}(x_j)$, and $\dyadic{B}$ is a $K\times\widetilde{d}$ approximate recovery matrix that gives the set of $K$-dimensional columns $\{\rvec{\phi}'(x_j)=\dyadic{B}\,\widetilde{\rvec{\phi}}(x_j)\}$. Henceforth, we shall assume that $\dyadic{B}$ is an isometry ($\dyadic{B}^\top\dyadic{B}=\dyadic{1}$). Then a reasonable prescription for defining the projected feature columns could be one which minimizes the average squared-error
\begin{equation}
	\mathcal{E}=\dfrac{1}{|X|}\sum_{x_j\in X}\Vert\rvec{\phi}(x_j)-\rvec{\phi}'(x_j)\Vert^2\,.
	\label{eq:E1}
\end{equation}
Setting the variation
\begin{equation}
	\updelta\mathcal{E}=\dfrac{1}{|X|}\sum_{x_j\in X}[\rvec{\phi}(x_j)-\rvec{\phi}'(x_j)]^\top\dyadic{B}\,\updelta\widetilde{\rvec{\phi}}(x_j)
\end{equation} 
with respect to $\widetilde{\rvec{\phi}}(x_j)$ to zero supplies us the extremal equation
\begin{equation}
	\dyadic{B}^\top\,\rvec{\phi}(x_j)=\dyadic{B}^\top\,\rvec{\phi}'(x_j)\,,
\end{equation}
implying that the optimal projected feature columns are given by $\widetilde{\rvec{\phi}}(x_j)=\dyadic{B}^\top\rvec{\phi}(x_j)$.

The next task would be to minimize 
\begin{equation}
	\mathcal{E}=\dfrac{1}{|X|}\sum_{x_j\in X}\left\|(\dyadic{1}-\dyadic{B}\,\dyadic{B}^\top)\,\rvec{\phi}(x_j)\right\|^2=K-\Tr{\bm{\Sigma}\,\dyadic{B}\,\dyadic{B}^\top}
	\label{eq:E2}
\end{equation}
with respect to all isometries $\dyadic{B}$, where $\bm{\Sigma}=\sum_{x_j\in X}\rvec{\phi}(x_j)\rvec{\phi}(x_j)^\top/|X|$ and a simplification to the second equality makes use of the fact that $\dyadic{1}-\dyadic{B}\,\dyadic{B}^\top$ is a projector and $\left\|\rvec{\phi}(x_j)\right\|^2=K$. By assigning a Lagrange matrix $\bm{\Lambda}$ for the isometry constraint, the relevant Lagrange function reads
\begin{equation}
	\mathcal{D}=K-\Tr{\bm{\Sigma}\,\dyadic{B}\,\dyadic{B}^\top}+\Tr{\bm{\Lambda}\,(\dyadic{B}^\top\dyadic{B}-\dyadic{1})}\,.
\end{equation}
Setting the variation of $\mathcal{D}$ with respect to $\dyadic{B}$ to zero gives the extremal equation
\begin{equation}
	\dyadic{B}^\top\,\bm{\Sigma}=\bm{\Lambda}\,\dyadic{B}^\top\,,
\end{equation}
upon solving which for $\bm{\Lambda}$ finally leads to the eigenmatrix equation
\begin{equation}
	\dyadic{B}^\top\,\bm{\Sigma}=\dyadic{B}^\top\,\bm{\Sigma}\,\dyadic{B}\,\dyadic{B}^\top\,.
\end{equation}
The solution of $\dyadic{B}$ to this equation is then $\dyadic{B}=\left(\rvec{s}_1\,\,\rvec{s}_2\,\,\ldots\,\,\rvec{s}_{\widetilde{d}}\right)$, where the $\rvec{s}_j$ are $\widetilde{d}$ eigenvectors of $\bm{\Sigma}$.

Therefore, the complete recipe to construct the optimal $\widetilde{d}$-dimensional projection column vectors $\widetilde{\rvec{\phi}}(x_j)$ that minimizes $\mathcal{E}$ in either \eqref{eq:E1} or \eqref{eq:E2} is to search for the projection matrix $\dyadic{B}^\top$ that houses $\widetilde{d}$ eigenvectors of $\bm{\Sigma}$ corresponding to its $\widetilde{d}$ \emph{largest} eigenvalues as its rows. This is a version of the so-called principal-component-analysis~(PCA) projection~\cite{Pearson:1901lines,Hotelling:1936relations,Murphy:2012machine} that applies to our context, which preserves the approximately recovered columns $\rvec{\phi}'(x_j)$ with respect to $\rvec{\phi}(x_j)$ \emph{via} the minimization of $\mathcal{E}$. The result is also an under-parametrized $K$-dimensional CFFLM:
\begin{equation}
	\fC^{\text{(PCA-proj)}}(x_j) = \widetilde{\cC}^\top \widetilde{\rvec{\phi}}(x_j) = \widetilde{\cC}^\top \dyadic{B}^\top\rvec{\phi}(x_j)=(\dyadic{B}\,\widetilde{\cC})^\top\rvec{\phi}(x_j)\,.
\end{equation} 

In terms of computational resources, there are still the necessary precomputation steps one needs to carry out. For the PCA projection, this includes the computation of $\bm{\Sigma}$ that incurs $O(|X|K^2)=O(K^3)$ as $|X|=O(K)$ from the Nyquist--Shannon theorem, the computation of its $\widetilde{d}$ largest eigenvalues that takes a worst-case complexity of $O(K^3)$, and the multiplication of $\dyadic{B}^\top$ with $|X|$ $\rvec{\phi}(x_j)$s [$O(|X|K\widetilde{d})=O(K^2\widetilde{d})$]. Hence, we see that $O(K^3)$ is the bottleneck for the precomputation steps.

\subsubsection{Overall dot-product computational resources}

For both random and PCA projections, it is possible to perform dot-product calculations using a set of projected $\widetilde{\rvec{\phi}}(x_j)$s in place of the original $\rvec{\phi}(x_j)$s, which may be of a much smaller dimension than $K$. By disregarding the precomputation steps, one may claim a reduction in dot-product computational resources with such projection methods. However, care has to be taken in comparing computational resources between CFFLMs and QFFLMs. Strictly speaking, only data preprocessing steps that map the training inputs $x_j$ to the Fourier feature columns $\rvec{\phi}(x_j)$ are common to these two different models, and may therefore be disregarded in the computational-resource comparisons. Otherwise, for a fair comparison between CFFLMs and QFFLMs, \emph{any} additional precomputation complexities must be accounted for in analyzing any method applied to these models. In view of this, there is really no advantage in using projection methods to approximate dot-product computations.

\subsection{General loss functions and non-gradient-based optimization}

Here we consider general loss functions of the form $\mathcal{L}_{\rvec{\theta}}\propto\sum_j\mathcal{L}[\fCQ(x_j;\rvec{\theta})]$ given a training dataset $\{x_j,y_j\}$, where $\fCQ$ refers to the model function from either a CFFLM or QFFLM. Its partial derivative with respect to $\theta_k$ reads 
\begin{align}
	\dfrac{\partial \mathcal{L}_{\rvec{\theta}}}{\partial \theta_k} \propto&\, \sum_j\dfrac{\partial \mathcal{L}[\fCQ(x_j ; \rvec{\theta})]}{\partial \fCQ(x_j ; \rvec{\theta})} \dfrac{\partial \fCQ(x_j ; \rvec{\theta})}{\partial \theta_k}\nonumber\\
	=&\,\sum_j \mathcal{F}[\fCQ(x_j ; \rvec{\theta})]\dfrac{\partial \fCQ(x_j ; \rvec{\theta})}{\partial \theta_k}\,,
	\label{eq:generalloss}
\end{align}
where $\mathcal{F}=\mathcal{L}'$ is some (non-linear) functional of $\fCQ(x_j; \rvec{\theta})$. We note that some loss function such as log-likelihood function in unsupervised learning, does not need any \emph{answers} $y_j$'s. Therefore, following arguments covers the general machine learning algorithms which uses function values $\fCQ$ as inputs for loss function.

We see that such generality only modifies \textbf{Step~2} in Secs.~\ref{subsec:cfflm} and \ref{subsec:qfflm}, namely that data-output subtraction is generalized to computing the possibly nonlinear functional $\mathcal{F}$. Since any $\mathcal{F}$ is clearly a functional of $\fCQ$, its resource complexity is \emph{at least} that for evaluating $\fCQ$. In practical cases, it is sensible to consider an $\mathcal{L}_{\rvec{\theta}}$ such that its component gradient $\mathcal{L}'(y)$ and higher-order derivatives are themselves easily computable for any argument $y$, so that the computational resources for $\mathcal{F}$ reduces to just that for evaluating $\fCQ$. In other words, the overall gradient computation for a general $\mathcal{L}_{\rvec{\theta}}$ still takes $\Omega(K^M)$ basic operations for CFFLMs, and $O(N_{gt}^2/\epsilon^2)$ basic operations for QFFLMs for such choices of $\mathcal{L}_{\rvec{\theta}}$.

\section{Structure of $C_{K^M}$}
\label{sec:CKM_struct}

Upon identifying that $C_{K^M}=\{\rvec{c}\,|\,-1\leq f(\rvec{x})=\rvec{c}^\top\rvec{\phi}(\rvec{x})\leq1\text{ for all }\rvec{x}\in[0,2\pi)^M]\}$ is the set of admissible columns $\rvec{c}$ such that the degree-$\DF$ Fourier-series function $f(\rvec{x})$ is bounded between $-1$ and 1, we may first notice that $C_{K^M}$ is convex. To understand this, we suppose that $\rvec{c}_1,\rvec{c}_2\in C_{K^M}$. Then if $\rvec{c}=\mu\rvec{c}_1+(1-\mu)\rvec{c}_2$ is any convex sum of $\rvec{c}_1$ and $\rvec{c}_2$ defined by $0\leq\mu\leq1$, by the triangular inequality,
\begin{equation}
	\left|\rvec{c}^\top\rvec{\phi}(\rvec{x})\right|\leq\mu\left|\rvec{c}^\top_1\rvec{\phi}(\rvec{x})\right|+(1-\mu)\left|\rvec{c}^\top_2\rvec{\phi}(\rvec{x})\right|\leq1\,,
\end{equation}
saying that $\rvec{c}\in C_{K^M}$.

Unfortunately, the general geometrical structure of the convex body $C_{K^M}$ is analytically hard to ascertain. For univariate degree-1 Fourier series [$M=1$ and $\DF=1$ (or $K=3$)], however, it is straightforward to find out the three-dimensional shape of $C_3$. In this case, all elements $\rvec{c}=(c_1\,\,c_2\,\,c_3)^\top$ in $C_3$ are such that the magnitude of $f(x)=c_1+\sqrt{2}\,c_2\cos x+\sqrt{2}\,c_3\sin x$ is less than or equal to unity. To exhaust all necessary and sufficient constraints on $\rvec{c}$, we first search for the one that imposes $|f(x)|\leq 1$ for all $x\in[0,2\pi)$. The boundary of $C_3$ contains all $\rvec{c}$s that satisfy the equation $\max_{x\in[0,2\pi)}|f(x)|=1$.

\begin{figure}[t]
	\centering
	\includegraphics[width = 1\columnwidth]{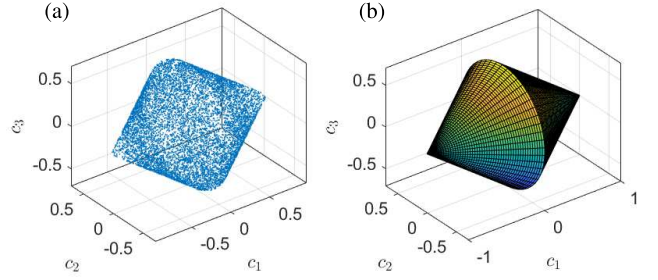}
	\caption{\label{fig:bicone}(a)~Monte~Carlo simulation of the boundary of $C_3$ and (b)~the exact surface.}
\end{figure}

When $0\leq c_1\leq1$, 
\begin{align}
	&\,\max_{x\in[0,2\pi)} |f(x)|\nonumber\\
	=&\,\max_{x\in[0,2\pi)}\left\{\left|c_1+\sqrt{2(c^2_2+c^2_3)}\,\sin\!\left(x+\tan^{-1}\!\dfrac{c_3}{c_2}\right)\right|\right\}\nonumber\\
	=&\,c_1+\sqrt{2(c^2_2+c^2_3)}=1\,.
\end{align}
By repeating the same exercise for $-1\leq c_1<0$, we obtain $-c_1+\sqrt{2(c^2_2+c^2_3)}=1$. Hence, the $C_3$ boundary is the biconical surface described by $(1-|c_1|)^2=2(c^2_2+c^2_3)$, where the common base has radius~$1/\sqrt{2}$ and length~$2$ (see Fig.~\ref{fig:bicone}).

\section{Technical details of numerical simulations concerning Fig.~\ref{fig:sim} and its extension}
\label{sec:last_sec}

\begin{figure}[t]
	\centering
	\includegraphics[width = 1\columnwidth]{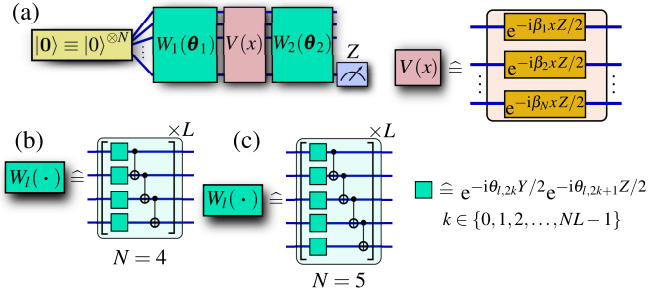}
	\caption{\label{fig:last_circuits}(a)~General circuit for the demonstrations discussed in Sec.~\ref{sec:last_sec} of this Supplemental Material, which deal with univariate target-function learning. The internal structures of the trainable modules in (b) and (c), which respectively apply for Figs.~\ref{fig:sim} and \ref{fig:inside}.}
\end{figure}

\begin{figure}[t]
	\centering
	\includegraphics[width = 1\columnwidth]{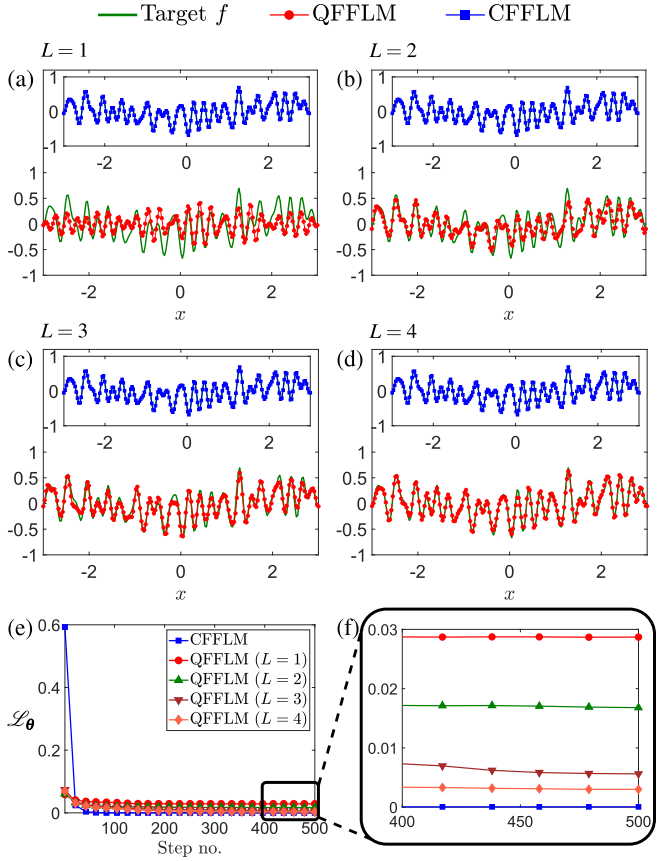}
	\caption{\label{fig:inside}Expressivity of QFFLMs with respect to the CFFLM and loss-function ($\mathcal{L}_{\rvec{\theta}}$) minimization performances (averaged over five runs per plot marker). A monotonous-drop in the saturated $\mathcal{L}_{\rvec{\theta}}$ values with increasing $L$ implies that the QFFLM expressivity improves as one uses deeper training modules. Note that all single-qubit rotation gates are still encoded with only two parameters.}
\end{figure}

All simulations concerning QFFLM training are performed with the \texttt{Pennylane} Python library, where the ``hardware-efficient'' circuit \emph{ansatz} described in the main text is employed. Those concerning CFFLM training are run with the \texttt{Pytorch} Python package for loss-function gradient computation purposes. We employ the Adam routine with a learning rate of~0.03 for both QFFLM and CFFLM optimization. A set of 200 equally-spaced training data points in the interval $[-\pi, \pi)$ are used to train both types of FFLMs, where full-batch learning is carried out with the complete training dataset. Iterative model training (gradient updates) is performed for a total of 500~training steps. The MSE $\mathcal{L}_{\rvec{\theta}}$ is used to measure the training accuracy, which in this case is also a direct measure of expressivity since the training dataset spans the entire function period with more than enough data points to accurately reconstruct the Fourier-series functions according to the Nyquist--Shannon theorem. Put differently, for sufficiently dense data points in the complete period, $\mathcal{L}_{\rvec{\theta}}$ is the discrete approximation of the average $L^2$ distance between the target function and trained FFLM stated in Eq~\eqref{eq:MSEcontinuous}.

All QFFLMs employ a four-qubit circuit (see Fig.~\ref{fig:last_circuits} for reference), where all basic data-encoding gates are arranged in parallel and collectively denoted by the unitary operator $V(x)$. This is accompanied by two trainable unitary operators $W_1(\rvec{\theta}_1)$ and $W_2(\rvec{\theta}_2)$ sandwiching $V(x)$. The measurement observable is set as the single-qubit Pauli-$Z$ operator acting on the last qubit. Both $W_1(\rvec{\theta}_1)$ and $W_2(\rvec{\theta}_2)$ are made up of $L$ layers of single-qubit rotation gate and a nearest-neighbor CNOT-gate array. Each single-qubit rotation is encoded with two training parameters; for instance, in $W_l(\rvec{\theta}_l)$ ($l=1,2$), if $l'$ labels the layer number, then the training-parameter column $\rvec{\theta}_{ll'}=(\theta_{l,l',0}\,\,\theta_{l,l',2}\,\,\ldots\,\,\theta_{l,l',7})^\top$ consisting of eight parameters with respect to the $l'$th layer in $W_l(\rvec{\theta}_l)$ is encoded according to $\rvec{\theta}_{ll'}\rightarrow\bigotimes_{k=0}^{3} [R_Z(\theta_{l,l',2k})R_Y(\theta_{l,l',2k+1})]$. By denoting $U^{(s,s')}_\textsc{CNOT}$ as the two-qubit CNOT unitary operator acting on qubits~$s$ and $s'$, the nearest-neighbor CNOT-gate-array unitary operator is given by $U_\textsc{CNOTarr}=\prod^3_{s=1} U^{(s,s+1)}_\textsc{CNOT}$.

While the crucial numerical results are presented in Fig.~\ref{fig:sim}, we would like to elaborate on the expressivity of QFFLMs in relation to the number of layers $L$ per training module. In the main text, only single-layered ($L=1$) trainable modules are used for the purpose of a fair comparison with the CFFLM under a similar $\cmplx$ order, and this clearly limits the expressivity of the QFFLM in general. Most notably, we revisit the case of $r=55.5$, where the target function $f$ is well-expressible with CFFLMs. Figure~\ref{fig:inside} in this subsection shows the performances for $L=1$ through~4. For the randomly-generated target Fourier-series function $f$ of dimension $\kappa=81$ considered, and also other tested random target functions, the figure confirms the expectation that increasing $L$ improves the QFFLM expressivity.

\begin{figure*}[t]
	\centering
	\includegraphics[width = 1.6\columnwidth]{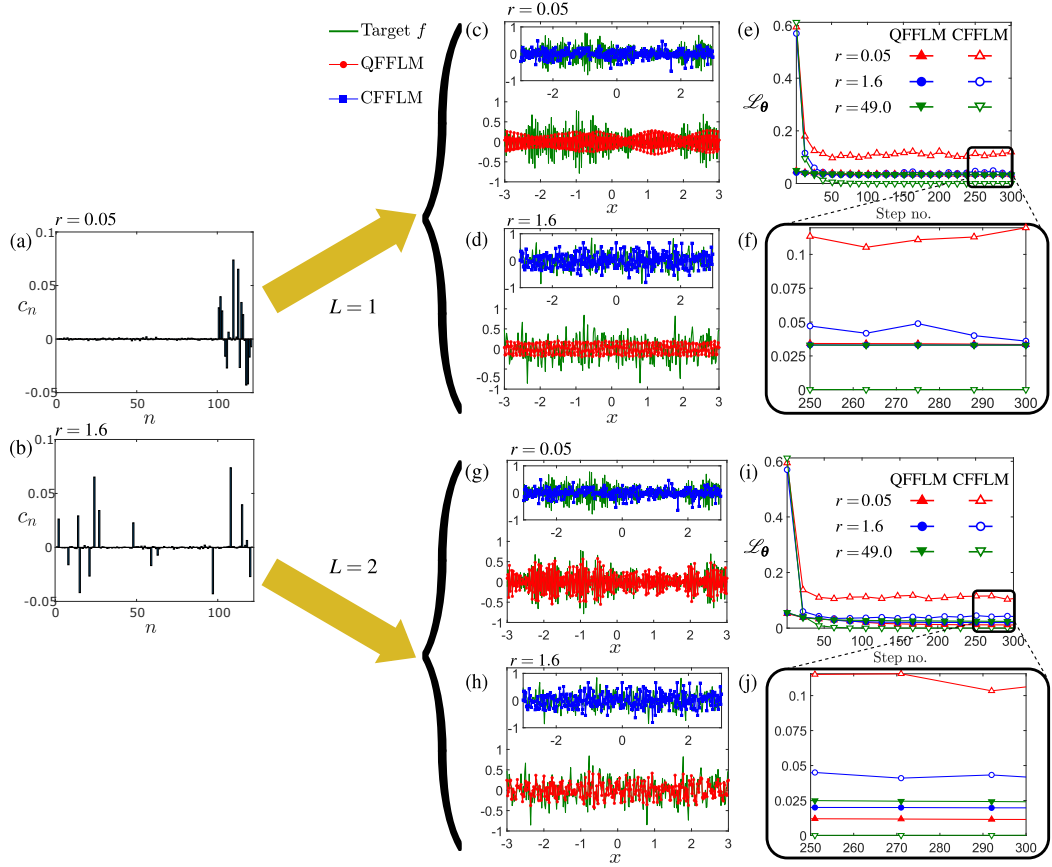}
	\caption{\label{fig:5qubit}Performances of CFFLMs and QFFLMs for $M=1$ and $\kappa=243$.}
\end{figure*}

Furthermore, we also provide Fig.~\ref{fig:5qubit} that shows similar behaviors in the CFFLM and QFFLM performances under the same resource-constraint conditions described in the main text, this time with a larger target Fourier-series-function dimension~$\kappa=243$. In this case, the CFFLM is $(\Ntp^\mathrm{CFFLM}=100)$-dimensional and universal in $C_{\Ntp^\mathrm{CFFLM}}$, whereas the QFFLMs considered here have either $L=1$ and $L=2$ training layers that are both $\kappa$-dimensional and nonuniversal with respect to $C_\kappa$. As mentioned in the main text, all single-qubit rotation gates are encoded with two training parameters respectively on the $Y$ and $Z$ Pauli gates. Figure~\ref{fig:5qubit}, therefore, supplies more examples demonstrating that a full-dimensional QFFLM that is nonuniversal can express functions outside the classically-expressible region, where a universal but lower-dimensional CFFLM shows signs of limited learning capacity.


\begin{thebibliography}{70}%
\makeatletter
\providecommand \@ifxundefined [1]{%
 \@ifx{#1\undefined}
}%
\providecommand \@ifnum [1]{%
 \ifnum #1\expandafter \@firstoftwo
 \else \expandafter \@secondoftwo
 \fi
}%
\providecommand \@ifx [1]{%
 \ifx #1\expandafter \@firstoftwo
 \else \expandafter \@secondoftwo
 \fi
}%
\providecommand \natexlab [1]{#1}%
\providecommand \enquote  [1]{``#1''}%
\providecommand \bibnamefont  [1]{#1}%
\providecommand \bibfnamefont [1]{#1}%
\providecommand \citenamefont [1]{#1}%
\providecommand \href@noop [0]{\@secondoftwo}%
\providecommand \href [0]{\begingroup \@sanitize@url \@href}%
\providecommand \@href[1]{\@@startlink{#1}\@@href}%
\providecommand \@@href[1]{\endgroup#1\@@endlink}%
\providecommand \@sanitize@url [0]{\catcode `\\12\catcode `\$12\catcode
  `\&12\catcode `\#12\catcode `\^12\catcode `\_12\catcode `\%12\relax}%
\providecommand \@@startlink[1]{}%
\providecommand \@@endlink[0]{}%
\providecommand \url  [0]{\begingroup\@sanitize@url \@url }%
\providecommand \@url [1]{\endgroup\@href {#1}{\urlprefix }}%
\providecommand \urlprefix  [0]{URL }%
\providecommand \Eprint [0]{\href }%
\providecommand \doibase [0]{https://doi.org/}%
\providecommand \selectlanguage [0]{\@gobble}%
\providecommand \bibinfo  [0]{\@secondoftwo}%
\providecommand \bibfield  [0]{\@secondoftwo}%
\providecommand \translation [1]{[#1]}%
\providecommand \BibitemOpen [0]{}%
\providecommand \bibitemStop [0]{}%
\providecommand \bibitemNoStop [0]{.\EOS\space}%
\providecommand \EOS [0]{\spacefactor3000\relax}%
\providecommand \BibitemShut  [1]{\csname bibitem#1\endcsname}%
\let\auto@bib@innerbib\@empty
%</preamble>
\bibitem [{\citenamefont {Preskill}(2018)}]{Preskill2018quantumcomputingin}%
  \BibitemOpen
  \bibfield  {author} {\bibinfo {author} {\bibfnamefont {J.}~\bibnamefont
  {Preskill}},\ }\bibfield  {title} {\bibinfo {title} {Quantum {C}omputing in
  the {NISQ} era and beyond},\ }\href
  {https://doi.org/10.22331/q-2018-08-06-79} {\bibfield  {journal} {\bibinfo
  {journal} {{Quantum}}\ }\textbf {\bibinfo {volume} {2}},\ \bibinfo {pages}
  {79} (\bibinfo {year} {2018})}\BibitemShut {NoStop}%
\bibitem [{\citenamefont {Bromley}\ \emph {et~al.}(2020)\citenamefont
  {Bromley}, \citenamefont {Arrazola}, \citenamefont {Jahangiri}, \citenamefont
  {Izaac}, \citenamefont {Quesada}, \citenamefont {Gran}, \citenamefont
  {Schuld}, \citenamefont {Swinarton}, \citenamefont {Zabaneh},\ and\
  \citenamefont {Killoran}}]{Bromley:2020applications}%
  \BibitemOpen
  \bibfield  {author} {\bibinfo {author} {\bibfnamefont {T.~R.}\ \bibnamefont
  {Bromley}}, \bibinfo {author} {\bibfnamefont {J.~M.}\ \bibnamefont
  {Arrazola}}, \bibinfo {author} {\bibfnamefont {S.}~\bibnamefont {Jahangiri}},
  \bibinfo {author} {\bibfnamefont {J.}~\bibnamefont {Izaac}}, \bibinfo
  {author} {\bibfnamefont {N.}~\bibnamefont {Quesada}}, \bibinfo {author}
  {\bibfnamefont {A.~D.}\ \bibnamefont {Gran}}, \bibinfo {author}
  {\bibfnamefont {M.}~\bibnamefont {Schuld}}, \bibinfo {author} {\bibfnamefont
  {J.}~\bibnamefont {Swinarton}}, \bibinfo {author} {\bibfnamefont
  {Z.}~\bibnamefont {Zabaneh}},\ and\ \bibinfo {author} {\bibfnamefont
  {N.}~\bibnamefont {Killoran}},\ }\bibfield  {title} {\bibinfo {title}
  {Applications of near-term photonic quantum computers: software and
  algorithms},\ }\href {https://doi.org/10.1088/2058-9565/ab8504} {\bibfield
  {journal} {\bibinfo  {journal} {Quantum Sci. Technol.}\ }\textbf {\bibinfo
  {volume} {5}},\ \bibinfo {pages} {034010} (\bibinfo {year}
  {2020})}\BibitemShut {NoStop}%
\bibitem [{\citenamefont {Bharti}\ \emph {et~al.}(2022)\citenamefont {Bharti},
  \citenamefont {Cervera-Lierta}, \citenamefont {Kyaw}, \citenamefont {Haug},
  \citenamefont {Alperin-Lea}, \citenamefont {Anand}, \citenamefont {Degroote},
  \citenamefont {Heimonen}, \citenamefont {Kottmann}, \citenamefont {Menke},
  \citenamefont {Mok}, \citenamefont {Sim}, \citenamefont {Kwek},\ and\
  \citenamefont {Aspuru-Guzik}}]{Bharti:2022noisy}%
  \BibitemOpen
  \bibfield  {author} {\bibinfo {author} {\bibfnamefont {K.}~\bibnamefont
  {Bharti}}, \bibinfo {author} {\bibfnamefont {A.}~\bibnamefont
  {Cervera-Lierta}}, \bibinfo {author} {\bibfnamefont {T.~H.}\ \bibnamefont
  {Kyaw}}, \bibinfo {author} {\bibfnamefont {T.}~\bibnamefont {Haug}}, \bibinfo
  {author} {\bibfnamefont {S.}~\bibnamefont {Alperin-Lea}}, \bibinfo {author}
  {\bibfnamefont {A.}~\bibnamefont {Anand}}, \bibinfo {author} {\bibfnamefont
  {M.}~\bibnamefont {Degroote}}, \bibinfo {author} {\bibfnamefont
  {H.}~\bibnamefont {Heimonen}}, \bibinfo {author} {\bibfnamefont {J.~S.}\
  \bibnamefont {Kottmann}}, \bibinfo {author} {\bibfnamefont {T.}~\bibnamefont
  {Menke}}, \bibinfo {author} {\bibfnamefont {W.-K.}\ \bibnamefont {Mok}},
  \bibinfo {author} {\bibfnamefont {S.}~\bibnamefont {Sim}}, \bibinfo {author}
  {\bibfnamefont {L.-C.}\ \bibnamefont {Kwek}},\ and\ \bibinfo {author}
  {\bibfnamefont {A.}~\bibnamefont {Aspuru-Guzik}},\ }\bibfield  {title}
  {\bibinfo {title} {Noisy intermediate-scale quantum algorithms},\ }\href
  {https://doi.org/10.1103/RevModPhys.94.015004} {\bibfield  {journal}
  {\bibinfo  {journal} {Rev. Mod. Phys.}\ }\textbf {\bibinfo {volume} {94}},\
  \bibinfo {pages} {015004} (\bibinfo {year} {2022})}\BibitemShut {NoStop}%
\bibitem [{\citenamefont {Finnila}\ \emph {et~al.}(1994)\citenamefont
  {Finnila}, \citenamefont {Gomez}, \citenamefont {Sebenik}, \citenamefont
  {Stenson},\ and\ \citenamefont {Doll}}]{Finnila:1994quantum}%
  \BibitemOpen
  \bibfield  {author} {\bibinfo {author} {\bibfnamefont {A.}~\bibnamefont
  {Finnila}}, \bibinfo {author} {\bibfnamefont {M.}~\bibnamefont {Gomez}},
  \bibinfo {author} {\bibfnamefont {C.}~\bibnamefont {Sebenik}}, \bibinfo
  {author} {\bibfnamefont {C.}~\bibnamefont {Stenson}},\ and\ \bibinfo {author}
  {\bibfnamefont {J.}~\bibnamefont {Doll}},\ }\bibfield  {title} {\bibinfo
  {title} {Quantum annealing: A new method for minimizing multidimensional
  functions},\ }\href
  {https://doi.org/https://doi.org/10.1016/0009-2614(94)00117-0} {\bibfield
  {journal} {\bibinfo  {journal} {Chemical Physics Letters}\ }\textbf {\bibinfo
  {volume} {219}},\ \bibinfo {pages} {343} (\bibinfo {year}
  {1994})}\BibitemShut {NoStop}%
\bibitem [{\citenamefont {Kadowaki}\ and\ \citenamefont
  {Nishimori}(1998)}]{Kadowaki:1998quantum}%
  \BibitemOpen
  \bibfield  {author} {\bibinfo {author} {\bibfnamefont {T.}~\bibnamefont
  {Kadowaki}}\ and\ \bibinfo {author} {\bibfnamefont {H.}~\bibnamefont
  {Nishimori}},\ }\bibfield  {title} {\bibinfo {title} {Quantum annealing in
  the transverse ising model},\ }\href
  {https://doi.org/10.1103/PhysRevE.58.5355} {\bibfield  {journal} {\bibinfo
  {journal} {Phys. Rev. E}\ }\textbf {\bibinfo {volume} {58}},\ \bibinfo
  {pages} {5355} (\bibinfo {year} {1998})}\BibitemShut {NoStop}%
\bibitem [{\citenamefont {Aaronson}\ and\ \citenamefont
  {Arkhipov}(2011)}]{Aaronson:2011computational}%
  \BibitemOpen
  \bibfield  {author} {\bibinfo {author} {\bibfnamefont {S.}~\bibnamefont
  {Aaronson}}\ and\ \bibinfo {author} {\bibfnamefont {A.}~\bibnamefont
  {Arkhipov}},\ }\bibfield  {title} {\bibinfo {title} {The computational
  complexity of linear optics},\ }in\ \href
  {https://doi.org/10.1145/1993636.1993682} {\emph {\bibinfo {booktitle}
  {Proceedings of the Forty-Third Annual ACM Symposium on Theory of
  Computing}}},\ \bibinfo {series and number} {STOC '11}\ (\bibinfo
  {publisher} {Association for Computing Machinery},\ \bibinfo {address} {New
  York, NY, USA},\ \bibinfo {year} {2011})\ p.\ \bibinfo {pages}
  {333–342}\BibitemShut {NoStop}%
\bibitem [{\citenamefont {Aaronson}(2011)}]{Aaronson:2011linear-optical}%
  \BibitemOpen
  \bibfield  {author} {\bibinfo {author} {\bibfnamefont {S.}~\bibnamefont
  {Aaronson}},\ }\bibfield  {title} {\bibinfo {title} {A linear-optical proof
  that the permanent is~\#{P}-hard},\ }\href
  {https://doi.org/10.1098/rspa.2011.0232} {\bibfield  {journal} {\bibinfo
  {journal} {Proceedings of the Royal Society A: Mathematical, Physical and
  Engineering Sciences}\ }\textbf {\bibinfo {volume} {467}},\ \bibinfo {pages}
  {3393} (\bibinfo {year} {2011})},\ \Eprint
  {https://arxiv.org/abs/https://royalsocietypublishing.org/doi/pdf/10.1098/rspa.2011.0232}
  {https://royalsocietypublishing.org/doi/pdf/10.1098/rspa.2011.0232}
  \BibitemShut {NoStop}%
\bibitem [{\citenamefont {Hamilton}\ \emph {et~al.}(2017)\citenamefont
  {Hamilton}, \citenamefont {Kruse}, \citenamefont {Sansoni}, \citenamefont
  {Barkhofen}, \citenamefont {Silberhorn},\ and\ \citenamefont
  {Jex}}]{Hamilton:2017gaussian}%
  \BibitemOpen
  \bibfield  {author} {\bibinfo {author} {\bibfnamefont {C.~S.}\ \bibnamefont
  {Hamilton}}, \bibinfo {author} {\bibfnamefont {R.}~\bibnamefont {Kruse}},
  \bibinfo {author} {\bibfnamefont {L.}~\bibnamefont {Sansoni}}, \bibinfo
  {author} {\bibfnamefont {S.}~\bibnamefont {Barkhofen}}, \bibinfo {author}
  {\bibfnamefont {C.}~\bibnamefont {Silberhorn}},\ and\ \bibinfo {author}
  {\bibfnamefont {I.}~\bibnamefont {Jex}},\ }\bibfield  {title} {\bibinfo
  {title} {Gaussian boson sampling},\ }\href
  {https://doi.org/10.1103/PhysRevLett.119.170501} {\bibfield  {journal}
  {\bibinfo  {journal} {Phys. Rev. Lett.}\ }\textbf {\bibinfo {volume} {119}},\
  \bibinfo {pages} {170501} (\bibinfo {year} {2017})}\BibitemShut {NoStop}%
\bibitem [{\citenamefont {Trabesinger}(2012)}]{Trabesinger:2012quantum}%
  \BibitemOpen
  \bibfield  {author} {\bibinfo {author} {\bibfnamefont {A.}~\bibnamefont
  {Trabesinger}},\ }\bibfield  {title} {\bibinfo {title} {Quantum simulation},\
  }\href {https://doi.org/10.1038/nphys2258} {\bibfield  {journal} {\bibinfo
  {journal} {Nature Physics}\ }\textbf {\bibinfo {volume} {8}},\ \bibinfo
  {pages} {263} (\bibinfo {year} {2012})}\BibitemShut {NoStop}%
\bibitem [{\citenamefont {Georgescu}\ \emph {et~al.}(2014)\citenamefont
  {Georgescu}, \citenamefont {Ashhab},\ and\ \citenamefont
  {Nori}}]{Georgescu:2014quantum}%
  \BibitemOpen
  \bibfield  {author} {\bibinfo {author} {\bibfnamefont {I.~M.}\ \bibnamefont
  {Georgescu}}, \bibinfo {author} {\bibfnamefont {S.}~\bibnamefont {Ashhab}},\
  and\ \bibinfo {author} {\bibfnamefont {F.}~\bibnamefont {Nori}},\ }\bibfield
  {title} {\bibinfo {title} {Quantum simulation},\ }\href
  {https://doi.org/10.1103/RevModPhys.86.153} {\bibfield  {journal} {\bibinfo
  {journal} {Rev. Mod. Phys.}\ }\textbf {\bibinfo {volume} {86}},\ \bibinfo
  {pages} {153} (\bibinfo {year} {2014})}\BibitemShut {NoStop}%
\bibitem [{\citenamefont {Schuld}\ \emph {et~al.}(2015)\citenamefont {Schuld},
  \citenamefont {Sinayskiy},\ and\ \citenamefont
  {Petruccione}}]{Schuld:2015introduction}%
  \BibitemOpen
  \bibfield  {author} {\bibinfo {author} {\bibfnamefont {M.}~\bibnamefont
  {Schuld}}, \bibinfo {author} {\bibfnamefont {I.}~\bibnamefont {Sinayskiy}},\
  and\ \bibinfo {author} {\bibfnamefont {F.}~\bibnamefont {Petruccione}},\
  }\bibfield  {title} {\bibinfo {title} {An introduction to quantum machine
  learning},\ }\href {https://doi.org/10.1080/00107514.2014.964942} {\bibfield
  {journal} {\bibinfo  {journal} {Contemporary Physics}\ }\textbf {\bibinfo
  {volume} {56}},\ \bibinfo {pages} {172} (\bibinfo {year} {2015})},\ \Eprint
  {https://arxiv.org/abs/https://doi.org/10.1080/00107514.2014.964942}
  {https://doi.org/10.1080/00107514.2014.964942} \BibitemShut {NoStop}%
\bibitem [{\citenamefont {Schuld}\ and\ \citenamefont
  {Killoran}(2019)}]{Schuld:2019quantum}%
  \BibitemOpen
  \bibfield  {author} {\bibinfo {author} {\bibfnamefont {M.}~\bibnamefont
  {Schuld}}\ and\ \bibinfo {author} {\bibfnamefont {N.}~\bibnamefont
  {Killoran}},\ }\bibfield  {title} {\bibinfo {title} {Quantum machine learning
  in feature hilbert spaces},\ }\href
  {https://doi.org/10.1103/PhysRevLett.122.040504} {\bibfield  {journal}
  {\bibinfo  {journal} {Phys. Rev. Lett.}\ }\textbf {\bibinfo {volume} {122}},\
  \bibinfo {pages} {040504} (\bibinfo {year} {2019})}\BibitemShut {NoStop}%
\bibitem [{\citenamefont {Carleo}\ \emph {et~al.}(2019)\citenamefont {Carleo},
  \citenamefont {Cirac}, \citenamefont {Cranmer}, \citenamefont {Daudet},
  \citenamefont {Schuld}, \citenamefont {Tishby}, \citenamefont
  {Vogt-Maranto},\ and\ \citenamefont {Zdeborov\'a}}]{Carleo:2019machine}%
  \BibitemOpen
  \bibfield  {author} {\bibinfo {author} {\bibfnamefont {G.}~\bibnamefont
  {Carleo}}, \bibinfo {author} {\bibfnamefont {I.}~\bibnamefont {Cirac}},
  \bibinfo {author} {\bibfnamefont {K.}~\bibnamefont {Cranmer}}, \bibinfo
  {author} {\bibfnamefont {L.}~\bibnamefont {Daudet}}, \bibinfo {author}
  {\bibfnamefont {M.}~\bibnamefont {Schuld}}, \bibinfo {author} {\bibfnamefont
  {N.}~\bibnamefont {Tishby}}, \bibinfo {author} {\bibfnamefont
  {L.}~\bibnamefont {Vogt-Maranto}},\ and\ \bibinfo {author} {\bibfnamefont
  {L.}~\bibnamefont {Zdeborov\'a}},\ }\bibfield  {title} {\bibinfo {title}
  {Machine learning and the physical sciences},\ }\href
  {https://doi.org/10.1103/RevModPhys.91.045002} {\bibfield  {journal}
  {\bibinfo  {journal} {Rev. Mod. Phys.}\ }\textbf {\bibinfo {volume} {91}},\
  \bibinfo {pages} {045002} (\bibinfo {year} {2019})}\BibitemShut {NoStop}%
\bibitem [{\citenamefont {Date}(2020)}]{date2020quantum}%
  \BibitemOpen
  \bibfield  {author} {\bibinfo {author} {\bibfnamefont {P.}~\bibnamefont
  {Date}},\ }\href@noop {} {\bibinfo {title} {Quantum discriminator for binary
  classification}} (\bibinfo {year} {2020}),\ \Eprint
  {https://arxiv.org/abs/2009.01235} {arXiv:2009.01235 [quant-ph]} \BibitemShut
  {NoStop}%
\bibitem [{\citenamefont {P{\'e}rez-Salinas}\ \emph {et~al.}(2020)\citenamefont
  {P{\'e}rez-Salinas}, \citenamefont {Cervera-Lierta}, \citenamefont
  {Gil-Fuster},\ and\ \citenamefont {Latorre}}]{Perez-Salinas:2020aa}%
  \BibitemOpen
  \bibfield  {author} {\bibinfo {author} {\bibfnamefont {A.}~\bibnamefont
  {P{\'e}rez-Salinas}}, \bibinfo {author} {\bibfnamefont {A.}~\bibnamefont
  {Cervera-Lierta}}, \bibinfo {author} {\bibfnamefont {E.}~\bibnamefont
  {Gil-Fuster}},\ and\ \bibinfo {author} {\bibfnamefont {J.~I.}\ \bibnamefont
  {Latorre}},\ }\bibfield  {title} {\bibinfo {title} {Data re-uploading for a
  universal quantum classifier},\ }\href
  {https://doi.org/10.22331/q-2020-02-06-226} {\bibfield  {journal} {\bibinfo
  {journal} {Quantum}\ }\textbf {\bibinfo {volume} {4}},\ \bibinfo {pages}
  {226} (\bibinfo {year} {2020})}\BibitemShut {NoStop}%
\bibitem [{\citenamefont {Dutta}\ \emph {et~al.}(2021)\citenamefont {Dutta},
  \citenamefont {P{\'e}rez-Salinas}, \citenamefont {Cheng}, \citenamefont
  {Latorre},\ and\ \citenamefont {Mukherjee}}]{dutta2021singlequbit}%
  \BibitemOpen
  \bibfield  {author} {\bibinfo {author} {\bibfnamefont {T.}~\bibnamefont
  {Dutta}}, \bibinfo {author} {\bibfnamefont {A.}~\bibnamefont
  {P{\'e}rez-Salinas}}, \bibinfo {author} {\bibfnamefont {J.~P.~S.}\
  \bibnamefont {Cheng}}, \bibinfo {author} {\bibfnamefont {J.~I.}\ \bibnamefont
  {Latorre}},\ and\ \bibinfo {author} {\bibfnamefont {M.}~\bibnamefont
  {Mukherjee}},\ }\href@noop {} {\bibinfo {title} {Single-qubit universal
  classifier implemented on an ion-trap quantum device}} (\bibinfo {year}
  {2021}),\ \Eprint {https://arxiv.org/abs/2106.14059} {arXiv:2106.14059
  [quant-ph]} \BibitemShut {NoStop}%
\bibitem [{\citenamefont {Goto}\ \emph {et~al.}(2021)\citenamefont {Goto},
  \citenamefont {Tran},\ and\ \citenamefont {Nakajima}}]{Goto:2021universal}%
  \BibitemOpen
  \bibfield  {author} {\bibinfo {author} {\bibfnamefont {T.}~\bibnamefont
  {Goto}}, \bibinfo {author} {\bibfnamefont {Q.~H.}\ \bibnamefont {Tran}},\
  and\ \bibinfo {author} {\bibfnamefont {K.}~\bibnamefont {Nakajima}},\
  }\bibfield  {title} {\bibinfo {title} {Universal approximation property of
  quantum machine learning models in quantum-enhanced feature spaces},\ }\href
  {https://doi.org/10.1103/PhysRevLett.127.090506} {\bibfield  {journal}
  {\bibinfo  {journal} {Phys. Rev. Lett.}\ }\textbf {\bibinfo {volume} {127}},\
  \bibinfo {pages} {090506} (\bibinfo {year} {2021})}\BibitemShut {NoStop}%
\bibitem [{\citenamefont {Schuld}\ and\ \citenamefont
  {Killoran}(2022)}]{Schuld.2203.01340}%
  \BibitemOpen
  \bibfield  {author} {\bibinfo {author} {\bibfnamefont {M.}~\bibnamefont
  {Schuld}}\ and\ \bibinfo {author} {\bibfnamefont {N.}~\bibnamefont
  {Killoran}},\ }\bibfield  {title} {\bibinfo {title} {Is quantum advantage the
  right goal for quantum machine learning?},\ }\href@noop {} {\  (\bibinfo
  {year} {2022})},\ \Eprint {https://arxiv.org/abs/arXiv:2203.01340}
  {arXiv:2203.01340} \BibitemShut {NoStop}%
\bibitem [{\citenamefont {Rebentrost}\ \emph {et~al.}(2014)\citenamefont
  {Rebentrost}, \citenamefont {Mohseni},\ and\ \citenamefont
  {Lloyd}}]{RebentrostQSVM.2014}%
  \BibitemOpen
  \bibfield  {author} {\bibinfo {author} {\bibfnamefont {P.}~\bibnamefont
  {Rebentrost}}, \bibinfo {author} {\bibfnamefont {M.}~\bibnamefont
  {Mohseni}},\ and\ \bibinfo {author} {\bibfnamefont {S.}~\bibnamefont
  {Lloyd}},\ }\bibfield  {title} {\bibinfo {title} {Quantum support vector
  machine for big data classification},\ }\href
  {https://doi.org/10.1103/PhysRevLett.113.130503} {\bibfield  {journal}
  {\bibinfo  {journal} {Phys. Rev. Lett.}\ }\textbf {\bibinfo {volume} {113}},\
  \bibinfo {pages} {130503} (\bibinfo {year} {2014})}\BibitemShut {NoStop}%
\bibitem [{\citenamefont {Cong}\ \emph {et~al.}(2019)\citenamefont {Cong},
  \citenamefont {Choi},\ and\ \citenamefont {Lukin}}]{cong2019quantum}%
  \BibitemOpen
  \bibfield  {author} {\bibinfo {author} {\bibfnamefont {I.}~\bibnamefont
  {Cong}}, \bibinfo {author} {\bibfnamefont {S.}~\bibnamefont {Choi}},\ and\
  \bibinfo {author} {\bibfnamefont {M.~D.}\ \bibnamefont {Lukin}},\ }\bibfield
  {title} {\bibinfo {title} {Quantum convolutional neural networks},\
  }\href@noop {} {\bibfield  {journal} {\bibinfo  {journal} {Nature Physics}\
  }\textbf {\bibinfo {volume} {15}},\ \bibinfo {pages} {1273} (\bibinfo {year}
  {2019})}\BibitemShut {NoStop}%
\bibitem [{\citenamefont {Lloyd}\ and\ \citenamefont
  {Weedbrook}(2018)}]{lloyd2018qgan}%
  \BibitemOpen
  \bibfield  {author} {\bibinfo {author} {\bibfnamefont {S.}~\bibnamefont
  {Lloyd}}\ and\ \bibinfo {author} {\bibfnamefont {C.}~\bibnamefont
  {Weedbrook}},\ }\bibfield  {title} {\bibinfo {title} {Quantum generative
  adversarial learning},\ }\href
  {https://doi.org/10.1103/PhysRevLett.121.040502} {\bibfield  {journal}
  {\bibinfo  {journal} {Phys. Rev. Lett.}\ }\textbf {\bibinfo {volume} {121}},\
  \bibinfo {pages} {040502} (\bibinfo {year} {2018})}\BibitemShut {NoStop}%
\bibitem [{\citenamefont {Liu}\ and\ \citenamefont
  {Rebentrost}(2018)}]{Liu2018anomaly}%
  \BibitemOpen
  \bibfield  {author} {\bibinfo {author} {\bibfnamefont {N.}~\bibnamefont
  {Liu}}\ and\ \bibinfo {author} {\bibfnamefont {P.}~\bibnamefont
  {Rebentrost}},\ }\bibfield  {title} {\bibinfo {title} {Quantum machine
  learning for quantum anomaly detection},\ }\href
  {https://doi.org/10.1103/PhysRevA.97.042315} {\bibfield  {journal} {\bibinfo
  {journal} {Phys. Rev. A}\ }\textbf {\bibinfo {volume} {97}},\ \bibinfo
  {pages} {042315} (\bibinfo {year} {2018})}\BibitemShut {NoStop}%
\bibitem [{\citenamefont {Dunjko}\ \emph {et~al.}(2016)\citenamefont {Dunjko},
  \citenamefont {Taylor},\ and\ \citenamefont {Briegel}}]{dunjko2016qrl}%
  \BibitemOpen
  \bibfield  {author} {\bibinfo {author} {\bibfnamefont {V.}~\bibnamefont
  {Dunjko}}, \bibinfo {author} {\bibfnamefont {J.~M.}\ \bibnamefont {Taylor}},\
  and\ \bibinfo {author} {\bibfnamefont {H.~J.}\ \bibnamefont {Briegel}},\
  }\bibfield  {title} {\bibinfo {title} {Quantum-enhanced machine learning},\
  }\href {https://doi.org/10.1103/PhysRevLett.117.130501} {\bibfield  {journal}
  {\bibinfo  {journal} {Phys. Rev. Lett.}\ }\textbf {\bibinfo {volume} {117}},\
  \bibinfo {pages} {130501} (\bibinfo {year} {2016})}\BibitemShut {NoStop}%
\bibitem [{\citenamefont {Du}\ \emph {et~al.}(2020)\citenamefont {Du},
  \citenamefont {Hsieh}, \citenamefont {Liu},\ and\ \citenamefont
  {Tao}}]{Du:2020aa}%
  \BibitemOpen
  \bibfield  {author} {\bibinfo {author} {\bibfnamefont {Y.}~\bibnamefont
  {Du}}, \bibinfo {author} {\bibfnamefont {M.-H.}\ \bibnamefont {Hsieh}},
  \bibinfo {author} {\bibfnamefont {T.}~\bibnamefont {Liu}},\ and\ \bibinfo
  {author} {\bibfnamefont {D.}~\bibnamefont {Tao}},\ }\bibfield  {title}
  {\bibinfo {title} {Expressive power of parametrized quantum circuits},\
  }\href {https://doi.org/10.1103/PhysRevResearch.2.033125} {\bibfield
  {journal} {\bibinfo  {journal} {Phys. Rev. Research}\ }\textbf {\bibinfo
  {volume} {2}},\ \bibinfo {pages} {033125} (\bibinfo {year}
  {2020})}\BibitemShut {NoStop}%
\bibitem [{\citenamefont {Du}\ \emph {et~al.}(2022)\citenamefont {Du},
  \citenamefont {Tu}, \citenamefont {Yuan},\ and\ \citenamefont
  {Tao}}]{Du:2022.expressivity}%
  \BibitemOpen
  \bibfield  {author} {\bibinfo {author} {\bibfnamefont {Y.}~\bibnamefont
  {Du}}, \bibinfo {author} {\bibfnamefont {Z.}~\bibnamefont {Tu}}, \bibinfo
  {author} {\bibfnamefont {X.}~\bibnamefont {Yuan}},\ and\ \bibinfo {author}
  {\bibfnamefont {D.}~\bibnamefont {Tao}},\ }\bibfield  {title} {\bibinfo
  {title} {Efficient measure for the expressivity of variational quantum
  algorithms},\ }\href {https://doi.org/10.1103/PhysRevLett.128.080506}
  {\bibfield  {journal} {\bibinfo  {journal} {Phys. Rev. Lett.}\ }\textbf
  {\bibinfo {volume} {128}},\ \bibinfo {pages} {080506} (\bibinfo {year}
  {2022})}\BibitemShut {NoStop}%
\bibitem [{\citenamefont {Banchi}\ \emph {et~al.}(2021)\citenamefont {Banchi},
  \citenamefont {Pereira},\ and\ \citenamefont
  {Pirandola}}]{Banchi:2021generalization}%
  \BibitemOpen
  \bibfield  {author} {\bibinfo {author} {\bibfnamefont {L.}~\bibnamefont
  {Banchi}}, \bibinfo {author} {\bibfnamefont {J.}~\bibnamefont {Pereira}},\
  and\ \bibinfo {author} {\bibfnamefont {S.}~\bibnamefont {Pirandola}},\
  }\bibfield  {title} {\bibinfo {title} {Generalization in quantum machine
  learning: A quantum information standpoint},\ }\href
  {https://doi.org/10.1103/PRXQuantum.2.040321} {\bibfield  {journal} {\bibinfo
   {journal} {PRX Quantum}\ }\textbf {\bibinfo {volume} {2}},\ \bibinfo {pages}
  {040321} (\bibinfo {year} {2021})}\BibitemShut {NoStop}%
\bibitem [{\citenamefont {Caro}\ \emph
  {et~al.}(2021{\natexlab{a}})\citenamefont {Caro}, \citenamefont {Gil-Fuster},
  \citenamefont {Meyer}, \citenamefont {Eisert},\ and\ \citenamefont
  {Sweke}}]{Caro2021encodingdependent}%
  \BibitemOpen
  \bibfield  {author} {\bibinfo {author} {\bibfnamefont {M.~C.}\ \bibnamefont
  {Caro}}, \bibinfo {author} {\bibfnamefont {E.}~\bibnamefont {Gil-Fuster}},
  \bibinfo {author} {\bibfnamefont {J.~J.}\ \bibnamefont {Meyer}}, \bibinfo
  {author} {\bibfnamefont {J.}~\bibnamefont {Eisert}},\ and\ \bibinfo {author}
  {\bibfnamefont {R.}~\bibnamefont {Sweke}},\ }\bibfield  {title} {\bibinfo
  {title} {Encoding-dependent generalization bounds for parametrized quantum
  circuits},\ }\href {https://doi.org/10.22331/q-2021-11-17-582} {\bibfield
  {journal} {\bibinfo  {journal} {{Quantum}}\ }\textbf {\bibinfo {volume}
  {5}},\ \bibinfo {pages} {582} (\bibinfo {year}
  {2021}{\natexlab{a}})}\BibitemShut {NoStop}%
\bibitem [{\citenamefont {Huang}\ \emph {et~al.}(2021)\citenamefont {Huang},
  \citenamefont {Kueng},\ and\ \citenamefont {Preskill}}]{Huang2021general}%
  \BibitemOpen
  \bibfield  {author} {\bibinfo {author} {\bibfnamefont {H.-Y.}\ \bibnamefont
  {Huang}}, \bibinfo {author} {\bibfnamefont {R.}~\bibnamefont {Kueng}},\ and\
  \bibinfo {author} {\bibfnamefont {J.}~\bibnamefont {Preskill}},\ }\bibfield
  {title} {\bibinfo {title} {{Information-Theoretic Bounds on Quantum Advantage
  in Machine Learning}},\ }\href
  {https://doi.org/10.1103/physrevlett.126.190505} {\bibfield  {journal}
  {\bibinfo  {journal} {Physical Review Letters}\ }\textbf {\bibinfo {volume}
  {126}},\ \bibinfo {pages} {190505} (\bibinfo {year} {2021})},\ \Eprint
  {https://arxiv.org/abs/2101.02464} {2101.02464} \BibitemShut {NoStop}%
\bibitem [{\citenamefont {Arunachalam}\ and\ \citenamefont
  {de~Wolf}(2017)}]{Arunachalam:2017}%
  \BibitemOpen
  \bibfield  {author} {\bibinfo {author} {\bibfnamefont {S.}~\bibnamefont
  {Arunachalam}}\ and\ \bibinfo {author} {\bibfnamefont {R.}~\bibnamefont
  {de~Wolf}},\ }\bibfield  {title} {\bibinfo {title} {Guest column: A survey of
  quantum learning theory},\ }\href {https://doi.org/10.1145/3106700.3106710}
  {\bibfield  {journal} {\bibinfo  {journal} {SIGACT News}\ }\textbf {\bibinfo
  {volume} {48}},\ \bibinfo {pages} {41–67} (\bibinfo {year}
  {2017})}\BibitemShut {NoStop}%
\bibitem [{\citenamefont {Hornik}\ \emph {et~al.}(1989)\citenamefont {Hornik},
  \citenamefont {Stinchcombe},\ and\ \citenamefont
  {White}}]{hornik1989multilayer}%
  \BibitemOpen
  \bibfield  {author} {\bibinfo {author} {\bibfnamefont {K.}~\bibnamefont
  {Hornik}}, \bibinfo {author} {\bibfnamefont {M.}~\bibnamefont
  {Stinchcombe}},\ and\ \bibinfo {author} {\bibfnamefont {H.}~\bibnamefont
  {White}},\ }\bibfield  {title} {\bibinfo {title} {Multilayer feedforward
  networks are universal approximators},\ }\href@noop {} {\bibfield  {journal}
  {\bibinfo  {journal} {Neural networks}\ }\textbf {\bibinfo {volume} {2}},\
  \bibinfo {pages} {359} (\bibinfo {year} {1989})}\BibitemShut {NoStop}%
\bibitem [{\citenamefont {Schuld}\ \emph {et~al.}(2021)\citenamefont {Schuld},
  \citenamefont {Sweke},\ and\ \citenamefont {Meyer}}]{Schuld:2021aa}%
  \BibitemOpen
  \bibfield  {author} {\bibinfo {author} {\bibfnamefont {M.}~\bibnamefont
  {Schuld}}, \bibinfo {author} {\bibfnamefont {R.}~\bibnamefont {Sweke}},\ and\
  \bibinfo {author} {\bibfnamefont {J.~J.}\ \bibnamefont {Meyer}},\ }\bibfield
  {title} {\bibinfo {title} {Effect of data encoding on the expressive power of
  variational quantum-machine-learning models},\ }\href
  {https://doi.org/10.1103/PhysRevA.103.032430} {\bibfield  {journal} {\bibinfo
   {journal} {Phys. Rev. A}\ }\textbf {\bibinfo {volume} {103}},\ \bibinfo
  {pages} {032430} (\bibinfo {year} {2021})}\BibitemShut {NoStop}%
\bibitem [{\citenamefont {Weisz}(2012)}]{Weisz:2012aa}%
  \BibitemOpen
  \bibfield  {author} {\bibinfo {author} {\bibfnamefont {F.}~\bibnamefont
  {Weisz}},\ }\bibfield  {title} {\bibinfo {title} {Summability of
  multi-dimensional trigonometric fourier series},\ }\href
  {https://www.emis.de/journals/SAT/papers/17/17.pdf} {\bibfield  {journal}
  {\bibinfo  {journal} {Surveys in Approximation Theory}\ }\textbf {\bibinfo
  {volume} {7}},\ \bibinfo {pages} {1} (\bibinfo {year} {2012})}\BibitemShut
  {NoStop}%
\bibitem [{\citenamefont {Rahimi}\ and\ \citenamefont
  {Recht}(2007)}]{Rahimi:2007random}%
  \BibitemOpen
  \bibfield  {author} {\bibinfo {author} {\bibfnamefont {A.}~\bibnamefont
  {Rahimi}}\ and\ \bibinfo {author} {\bibfnamefont {B.}~\bibnamefont {Recht}},\
  }\bibfield  {title} {\bibinfo {title} {Random features for large-scale kernel
  machines},\ }in\ \href
  {https://proceedings.neurips.cc/paper/2007/file/013a006f03dbc5392effeb8f18fda755-Paper.pdf}
  {\emph {\bibinfo {booktitle} {Advances in Neural Information Processing
  Systems}}},\ Vol.~\bibinfo {volume} {20},\ \bibinfo {editor} {edited by\
  \bibinfo {editor} {\bibfnamefont {J.}~\bibnamefont {Platt}}, \bibinfo
  {editor} {\bibfnamefont {D.}~\bibnamefont {Koller}}, \bibinfo {editor}
  {\bibfnamefont {Y.}~\bibnamefont {Singer}},\ and\ \bibinfo {editor}
  {\bibfnamefont {S.}~\bibnamefont {Roweis}}}\ (\bibinfo  {publisher} {Curran
  Associates, Inc.},\ \bibinfo {year} {2007})\BibitemShut {NoStop}%
\bibitem [{\citenamefont {Tancik}\ \emph {et~al.}(2020)\citenamefont {Tancik},
  \citenamefont {Srinivasan}, \citenamefont {Mildenhall}, \citenamefont
  {Fridovich-Keil}, \citenamefont {Raghavan}, \citenamefont {Singhal},
  \citenamefont {Ramamoorthi}, \citenamefont {Barron},\ and\ \citenamefont
  {Ng}}]{Tancik:2020Fourier}%
  \BibitemOpen
  \bibfield  {author} {\bibinfo {author} {\bibfnamefont {M.}~\bibnamefont
  {Tancik}}, \bibinfo {author} {\bibfnamefont {P.}~\bibnamefont {Srinivasan}},
  \bibinfo {author} {\bibfnamefont {B.}~\bibnamefont {Mildenhall}}, \bibinfo
  {author} {\bibfnamefont {S.}~\bibnamefont {Fridovich-Keil}}, \bibinfo
  {author} {\bibfnamefont {N.}~\bibnamefont {Raghavan}}, \bibinfo {author}
  {\bibfnamefont {U.}~\bibnamefont {Singhal}}, \bibinfo {author} {\bibfnamefont
  {R.}~\bibnamefont {Ramamoorthi}}, \bibinfo {author} {\bibfnamefont
  {J.}~\bibnamefont {Barron}},\ and\ \bibinfo {author} {\bibfnamefont
  {R.}~\bibnamefont {Ng}},\ }\bibfield  {title} {\bibinfo {title} {Fourier
  features let networks learn high frequency functions in low dimensional
  domains},\ }in\ \href
  {https://proceedings.neurips.cc/paper/2020/file/55053683268957697aa39fba6f231c68-Paper.pdf}
  {\emph {\bibinfo {booktitle} {Advances in Neural Information Processing
  Systems}}},\ Vol.~\bibinfo {volume} {33},\ \bibinfo {editor} {edited by\
  \bibinfo {editor} {\bibfnamefont {H.}~\bibnamefont {Larochelle}}, \bibinfo
  {editor} {\bibfnamefont {M.}~\bibnamefont {Ranzato}}, \bibinfo {editor}
  {\bibfnamefont {R.}~\bibnamefont {Hadsell}}, \bibinfo {editor} {\bibfnamefont
  {M.~F.}\ \bibnamefont {Balcan}},\ and\ \bibinfo {editor} {\bibfnamefont
  {H.}~\bibnamefont {Lin}}}\ (\bibinfo  {publisher} {Curran Associates, Inc.},\
  \bibinfo {year} {2020})\ pp.\ \bibinfo {pages} {7537--7547}\BibitemShut
  {NoStop}%
\bibitem [{\citenamefont {Biamonte}(2021)}]{Biamonte:2021universal}%
  \BibitemOpen
  \bibfield  {author} {\bibinfo {author} {\bibfnamefont {J.}~\bibnamefont
  {Biamonte}},\ }\bibfield  {title} {\bibinfo {title} {Universal variational
  quantum computation},\ }\href {https://doi.org/10.1103/PhysRevA.103.L030401}
  {\bibfield  {journal} {\bibinfo  {journal} {Phys. Rev. A}\ }\textbf {\bibinfo
  {volume} {103}},\ \bibinfo {pages} {L030401} (\bibinfo {year}
  {2021})}\BibitemShut {NoStop}%
\bibitem [{\citenamefont {Cerezo}\ \emph
  {et~al.}(2021{\natexlab{a}})\citenamefont {Cerezo}, \citenamefont
  {Arrasmith}, \citenamefont {Babbush}, \citenamefont {Benjamin}, \citenamefont
  {Endo}, \citenamefont {Fujii}, \citenamefont {McClean}, \citenamefont
  {Mitarai}, \citenamefont {Yuan}, \citenamefont {Cincio},\ and\ \citenamefont
  {Coles}}]{Cerezo:2021variational}%
  \BibitemOpen
  \bibfield  {author} {\bibinfo {author} {\bibfnamefont {M.}~\bibnamefont
  {Cerezo}}, \bibinfo {author} {\bibfnamefont {A.}~\bibnamefont {Arrasmith}},
  \bibinfo {author} {\bibfnamefont {R.}~\bibnamefont {Babbush}}, \bibinfo
  {author} {\bibfnamefont {S.~C.}\ \bibnamefont {Benjamin}}, \bibinfo {author}
  {\bibfnamefont {S.}~\bibnamefont {Endo}}, \bibinfo {author} {\bibfnamefont
  {K.}~\bibnamefont {Fujii}}, \bibinfo {author} {\bibfnamefont {J.~R.}\
  \bibnamefont {McClean}}, \bibinfo {author} {\bibfnamefont {K.}~\bibnamefont
  {Mitarai}}, \bibinfo {author} {\bibfnamefont {X.}~\bibnamefont {Yuan}},
  \bibinfo {author} {\bibfnamefont {L.}~\bibnamefont {Cincio}},\ and\ \bibinfo
  {author} {\bibfnamefont {P.~J.}\ \bibnamefont {Coles}},\ }\bibfield  {title}
  {\bibinfo {title} {Variational quantum algorithms},\ }\href
  {https://doi.org/10.1038/s42254-021-00348-9} {\bibfield  {journal} {\bibinfo
  {journal} {Nature Reviews Physics}\ }\textbf {\bibinfo {volume} {3}},\
  \bibinfo {pages} {625} (\bibinfo {year} {2021}{\natexlab{a}})}\BibitemShut
  {NoStop}%
\bibitem [{\citenamefont {Cao}\ \emph {et~al.}(2019)\citenamefont {Cao},
  \citenamefont {Romero}, \citenamefont {Olson}, \citenamefont {Degroote},
  \citenamefont {Johnson}, \citenamefont {Kieferová}, \citenamefont
  {Kivlichan}, \citenamefont {Menke}, \citenamefont {Peropadre}, \citenamefont
  {Sawaya}, \citenamefont {Sim}, \citenamefont {Veis},\ and\ \citenamefont
  {Aspuru-Guzik}}]{Cao:2019quantum}%
  \BibitemOpen
  \bibfield  {author} {\bibinfo {author} {\bibfnamefont {Y.}~\bibnamefont
  {Cao}}, \bibinfo {author} {\bibfnamefont {J.}~\bibnamefont {Romero}},
  \bibinfo {author} {\bibfnamefont {J.~P.}\ \bibnamefont {Olson}}, \bibinfo
  {author} {\bibfnamefont {M.}~\bibnamefont {Degroote}}, \bibinfo {author}
  {\bibfnamefont {P.~D.}\ \bibnamefont {Johnson}}, \bibinfo {author}
  {\bibfnamefont {M.}~\bibnamefont {Kieferová}}, \bibinfo {author}
  {\bibfnamefont {I.~D.}\ \bibnamefont {Kivlichan}}, \bibinfo {author}
  {\bibfnamefont {T.}~\bibnamefont {Menke}}, \bibinfo {author} {\bibfnamefont
  {B.}~\bibnamefont {Peropadre}}, \bibinfo {author} {\bibfnamefont {N.~P.~D.}\
  \bibnamefont {Sawaya}}, \bibinfo {author} {\bibfnamefont {S.}~\bibnamefont
  {Sim}}, \bibinfo {author} {\bibfnamefont {L.}~\bibnamefont {Veis}},\ and\
  \bibinfo {author} {\bibfnamefont {A.}~\bibnamefont {Aspuru-Guzik}},\
  }\bibfield  {title} {\bibinfo {title} {Quantum chemistry in the age of
  quantum computing},\ }\href {https://doi.org/10.1021/acs.chemrev.8b00803}
  {\bibfield  {journal} {\bibinfo  {journal} {Chemical Reviews}\ }\textbf
  {\bibinfo {volume} {119}},\ \bibinfo {pages} {10856} (\bibinfo {year}
  {2019})},\ \bibinfo {note} {pMID: 31469277},\ \Eprint
  {https://arxiv.org/abs/https://doi.org/10.1021/acs.chemrev.8b00803}
  {https://doi.org/10.1021/acs.chemrev.8b00803} \BibitemShut {NoStop}%
\bibitem [{\citenamefont {Endo}\ \emph {et~al.}(2021)\citenamefont {Endo},
  \citenamefont {Cai}, \citenamefont {Benjamin},\ and\ \citenamefont
  {Yuan}}]{Endo:2021hybrid}%
  \BibitemOpen
  \bibfield  {author} {\bibinfo {author} {\bibfnamefont {S.}~\bibnamefont
  {Endo}}, \bibinfo {author} {\bibfnamefont {Z.}~\bibnamefont {Cai}}, \bibinfo
  {author} {\bibfnamefont {S.~C.}\ \bibnamefont {Benjamin}},\ and\ \bibinfo
  {author} {\bibfnamefont {X.}~\bibnamefont {Yuan}},\ }\bibfield  {title}
  {\bibinfo {title} {Hybrid quantum-classical algorithms and quantum error
  mitigation},\ }\href {https://doi.org/10.7566/jpsj.90.032001} {\bibfield
  {journal} {\bibinfo  {journal} {Journal of the Physical Society of Japan}\
  }\textbf {\bibinfo {volume} {90}},\ \bibinfo {pages} {032001} (\bibinfo
  {year} {2021})}\BibitemShut {NoStop}%
\bibitem [{\citenamefont {McArdle}\ \emph {et~al.}(2020)\citenamefont
  {McArdle}, \citenamefont {Endo}, \citenamefont {Aspuru-Guzik}, \citenamefont
  {Benjamin},\ and\ \citenamefont {Yuan}}]{McArdle:2020quantum}%
  \BibitemOpen
  \bibfield  {author} {\bibinfo {author} {\bibfnamefont {S.}~\bibnamefont
  {McArdle}}, \bibinfo {author} {\bibfnamefont {S.}~\bibnamefont {Endo}},
  \bibinfo {author} {\bibfnamefont {A.}~\bibnamefont {Aspuru-Guzik}}, \bibinfo
  {author} {\bibfnamefont {S.~C.}\ \bibnamefont {Benjamin}},\ and\ \bibinfo
  {author} {\bibfnamefont {X.}~\bibnamefont {Yuan}},\ }\bibfield  {title}
  {\bibinfo {title} {Quantum computational chemistry},\ }\href
  {https://doi.org/10.1103/RevModPhys.92.015003} {\bibfield  {journal}
  {\bibinfo  {journal} {Rev. Mod. Phys.}\ }\textbf {\bibinfo {volume} {92}},\
  \bibinfo {pages} {015003} (\bibinfo {year} {2020})}\BibitemShut {NoStop}%
\bibitem [{\citenamefont {Schuld}(2021)}]{Schuld:2021supervised}%
  \BibitemOpen
  \bibfield  {author} {\bibinfo {author} {\bibfnamefont {M.}~\bibnamefont
  {Schuld}},\ }\bibfield  {title} {\bibinfo {title} {{Supervised quantum
  machine learning models are kernel methods}},\ }\href@noop {} {\bibfield
  {journal} {\bibinfo  {journal} {arXiv}\ } (\bibinfo {year} {2021})},\ \Eprint
  {https://arxiv.org/abs/2101.11020} {2101.11020} \BibitemShut {NoStop}%
\bibitem [{\citenamefont {Mitarai}\ \emph
  {et~al.}(2018{\natexlab{a}})\citenamefont {Mitarai}, \citenamefont {Negoro},
  \citenamefont {Kitagawa},\ and\ \citenamefont {Fujii}}]{Mitarai:2018quantum}%
  \BibitemOpen
  \bibfield  {author} {\bibinfo {author} {\bibfnamefont {K.}~\bibnamefont
  {Mitarai}}, \bibinfo {author} {\bibfnamefont {M.}~\bibnamefont {Negoro}},
  \bibinfo {author} {\bibfnamefont {M.}~\bibnamefont {Kitagawa}},\ and\
  \bibinfo {author} {\bibfnamefont {K.}~\bibnamefont {Fujii}},\ }\bibfield
  {title} {\bibinfo {title} {Quantum circuit learning},\ }\href
  {https://doi.org/10.1103/PhysRevA.98.032309} {\bibfield  {journal} {\bibinfo
  {journal} {Phys. Rev. A}\ }\textbf {\bibinfo {volume} {98}},\ \bibinfo
  {pages} {032309} (\bibinfo {year} {2018}{\natexlab{a}})}\BibitemShut
  {NoStop}%
\bibitem [{\citenamefont {Schuld}\ \emph {et~al.}(2019)\citenamefont {Schuld},
  \citenamefont {Bergholm}, \citenamefont {Gogolin}, \citenamefont {Izaac},\
  and\ \citenamefont {Killoran}}]{Schuld:2019evaluating}%
  \BibitemOpen
  \bibfield  {author} {\bibinfo {author} {\bibfnamefont {M.}~\bibnamefont
  {Schuld}}, \bibinfo {author} {\bibfnamefont {V.}~\bibnamefont {Bergholm}},
  \bibinfo {author} {\bibfnamefont {C.}~\bibnamefont {Gogolin}}, \bibinfo
  {author} {\bibfnamefont {J.}~\bibnamefont {Izaac}},\ and\ \bibinfo {author}
  {\bibfnamefont {N.}~\bibnamefont {Killoran}},\ }\bibfield  {title} {\bibinfo
  {title} {Evaluating analytic gradients on quantum hardware},\ }\href
  {https://doi.org/10.1103/PhysRevA.99.032331} {\bibfield  {journal} {\bibinfo
  {journal} {Phys. Rev. A}\ }\textbf {\bibinfo {volume} {99}},\ \bibinfo
  {pages} {032331} (\bibinfo {year} {2019})}\BibitemShut {NoStop}%
\bibitem [{\citenamefont {Gudenberg}(1994)}]{gudenberg:inria-00074262}%
  \BibitemOpen
  \bibfield  {author} {\bibinfo {author} {\bibfnamefont {J.~W.~V.}\
  \bibnamefont {Gudenberg}},\ }\href {https://hal.inria.fr/inria-00074262}
  {\emph {\bibinfo {title} {{Comparison of Accurate Dot Product
  Algorithms}}}},\ \bibinfo {type} {Research Report}\ \bibinfo {number}
  {RR-2413}\ (\bibinfo  {institution} {{INRIA}},\ \bibinfo {year}
  {1994})\BibitemShut {NoStop}%
\bibitem [{\citenamefont {Johnson}\ and\ \citenamefont
  {Lindenstrauss}(1984)}]{Johnson:1982extensions}%
  \BibitemOpen
  \bibfield  {author} {\bibinfo {author} {\bibfnamefont {W.~B.}\ \bibnamefont
  {Johnson}}\ and\ \bibinfo {author} {\bibfnamefont {J.}~\bibnamefont
  {Lindenstrauss}},\ }\bibfield  {title} {\bibinfo {title} {Extensions of
  {L}ipschitz mappings into a {H}ilbert space},\ }in\ \href
  {https://doi.org/10.1090/conm/026/737400} {\emph {\bibinfo {booktitle}
  {Conference in modern analysis and probability ({N}ew {H}aven, {C}onn.,
  1982)}}},\ \bibinfo {series} {Contemp. Math.}, Vol.~\bibinfo {volume} {26}\
  (\bibinfo  {publisher} {Amer. Math. Soc., Providence, RI},\ \bibinfo {year}
  {1984})\ pp.\ \bibinfo {pages} {189--206}\BibitemShut {NoStop}%
\bibitem [{\citenamefont {riewank}(2008)}]{Griewank:2008}%
  \BibitemOpen
  \bibfield  {author} {\bibinfo {author} {\bibfnamefont {A.~C.}\ \bibnamefont
  {riewank}, \bibfnamefont {A.and~Walther}},\ }\href
  {https://doi.org/https://doi.org/10.1137/1.9780898717761} {\emph {\bibinfo
  {title} {Evaluating derivatives: Principles and techniques of algorithmic
  differentiation.}}}\ (\bibinfo  {publisher} {Society for Industrial and
  Applied Mathematics (SIAM), 2},\ \bibinfo {year} {2008})\BibitemShut
  {NoStop}%
\bibitem [{\citenamefont {Margossian}(2019)}]{margossianADreview}%
  \BibitemOpen
  \bibfield  {author} {\bibinfo {author} {\bibfnamefont {C.~C.}\ \bibnamefont
  {Margossian}},\ }\bibfield  {title} {\bibinfo {title} {{A review of automatic
  differentiation and its efficient implementation}},\ }\bibfield  {journal}
  {\bibinfo  {journal} {Wiley Interdisciplinary Reviews: Data Mining and
  Knowledge Discovery}\ }\textbf {\bibinfo {volume} {9}},\ \href
  {https://doi.org/10.1002/widm.1305} {10.1002/widm.1305} (\bibinfo {year}
  {2019}),\ \Eprint {https://arxiv.org/abs/1811.05031} {1811.05031}
  \BibitemShut {NoStop}%
\bibitem [{\citenamefont {Wu}\ \emph {et~al.}(2019)\citenamefont {Wu},
  \citenamefont {Guo}, \citenamefont {Simcha}, \citenamefont {Dopson},\ and\
  \citenamefont {Kumar}}]{wu2019efficient}%
  \BibitemOpen
  \bibfield  {author} {\bibinfo {author} {\bibfnamefont {X.}~\bibnamefont
  {Wu}}, \bibinfo {author} {\bibfnamefont {R.}~\bibnamefont {Guo}}, \bibinfo
  {author} {\bibfnamefont {D.}~\bibnamefont {Simcha}}, \bibinfo {author}
  {\bibfnamefont {D.}~\bibnamefont {Dopson}},\ and\ \bibinfo {author}
  {\bibfnamefont {S.}~\bibnamefont {Kumar}},\ }\bibfield  {title} {\bibinfo
  {title} {Efficient inner product approximation in hybrid spaces},\
  }\href@noop {} {\bibfield  {journal} {\bibinfo  {journal} {arXiv preprint
  arXiv:1903.08690}\ } (\bibinfo {year} {2019})}\BibitemShut {NoStop}%
\bibitem [{\citenamefont {Shrivastava}\ and\ \citenamefont
  {Li}(2014)}]{NIPS2014_310ce61c}%
  \BibitemOpen
  \bibfield  {author} {\bibinfo {author} {\bibfnamefont {A.}~\bibnamefont
  {Shrivastava}}\ and\ \bibinfo {author} {\bibfnamefont {P.}~\bibnamefont
  {Li}},\ }\bibfield  {title} {\bibinfo {title} {Asymmetric lsh (alsh) for
  sublinear time maximum inner product search (mips)},\ }in\ \href
  {https://proceedings.neurips.cc/paper/2014/file/310ce61c90f3a46e340ee8257bc70e93-Paper.pdf}
  {\emph {\bibinfo {booktitle} {Advances in Neural Information Processing
  Systems}}},\ Vol.~\bibinfo {volume} {27},\ \bibinfo {editor} {edited by\
  \bibinfo {editor} {\bibfnamefont {Z.}~\bibnamefont {Ghahramani}}, \bibinfo
  {editor} {\bibfnamefont {M.}~\bibnamefont {Welling}}, \bibinfo {editor}
  {\bibfnamefont {C.}~\bibnamefont {Cortes}}, \bibinfo {editor} {\bibfnamefont
  {N.}~\bibnamefont {Lawrence}},\ and\ \bibinfo {editor} {\bibfnamefont
  {K.~Q.}\ \bibnamefont {Weinberger}}}\ (\bibinfo  {publisher} {Curran
  Associates, Inc.},\ \bibinfo {year} {2014})\BibitemShut {NoStop}%
\bibitem [{\citenamefont {Guo}\ \emph {et~al.}(2016)\citenamefont {Guo},
  \citenamefont {Kumar}, \citenamefont {Choromanski},\ and\ \citenamefont
  {Simcha}}]{guo2016quantization}%
  \BibitemOpen
  \bibfield  {author} {\bibinfo {author} {\bibfnamefont {R.}~\bibnamefont
  {Guo}}, \bibinfo {author} {\bibfnamefont {S.}~\bibnamefont {Kumar}}, \bibinfo
  {author} {\bibfnamefont {K.}~\bibnamefont {Choromanski}},\ and\ \bibinfo
  {author} {\bibfnamefont {D.}~\bibnamefont {Simcha}},\ }\bibfield  {title}
  {\bibinfo {title} {Quantization based fast inner product search},\ }in\
  \href@noop {} {\emph {\bibinfo {booktitle} {Artificial intelligence and
  statistics}}}\ (\bibinfo {organization} {PMLR},\ \bibinfo {year} {2016})\
  pp.\ \bibinfo {pages} {482--490}\BibitemShut {NoStop}%
\bibitem [{\citenamefont {Murphy}(2012)}]{Murphy:2012machine}%
  \BibitemOpen
  \bibfield  {author} {\bibinfo {author} {\bibfnamefont {K.~P.}\ \bibnamefont
  {Murphy}},\ }\href@noop {} {\emph {\bibinfo {title} {{M}achine {L}earning:
  {A} {P}robabilistic {P}erspective}}}\ (\bibinfo  {publisher} {The MIT
  Press},\ \bibinfo {address} {London},\ \bibinfo {year} {2012})\BibitemShut
  {NoStop}%
\bibitem [{\citenamefont {Pearson}(1901)}]{Pearson:1901lines}%
  \BibitemOpen
  \bibfield  {author} {\bibinfo {author} {\bibfnamefont {K.}~\bibnamefont
  {Pearson}},\ }\bibfield  {title} {\bibinfo {title} {On lines and planes of
  closest fit to systems of points in space},\ }\href
  {https://doi.org/10.1080/14786440109462720} {\bibfield  {journal} {\bibinfo
  {journal} {The London, Edinburgh, and Dublin Philosophical Magazine and
  Journal of Science}\ }\textbf {\bibinfo {volume} {2}},\ \bibinfo {pages}
  {559} (\bibinfo {year} {1901})},\ \Eprint
  {https://arxiv.org/abs/https://doi.org/10.1080/14786440109462720}
  {https://doi.org/10.1080/14786440109462720} \BibitemShut {NoStop}%
\bibitem [{\citenamefont {Hotelling}(1936)}]{Hotelling:1936relations}%
  \BibitemOpen
  \bibfield  {author} {\bibinfo {author} {\bibfnamefont {H.}~\bibnamefont
  {Hotelling}},\ }\bibfield  {title} {\bibinfo {title} {Relations between two
  sets of variates},\ }\href {http://www.jstor.org/stable/2333955} {\bibfield
  {journal} {\bibinfo  {journal} {Biometrika}\ }\textbf {\bibinfo {volume}
  {28}},\ \bibinfo {pages} {321} (\bibinfo {year} {1936})}\BibitemShut
  {NoStop}%
\bibitem [{\citenamefont {McClean}\ \emph {et~al.}(2018)\citenamefont
  {McClean}, \citenamefont {Boixo}, \citenamefont {Smelyanskiy}, \citenamefont
  {Babbush},\ and\ \citenamefont {Neven}}]{McClean:2018barren}%
  \BibitemOpen
  \bibfield  {author} {\bibinfo {author} {\bibfnamefont {J.~R.}\ \bibnamefont
  {McClean}}, \bibinfo {author} {\bibfnamefont {S.}~\bibnamefont {Boixo}},
  \bibinfo {author} {\bibfnamefont {V.~N.}\ \bibnamefont {Smelyanskiy}},
  \bibinfo {author} {\bibfnamefont {R.}~\bibnamefont {Babbush}},\ and\ \bibinfo
  {author} {\bibfnamefont {H.}~\bibnamefont {Neven}},\ }\bibfield  {title}
  {\bibinfo {title} {Barren plateaus in quantum neural network training
  landscapes},\ }\href {https://doi.org/10.1038/s41467-018-07090-4} {\bibfield
  {journal} {\bibinfo  {journal} {Nature Communications}\ }\textbf {\bibinfo
  {volume} {9}},\ \bibinfo {pages} {4812} (\bibinfo {year} {2018})}\BibitemShut
  {NoStop}%
\bibitem [{\citenamefont {Arrasmith}\ \emph {et~al.}(2021)\citenamefont
  {Arrasmith}, \citenamefont {Cerezo}, \citenamefont {Czarnik}, \citenamefont
  {Cincio},\ and\ \citenamefont {Coles}}]{Arrasmith:2021effect}%
  \BibitemOpen
  \bibfield  {author} {\bibinfo {author} {\bibfnamefont {A.}~\bibnamefont
  {Arrasmith}}, \bibinfo {author} {\bibfnamefont {M.}~\bibnamefont {Cerezo}},
  \bibinfo {author} {\bibfnamefont {P.}~\bibnamefont {Czarnik}}, \bibinfo
  {author} {\bibfnamefont {L.}~\bibnamefont {Cincio}},\ and\ \bibinfo {author}
  {\bibfnamefont {P.~J.}\ \bibnamefont {Coles}},\ }\bibfield  {title} {\bibinfo
  {title} {Effect of barren plateaus on gradient-free optimization},\ }\href
  {https://doi.org/10.22331/q-2021-10-05-558} {\bibfield  {journal} {\bibinfo
  {journal} {Quantum}\ }\textbf {\bibinfo {volume} {5}},\ \bibinfo {pages}
  {558} (\bibinfo {year} {2021})}\BibitemShut {NoStop}%
\bibitem [{\citenamefont {Cerezo}\ \emph
  {et~al.}(2021{\natexlab{b}})\citenamefont {Cerezo}, \citenamefont {Sone},
  \citenamefont {Volkoff}, \citenamefont {Cincio},\ and\ \citenamefont
  {Coles}}]{Cerezo:2021cost}%
  \BibitemOpen
  \bibfield  {author} {\bibinfo {author} {\bibfnamefont {M.}~\bibnamefont
  {Cerezo}}, \bibinfo {author} {\bibfnamefont {A.}~\bibnamefont {Sone}},
  \bibinfo {author} {\bibfnamefont {T.}~\bibnamefont {Volkoff}}, \bibinfo
  {author} {\bibfnamefont {L.}~\bibnamefont {Cincio}},\ and\ \bibinfo {author}
  {\bibfnamefont {P.~J.}\ \bibnamefont {Coles}},\ }\bibfield  {title} {\bibinfo
  {title} {Cost function dependent barren plateaus in shallow parametrized
  quantum circuits},\ }\href {https://doi.org/10.1038/s41467-021-21728-w}
  {\bibfield  {journal} {\bibinfo  {journal} {Nature Communications}\ }\textbf
  {\bibinfo {volume} {12}},\ \bibinfo {pages} {1791} (\bibinfo {year}
  {2021}{\natexlab{b}})}\BibitemShut {NoStop}%
\bibitem [{\citenamefont {Holmes}\ \emph {et~al.}(2022)\citenamefont {Holmes},
  \citenamefont {Sharma}, \citenamefont {Cerezo},\ and\ \citenamefont
  {Coles}}]{Holmes:2022connecting}%
  \BibitemOpen
  \bibfield  {author} {\bibinfo {author} {\bibfnamefont {Z.}~\bibnamefont
  {Holmes}}, \bibinfo {author} {\bibfnamefont {K.}~\bibnamefont {Sharma}},
  \bibinfo {author} {\bibfnamefont {M.}~\bibnamefont {Cerezo}},\ and\ \bibinfo
  {author} {\bibfnamefont {P.~J.}\ \bibnamefont {Coles}},\ }\bibfield  {title}
  {\bibinfo {title} {Connecting ansatz expressibility to gradient magnitudes
  and barren plateaus},\ }\href {https://doi.org/10.1103/PRXQuantum.3.010313}
  {\bibfield  {journal} {\bibinfo  {journal} {PRX Quantum}\ }\textbf {\bibinfo
  {volume} {3}},\ \bibinfo {pages} {010313} (\bibinfo {year}
  {2022})}\BibitemShut {NoStop}%
\bibitem [{\citenamefont {Thanasilp}\ \emph {et~al.}(2021)\citenamefont
  {Thanasilp}, \citenamefont {Wang}, \citenamefont {Nghiem}, \citenamefont
  {Coles},\ and\ \citenamefont {Cerezo}}]{Thanasilp:2021subtleties}%
  \BibitemOpen
  \bibfield  {author} {\bibinfo {author} {\bibfnamefont {S.}~\bibnamefont
  {Thanasilp}}, \bibinfo {author} {\bibfnamefont {S.}~\bibnamefont {Wang}},
  \bibinfo {author} {\bibfnamefont {N.~A.}\ \bibnamefont {Nghiem}}, \bibinfo
  {author} {\bibfnamefont {P.~J.}\ \bibnamefont {Coles}},\ and\ \bibinfo
  {author} {\bibfnamefont {M.}~\bibnamefont {Cerezo}},\ }\href
  {https://doi.org/10.48550/ARXIV.2110.14753} {\bibinfo {title} {Subtleties in
  the trainability of quantum machine learning models}} (\bibinfo {year}
  {2021})\BibitemShut {NoStop}%
\bibitem [{\citenamefont {Harrow}\ and\ \citenamefont
  {Low}(2009)}]{harrow_random_2009}%
  \BibitemOpen
  \bibfield  {author} {\bibinfo {author} {\bibfnamefont {A.~W.}\ \bibnamefont
  {Harrow}}\ and\ \bibinfo {author} {\bibfnamefont {R.~A.}\ \bibnamefont
  {Low}},\ }\bibfield  {title} {\bibinfo {title} {Random {Quantum} {Circuits}
  are {Approximate} 2-designs},\ }\href
  {https://doi.org/10.1007/s00220-009-0873-6} {\bibfield  {journal} {\bibinfo
  {journal} {Communications in Mathematical Physics}\ }\textbf {\bibinfo
  {volume} {291}},\ \bibinfo {pages} {257} (\bibinfo {year}
  {2009})}\BibitemShut {NoStop}%
\bibitem [{\citenamefont {Pucha{\l}a}\ and\ \citenamefont
  {Miszczak}(2017)}]{Puchala_Z._Symbolic_2017}%
  \BibitemOpen
  \bibfield  {author} {\bibinfo {author} {\bibfnamefont {Z.}~\bibnamefont
  {Pucha{\l}a}}\ and\ \bibinfo {author} {\bibfnamefont {J.}~\bibnamefont
  {Miszczak}},\ }\bibfield  {title} {\bibinfo {title} {Symbolic integration
  with respect to the haar measure on the unitary groups},\ }\href
  {https://doi.org/10.1515/bpasts-2017-0003} {\bibfield  {journal} {\bibinfo
  {journal} {Bulletin of the Polish Academy of Sciences: Technical Sciences}\
  }\textbf {\bibinfo {volume} {65}},\ \bibinfo {pages} {21} (\bibinfo {year}
  {2017})}\BibitemShut {NoStop}%
\bibitem [{\citenamefont {Christensen}\ and\ \citenamefont
  {lilienfeld}(2020)}]{Christensen2020}%
  \BibitemOpen
  \bibfield  {author} {\bibinfo {author} {\bibfnamefont {A.~S.}\ \bibnamefont
  {Christensen}}\ and\ \bibinfo {author} {\bibfnamefont {A.~V.}\ \bibnamefont
  {lilienfeld}},\ }\bibfield  {title} {\bibinfo {title} {Revised md17 dataset
  (rmd17)}\ }\href {https://doi.org/10.6084/m9.figshare.12672038.v3}
  {10.6084/m9.figshare.12672038.v3} (\bibinfo {year} {2020})\BibitemShut
  {NoStop}%
\bibitem [{\citenamefont {Chmiela}\ \emph {et~al.}(2017)\citenamefont
  {Chmiela}, \citenamefont {Tkatchenko}, \citenamefont {Sauceda}, \citenamefont
  {Poltavsky}, \citenamefont {Schütt},\ and\ \citenamefont {Müller}}]{MD17}%
  \BibitemOpen
  \bibfield  {author} {\bibinfo {author} {\bibfnamefont {S.}~\bibnamefont
  {Chmiela}}, \bibinfo {author} {\bibfnamefont {A.}~\bibnamefont {Tkatchenko}},
  \bibinfo {author} {\bibfnamefont {H.~E.}\ \bibnamefont {Sauceda}}, \bibinfo
  {author} {\bibfnamefont {I.}~\bibnamefont {Poltavsky}}, \bibinfo {author}
  {\bibfnamefont {K.~T.}\ \bibnamefont {Schütt}},\ and\ \bibinfo {author}
  {\bibfnamefont {K.-R.}\ \bibnamefont {Müller}},\ }\bibfield  {title}
  {\bibinfo {title} {Machine learning of accurate energy-conserving molecular
  force fields},\ }\href {https://doi.org/10.1126/sciadv.1603015} {\bibfield
  {journal} {\bibinfo  {journal} {Science Advances}\ }\textbf {\bibinfo
  {volume} {3}},\ \bibinfo {pages} {e1603015} (\bibinfo {year} {2017})},\
  \Eprint
  {https://arxiv.org/abs/https://www.science.org/doi/pdf/10.1126/sciadv.1603015}
  {https://www.science.org/doi/pdf/10.1126/sciadv.1603015} \BibitemShut
  {NoStop}%
\bibitem [{\citenamefont {Rupp}\ \emph {et~al.}(2012)\citenamefont {Rupp},
  \citenamefont {Tkatchenko}, \citenamefont {M\"uller},\ and\ \citenamefont
  {von Lilienfeld}}]{Rupp:2012fast}%
  \BibitemOpen
  \bibfield  {author} {\bibinfo {author} {\bibfnamefont {M.}~\bibnamefont
  {Rupp}}, \bibinfo {author} {\bibfnamefont {A.}~\bibnamefont {Tkatchenko}},
  \bibinfo {author} {\bibfnamefont {K.-R.}\ \bibnamefont {M\"uller}},\ and\
  \bibinfo {author} {\bibfnamefont {O.~A.}\ \bibnamefont {von Lilienfeld}},\
  }\bibfield  {title} {\bibinfo {title} {Fast and accurate modeling of
  molecular atomization energies with machine learning},\ }\href
  {https://doi.org/10.1103/PhysRevLett.108.058301} {\bibfield  {journal}
  {\bibinfo  {journal} {Phys. Rev. Lett.}\ }\textbf {\bibinfo {volume} {108}},\
  \bibinfo {pages} {058301} (\bibinfo {year} {2012})}\BibitemShut {NoStop}%
\bibitem [{\citenamefont {Bergholm}\ \emph {et~al.}(2018)\citenamefont
  {Bergholm}, \citenamefont {Izaac}, \citenamefont {Schuld}, \citenamefont
  {Gogolin}, \citenamefont {Ahmed}, \citenamefont {Ajith}, \citenamefont
  {Alam}, \citenamefont {Alonso-Linaje}, \citenamefont {AkashNarayanan},
  \citenamefont {Asadi}, \citenamefont {Arrazola}, \citenamefont {Azad},
  \citenamefont {Banning}, \citenamefont {Blank}, \citenamefont {Bromley},
  \citenamefont {Cordier}, \citenamefont {Ceroni}, \citenamefont {Delgado},
  \citenamefont {Di~Matteo}, \citenamefont {Dusko}, \citenamefont {Garg},
  \citenamefont {Guala}, \citenamefont {Hayes}, \citenamefont {Hill},
  \citenamefont {Ijaz}, \citenamefont {Isacsson}, \citenamefont {Ittah},
  \citenamefont {Jahangiri}, \citenamefont {Jain}, \citenamefont {Jiang},
  \citenamefont {Khandelwal}, \citenamefont {Kottmann}, \citenamefont {Lang},
  \citenamefont {Lee}, \citenamefont {Loke}, \citenamefont {Lowe},
  \citenamefont {McKiernan}, \citenamefont {Meyer}, \citenamefont
  {Montañez-Barrera}, \citenamefont {Moyard}, \citenamefont {Niu},
  \citenamefont {O'Riordan}, \citenamefont {Oud}, \citenamefont {Panigrahi},
  \citenamefont {Park}, \citenamefont {Polatajko}, \citenamefont {Quesada},
  \citenamefont {Roberts}, \citenamefont {Sá}, \citenamefont {Schoch},
  \citenamefont {Shi}, \citenamefont {Shu}, \citenamefont {Sim}, \citenamefont
  {Singh}, \citenamefont {Strandberg}, \citenamefont {Soni}, \citenamefont
  {Száva}, \citenamefont {Thabet}, \citenamefont {Vargas-Hernández},
  \citenamefont {Vincent}, \citenamefont {Vitucci}, \citenamefont {Weber},
  \citenamefont {Wierichs}, \citenamefont {Wiersema}, \citenamefont {Willmann},
  \citenamefont {Wong}, \citenamefont {Zhang},\ and\ \citenamefont
  {Killoran}}]{pennylane}%
  \BibitemOpen
  \bibfield  {author} {\bibinfo {author} {\bibfnamefont {V.}~\bibnamefont
  {Bergholm}}, \bibinfo {author} {\bibfnamefont {J.}~\bibnamefont {Izaac}},
  \bibinfo {author} {\bibfnamefont {M.}~\bibnamefont {Schuld}}, \bibinfo
  {author} {\bibfnamefont {C.}~\bibnamefont {Gogolin}}, \bibinfo {author}
  {\bibfnamefont {S.}~\bibnamefont {Ahmed}}, \bibinfo {author} {\bibfnamefont
  {V.}~\bibnamefont {Ajith}}, \bibinfo {author} {\bibfnamefont {M.~S.}\
  \bibnamefont {Alam}}, \bibinfo {author} {\bibfnamefont {G.}~\bibnamefont
  {Alonso-Linaje}}, \bibinfo {author} {\bibfnamefont {B.}~\bibnamefont
  {AkashNarayanan}}, \bibinfo {author} {\bibfnamefont {A.}~\bibnamefont
  {Asadi}}, \bibinfo {author} {\bibfnamefont {J.~M.}\ \bibnamefont {Arrazola}},
  \bibinfo {author} {\bibfnamefont {U.}~\bibnamefont {Azad}}, \bibinfo {author}
  {\bibfnamefont {S.}~\bibnamefont {Banning}}, \bibinfo {author} {\bibfnamefont
  {C.}~\bibnamefont {Blank}}, \bibinfo {author} {\bibfnamefont {T.~R.}\
  \bibnamefont {Bromley}}, \bibinfo {author} {\bibfnamefont {B.~A.}\
  \bibnamefont {Cordier}}, \bibinfo {author} {\bibfnamefont {J.}~\bibnamefont
  {Ceroni}}, \bibinfo {author} {\bibfnamefont {A.}~\bibnamefont {Delgado}},
  \bibinfo {author} {\bibfnamefont {O.}~\bibnamefont {Di~Matteo}}, \bibinfo
  {author} {\bibfnamefont {A.}~\bibnamefont {Dusko}}, \bibinfo {author}
  {\bibfnamefont {T.}~\bibnamefont {Garg}}, \bibinfo {author} {\bibfnamefont
  {D.}~\bibnamefont {Guala}}, \bibinfo {author} {\bibfnamefont
  {A.}~\bibnamefont {Hayes}}, \bibinfo {author} {\bibfnamefont
  {R.}~\bibnamefont {Hill}}, \bibinfo {author} {\bibfnamefont {A.}~\bibnamefont
  {Ijaz}}, \bibinfo {author} {\bibfnamefont {T.}~\bibnamefont {Isacsson}},
  \bibinfo {author} {\bibfnamefont {D.}~\bibnamefont {Ittah}}, \bibinfo
  {author} {\bibfnamefont {S.}~\bibnamefont {Jahangiri}}, \bibinfo {author}
  {\bibfnamefont {P.}~\bibnamefont {Jain}}, \bibinfo {author} {\bibfnamefont
  {E.}~\bibnamefont {Jiang}}, \bibinfo {author} {\bibfnamefont
  {A.}~\bibnamefont {Khandelwal}}, \bibinfo {author} {\bibfnamefont
  {K.}~\bibnamefont {Kottmann}}, \bibinfo {author} {\bibfnamefont {R.~A.}\
  \bibnamefont {Lang}}, \bibinfo {author} {\bibfnamefont {C.}~\bibnamefont
  {Lee}}, \bibinfo {author} {\bibfnamefont {T.}~\bibnamefont {Loke}}, \bibinfo
  {author} {\bibfnamefont {A.}~\bibnamefont {Lowe}}, \bibinfo {author}
  {\bibfnamefont {K.}~\bibnamefont {McKiernan}}, \bibinfo {author}
  {\bibfnamefont {J.~J.}\ \bibnamefont {Meyer}}, \bibinfo {author}
  {\bibfnamefont {J.~A.}\ \bibnamefont {Montañez-Barrera}}, \bibinfo {author}
  {\bibfnamefont {R.}~\bibnamefont {Moyard}}, \bibinfo {author} {\bibfnamefont
  {Z.}~\bibnamefont {Niu}}, \bibinfo {author} {\bibfnamefont {L.~J.}\
  \bibnamefont {O'Riordan}}, \bibinfo {author} {\bibfnamefont {S.}~\bibnamefont
  {Oud}}, \bibinfo {author} {\bibfnamefont {A.}~\bibnamefont {Panigrahi}},
  \bibinfo {author} {\bibfnamefont {C.-Y.}\ \bibnamefont {Park}}, \bibinfo
  {author} {\bibfnamefont {D.}~\bibnamefont {Polatajko}}, \bibinfo {author}
  {\bibfnamefont {N.}~\bibnamefont {Quesada}}, \bibinfo {author} {\bibfnamefont
  {C.}~\bibnamefont {Roberts}}, \bibinfo {author} {\bibfnamefont
  {N.}~\bibnamefont {Sá}}, \bibinfo {author} {\bibfnamefont {I.}~\bibnamefont
  {Schoch}}, \bibinfo {author} {\bibfnamefont {B.}~\bibnamefont {Shi}},
  \bibinfo {author} {\bibfnamefont {S.}~\bibnamefont {Shu}}, \bibinfo {author}
  {\bibfnamefont {S.}~\bibnamefont {Sim}}, \bibinfo {author} {\bibfnamefont
  {A.}~\bibnamefont {Singh}}, \bibinfo {author} {\bibfnamefont
  {I.}~\bibnamefont {Strandberg}}, \bibinfo {author} {\bibfnamefont
  {J.}~\bibnamefont {Soni}}, \bibinfo {author} {\bibfnamefont {A.}~\bibnamefont
  {Száva}}, \bibinfo {author} {\bibfnamefont {S.}~\bibnamefont {Thabet}},
  \bibinfo {author} {\bibfnamefont {R.~A.}\ \bibnamefont {Vargas-Hernández}},
  \bibinfo {author} {\bibfnamefont {T.}~\bibnamefont {Vincent}}, \bibinfo
  {author} {\bibfnamefont {N.}~\bibnamefont {Vitucci}}, \bibinfo {author}
  {\bibfnamefont {M.}~\bibnamefont {Weber}}, \bibinfo {author} {\bibfnamefont
  {D.}~\bibnamefont {Wierichs}}, \bibinfo {author} {\bibfnamefont
  {R.}~\bibnamefont {Wiersema}}, \bibinfo {author} {\bibfnamefont
  {M.}~\bibnamefont {Willmann}}, \bibinfo {author} {\bibfnamefont
  {V.}~\bibnamefont {Wong}}, \bibinfo {author} {\bibfnamefont {S.}~\bibnamefont
  {Zhang}},\ and\ \bibinfo {author} {\bibfnamefont {N.}~\bibnamefont
  {Killoran}},\ }\href {https://doi.org/10.48550/ARXIV.1811.04968} {\bibinfo
  {title} {Pennylane: Automatic differentiation of hybrid quantum-classical
  computations}} (\bibinfo {year} {2018}),\ \Eprint
  {https://arxiv.org/abs/arXiv:1811.04968} {arXiv:1811.04968} \BibitemShut
  {NoStop}%
\bibitem [{\citenamefont {Peters}\ and\ \citenamefont
  {Schuld}()}]{Peters:2022generalization}%
  \BibitemOpen
  \bibfield  {author} {\bibinfo {author} {\bibfnamefont {E.}~\bibnamefont
  {Peters}}\ and\ \bibinfo {author} {\bibfnamefont {M.}~\bibnamefont
  {Schuld}},\ }\href {https://doi.org/10.48550/ARXIV.2209.05523} {\bibinfo
  {title} {Generalization despite overfitting in quantum machine learning
  models}},\ \Eprint {https://arxiv.org/abs/arXiv:2209.05523}
  {arXiv:2209.05523} \BibitemShut {NoStop}%
\bibitem [{\citenamefont {Franz J.~Schreiber}()}]{classicalsurrogate}%
  \BibitemOpen
  \bibfield  {author} {\bibinfo {author} {\bibfnamefont {J.~J.~M.}\
  \bibnamefont {Franz J.~Schreiber}, \bibfnamefont {Jens~Eisert}},\ }\bibfield
  {title} {\bibinfo {title} {Classical surrogates for quantum learning models}\
  }\href {https://doi.org/10.48550/arXiv.2206.11740}
  {10.48550/arXiv.2206.11740}\BibitemShut {NoStop}%
\bibitem [{\citenamefont {Mitarai}\ \emph
  {et~al.}(2018{\natexlab{b}})\citenamefont {Mitarai}, \citenamefont {Negoro},
  \citenamefont {Kitagawa},\ and\ \citenamefont {Fujii}}]{Mitarai2018QCL}%
  \BibitemOpen
  \bibfield  {author} {\bibinfo {author} {\bibfnamefont {K.}~\bibnamefont
  {Mitarai}}, \bibinfo {author} {\bibfnamefont {M.}~\bibnamefont {Negoro}},
  \bibinfo {author} {\bibfnamefont {M.}~\bibnamefont {Kitagawa}},\ and\
  \bibinfo {author} {\bibfnamefont {K.}~\bibnamefont {Fujii}},\ }\bibfield
  {title} {\bibinfo {title} {Quantum circuit learning},\ }\href
  {https://doi.org/10.1103/PhysRevA.98.032309} {\bibfield  {journal} {\bibinfo
  {journal} {Phys. Rev. A}\ }\textbf {\bibinfo {volume} {98}},\ \bibinfo
  {pages} {032309} (\bibinfo {year} {2018}{\natexlab{b}})}\BibitemShut
  {NoStop}%
\bibitem [{\citenamefont {Caro}\ \emph
  {et~al.}(2021{\natexlab{b}})\citenamefont {Caro}, \citenamefont {Gil-Fuster},
  \citenamefont {Meyer}, \citenamefont {Eisert},\ and\ \citenamefont
  {Sweke}}]{Caro:2021encoding-dependent}%
  \BibitemOpen
  \bibfield  {author} {\bibinfo {author} {\bibfnamefont {M.~C.}\ \bibnamefont
  {Caro}}, \bibinfo {author} {\bibfnamefont {E.}~\bibnamefont {Gil-Fuster}},
  \bibinfo {author} {\bibfnamefont {J.~J.}\ \bibnamefont {Meyer}}, \bibinfo
  {author} {\bibfnamefont {J.}~\bibnamefont {Eisert}},\ and\ \bibinfo {author}
  {\bibfnamefont {R.}~\bibnamefont {Sweke}},\ }\bibfield  {title} {\bibinfo
  {title} {Encoding-dependent generalization bounds for parametrized quantum
  circuits},\ }\href {https://doi.org/10.22331/q-2021-11-17-582} {\bibfield
  {journal} {\bibinfo  {journal} {Quantum}\ }\textbf {\bibinfo {volume} {5}},\
  \bibinfo {pages} {582} (\bibinfo {year} {2021}{\natexlab{b}})}\BibitemShut
  {NoStop}%
\bibitem [{Note1()}]{Note1}%
  \BibitemOpen
  \bibinfo {note} {Visit \protect \url {https://pennylane.ai/}.}\BibitemShut
  {Stop}%
\bibitem [{Note2()}]{Note2}%
  \BibitemOpen
  \bibinfo {note} {Visit the official MNIST website at \protect \url
  {http://yann.lecun.com/exdb/mnist/}.}\BibitemShut {Stop}%
\bibitem [{JLn()}]{JLnotes}%
  \BibitemOpen
  \href@noop {} {\bibinfo {title} {Lecture notes on this lemma may be found in
  the following url:
  \url{https://www.cs.princeton.edu/~smattw/Teaching/Fa19Lectures/lec9/lec9.pdf}}}\BibitemShut
  {NoStop}%
\end{thebibliography}
\end{document}